\documentclass{aa}  
\usepackage{graphicx}
\usepackage{txfonts}
\def\3cm{\rm {cm^{-3}}}
\def\2cm{\rm {cm^{-2}}}
\def\s-1{\rm {s^{-1}}}
\def\etal {et al.}
\def\kms {\hbox{${\rm km\,s}^{-1}$}}

\begin{document}
\title{HNC, HCN and CN in Seyfert galaxies\thanks{Research supported by the Swedish Institute as a scholarship granted to J.P. P\'erez-Beaupuits to pursue his M.Sc. degree at Chalmers Tekniska H\"ogskola, Sweden}}
\subtitle{}
\author{J.P. P\'{e}rez-Beaupuits\inst{1,2} \and 
        S.~Aalto\inst{2} \and 
	H. Gerebro\inst{2}}
\offprints{J.P. P\'erez-Beaupuits}
\institute{
 Kapteyn Astronomical Institute, Rijksuniversiteit Groningen, 9700 AV Groningen, The Netherlands -  
 \email{jp@astro.rug.nl}
 \and
 Onsala Rymdobservatorium, Chalmers Tekniska H\"ogskola, S - 439 92 Onsala, Sweden - 
 \email{susanne@oso.chalmers.se}
}
\date{Received 14 August 2007 / Accepted  27 September 2007}
\titlerunning{HNC, HCN and CN in Seyferts}

%
\abstract
  {}    
  {Bright HNC 1--0 emission, rivalling that of HCN 1--0, 
  has been found towards several Seyfert galaxies. This is 
  unexpected since traditionally HNC is a tracer of cold (10 K) gas, and the 
  molecular gas of luminous galaxies like Seyferts is thought to have bulk 
  kinetic temperatures surpassing 50 K. There are four possible explanations for the bright HNC:
  (a) Large masses of hidden cold gas; (b) chemistry dominated by ion-neutral reactions;
  (c) chemistry dominated by X-ray radiation; and (d) HNC enhanced through mid-IR pumping.
  In this work we aim to distinguish the cause of the bright HNC and to model the physical conditions of the 
  HNC and HCN emitting gas.}
  {We have used SEST, JCMT and IRAM 30m telescopes to observe HNC 3--2 and HCN 3--2 line emission 
  in a selection of 5 HNC-luminous Seyfert galaxies. We estimate and discuss
  the excitation conditions of HCN and HNC in NGC~1068, NGC~3079, NGC~2623 and NGC~7469, 
  based on the observed 3--2/1--0 line intensity ratios. We also observed 
  CN 1--0 and 2--1 emission and discuss its role in photon and X-ray dominated regions.}
  {HNC 3--2 was detected in 3 galaxies (NGC~3079, NGC~1068 and NGC~2623). Not detected in NGC~7469. 
  HCN 3--2 was detected in NGC~3079, NGC~1068 and NGC~1365, it was not detected in NGC~2623. 
  The HCN 3--2/1--0 ratio is lower than 0.3 only in NGC~3079, whereas the HNC 3--2/1--0
  ratio is larger than 0.3 only in NGC~2623. The HCN/HNC 1--0 and 3--2 line ratios are larger 
  than unity in all the galaxies.
  The HCN/HNC 3--2 line ratio is lower than unity only in NGC~2623, which makes it comparable to
  galaxies like Arp~220, Mrk~231 and NGC~4418.
  }
  {We conclude that in three of the galaxies the HNC emissions emerge 
  from gas of densities $n\lesssim10^5~\3cm$, where the chemistry is 
  dominated by ion-neutral reactions.
  The line shapes observed in NGC~1365 and NGC~3079 show that these 
  galaxies have no circumnuclear disk. 
  In NGC~1068 the emission of HNC emerges from lower ($<10^5~\3cm$) density  
  gas than HCN ($>10^5~\3cm$). Instead, we conclude that the emissions of HNC 
  and HCN emerge from the same gas in NGC~3079. 
  The observed HCN/HNC and CN/HCN line ratios favor a PDR scenario, rather 
  than an XDR one, which is consistent with previous indications of a starburst 
  component in the central regions of these galaxies. However, the 
  $N({\rm HNC})/N({\rm HCN})$ column density ratios obtained for 
  NGC~3079 can be found only in XDR environments.}
  

\maketitle

\section{Introduction}

The hydrogene cyanide, HCN molecule, is commonly used as an extragalactic tracer 
of molecular gas with densities $n$(H$_2$) larger than $10^4~\3cm$ 
(e.g. Solomon et al. \cite{solomon92}; Curran et al. \cite{curran00}; 
Kohno \cite{kohno05}). The HCN to CO intensity ratio varies significantly, 
from 1/3 to 1/40 in starburst galaxies, and it has not been determined 
whether this variation depends on the dense molecular gas content or on 
the abundance and/or excitation conditions. In addition, recent results 
seems to indicate that HCN may not be an unbiased tracer of the dense 
molecular gas content in LIRGs and ULIRGs (Graci\'a-Carpio et al. \cite{gracia06}). 
It is essential, therefore, to use other molecular tracers than HCN, in 
order to understand the physical conditions in the dense gas.

A molecule of particular interest, for comparison with HCN, is its isomer 
HNC. The detection of interstellar HNC supports the theory of dominant 
ion-molecule chemistry in dark molecular clouds. Both species are thought 
to be created by the same dissociative recombination of HCNH$^+$. This ion 
can produce HCN and HNC, with approximately equal abundances. Models based 
only on this scheme would predict then an HNC/HCN ratio $\approx 1$. However, the 
observed HNC/HCN abundance ratios vary significantly between different kinds 
of molecular clouds - the ratio ranges from 0.03 to 0.4 in warm cores ($T_k > 15$ K), 
and can be as high as 4.4 in cold cores ($T_k < 15$ K).

The CN (cyanogen radical) molecule is another tracer of dense gas, with a lower 
(by a factor of 5) critical density than HCN. CN is also chemically linked to 
HCN and HNC by photodissociations (e.g. Hirota \& Yamamoto \cite{hirota99}). Surveys 
of the 1--0 transition of CN and HNC have been done in order to trace a cold, 
dense phase of the gas in luminous galaxies (Aalto et al. \cite{aalto02}). It was 
found that the HNC 1--0 luminosities often rivalled those of HCN 1--0. 
These results seem to contradict the idea of warm ($T_k \gtrsim 50$ K) gas in the 
centers of luminous galaxies (e.g. Wild et al. \cite{wild92}; Wall et al. 
\cite{wall93}) whose IR luminosities were suggested to originate from star formation 
rather than AGN activity (Solomon et al. \cite{solomon92}).

\begin{table*}[!t]
   
   \begin{minipage}{18cm}   
      \caption[]{Sample of galaxies$^{~\rm a}$.}
         \label{tab:galaxies}
	\centering
         \begin{tabular}{lccccccc}
            \hline
            \noalign{\smallskip}
            Galaxy & Seyfert  & RA & DEC & v$_{sys}$ & Distance$^{~\rm b}$ & $\Omega_S$(CO)$^{~\rm c}$ & $\Omega_S$(HCN)$^{~\rm d}$\\
                   &     & [hh mm ss] & [$^{\circ}~{}'~{}''$] & [\kms] & [Mpc] & [$''~^2$] & [$''~^2$]\\  
            \noalign{\smallskip}
            \hline
            \noalign{\smallskip}
            NGC~3079 & 2   & 10 01 57.805 & +55 40 47.20 & 1116$\pm$1 & 15.0$\pm$1.1 & $15\times 7.5$ & $5\times 5$ \\
            NGC~1068 & 2   & 02 42 40.711 & -00 00 47.81 & 1137$\pm$3 & 15.3$\pm$1.1 & $30\times 30$ & $10\times 10$ \\
            NGC~2623 & 2   & 08 38 24.090 & +25 45 16.80 & 5549$\pm$1 & 74.6$\pm$5.4 & $8\times 8$ & $2.6\times 2.6$ \\
            NGC~1365 & 1.8 & 03 33 36.371 & -36 08 25.45 & 1636$\pm$1 & 22.0$\pm$1.6 & $50\times 50$ & $16.5\times 16.5$ \\
            NGC~7469 & 1.2 & 23 03 15.623 & +08 52 26.39 & 4892$\pm$2 & 65.8$\pm$4.8 & $8\times 8$ & $4\times 6$ \\
            \noalign{\smallskip}
            \hline
         \end{tabular}
\begin{list}{}{}
\item[${\mathrm{a}}$)] The Seyfert classification, positions (in equatorial 
J2000 coordinates) and heliocentric radial velocities were taken from NED.

\item[${\mathrm{b}}$)] The distances were calculated using the Hubble constant 
(H$_0 \approx 74.37$ \kms~Mpc$^{-1}$) estimated by Ngeow and Kanbur (\cite{ngeow06}).

\item[${\mathrm{c}}$)] The source sizes of the CO 1--0 transition line were 
estimated from the maps presented in (Koda et al. \cite{koda02}) for NGC~3079, 
(Schinnerer et al. \cite{schinnerer00}) for NGC~1068, (Bryant et al. \cite{bryant99}) 
for NGC~2623, (Sandqvist \cite{sandqvist99}) for NGC~1365, and (Papadopoulos \& 
Allen \cite{papadopoulos00}) for NGC~7469. The source sizes for the $J = 2-1$ 
transition line were assumed equal to that of the $J = 1-0$ line. For NGC~7469, 
the source size of the CO 2--1 emission estimated from the corresponding map 
presented by Davies et al. (\cite{davies04}) agrees well with the source size
estimated for the CO 1--0 line.

\item[${\mathrm{d}}$)] Source sizes of HCN 1--0 were estimated from the corresponding 
maps published in (Kohno et al. \cite{kohno00}) for NGC~3079, (Kohno et al. 
\cite{kohno01} and Helfer \& Blitz \cite{helfer95}) for NGC~1068, (Davies 
et al. \cite{davies04}) for NGC~7469. The source sizes of NGC~2623 and NGC~1365 
were estimated using proportions found in NGC~1068 (read text in \S3.5). 
Because of their chemical link, the source sizes of the CN and HNC molecules 
were assumed the same as that of HCN. Due to the lack of high resolution maps, 
the source sizes corresponding to the emission of the higher transition lines 
were assumed equal to that of the $J = 1-0$ line.
\end{list}          

\end{minipage}
   
\end{table*}

According to observations in the vicinity of the hot core of Orion KL, experimental 
data, chemical steady state and shock models, the HNC/HCN ratio decreases as the 
temperature and density increase (e.g. Schilke et al. \cite{schilke92}; Talbi et al. 
\cite{talbi96}; Tachikawa et al. \cite{tachikawa03}). If a bright HNC 1--0 transition 
line is nevertheless detected under these conditions, it could be due to the following 
possible explanations: (a) the presence of large masses of hidden cold gas and dust at 
high densities ($n > 10^5~\3cm$); (b) chemistry dominated by ion-molecule reactions 
with HCNH$^+$ at low density ($n \approx 10^4~\3cm$) in regions where the temperature 
dependence of the HNC abundance becomes weaker; (c) enhancement by mid-IR pumping, also 
in low density regions where the lines would not be collisionally excited; and (d) the 
influence of UV-rays in Photon Dominated Regions (PDRs) and/or X-rays in X-ray Dominated 
Regions (XDRs) at densities $n \gtrsim 10^4~\3cm$ and at total column densities 
$N_{\rm H} > 3\times 10^{21}~\2cm$ (Meijerink \& Spaans \cite{meijerink05}).

In the case of CN, observations of its emission towards the Orion A molecular 
complex (Rodr\'iguez-Franco et al. \cite{rodriguez98}) suggest that this 
molecule is also enhanced in PDRs, but particularly in XDRs, where a CN/HCN 
abundance ratio larger than unity is expected (e.g., Lepp \& Dalgarno \cite{lepp96} and
Meijerink, Spaans, Israel \cite{meijerink07}).

We have observed low and high transition lines of the HCN, HNC and CN molecules 
in a group of Seyfert galaxies, which are supposed to host both sources of power, 
AGN and starburst activity, in their central region. Our interest is to assess 
the excitation conditions of HCN and HNC, distinguish between the above 
possible causes of the bright HNC, and to explore the relation between the 
CN emission, XDRs and dense PDRs in these sources.

In \S2 we describe the observations. The results (spectral lines, line 
intensities and line ratios) are presented in \S3. The interpretation of 
line shapes and gas distribution in the most relevant cases, as well as 
the possible explanations for the bright HNC and the modelling of the 
excitation conditions of HCN and HNC are discussed in \S4. 
The conclusions and final remarks of this work are presented in \S5.

\section{Observations}

We have used the James Clerk Maxwell Telescope (JCMT) in 2005 to observe 
the HNC $J$=3--2 (271~GHz) and the HCN $J$=3--2 (267~GHz) lines towards a 
sample of Seyfert galaxies. Observations of CN and HNC $J$=1--0 (90~GHz) 
were made in 2002 using the Swedish ESO Southern Telescope (SEST). HCN 
$J$=1--0 (88~GHz) data from literature were also used. The system temperatures 
ranged between 350 K and 430 K. In the case of CN $J$=1--0 (226~GHz) the 
weather conditions were not so good, making the system temperature range 
between 490 K and 760 K. Pointing was checked regularly on SiO masers and 
the rms was found to be about 3$''$.

   \begin{table}[!t]
      \caption[]{Beamsizes \& efficiencies. }
         \label{tab:beams}
         \begin{tabular}{lcccc}
            \hline
            \noalign{\smallskip}
            Transition & $\nu$ & HPBW$^{~\rm a}$ & $\eta_{\rm mb}$~$^{~\rm a}$ & Telescope$^{~\rm a}$\\
	               & [GHz] &    [$''$]       &                             &   \\
            \noalign{\smallskip}
            \hline
            \noalign{\smallskip}
            HCN~1-0 &  88.632 & 44 ~28& 0.65 ~0.77& OSO  ~IRAM\\
            HNC~1-0 &  90.663 & 55 ~27& 0.74 ~0.77& SEST ~IRAM\\
            CN~1-0  & 113.491 & 45 ~22& 0.71 ~0.74& SEST ~IRAM\\
            CO~1-0  & 115.271 & 44 ~21& 0.70 ~0.74& SEST ~IRAM\\
            CN~2-1  & 226.875 & 23 ~11& 0.51 ~0.53& SEST ~IRAM\\
            CO~2-1  & 230.538 & 20 ~11& 0.66 ~0.52& JCMT ~IRAM\\
            HCN~3-2 & 265.886 & 18 ~9 & 0.60 ~0.45& JCMT ~IRAM\\
            HNC~3-2 & 271.981 & 18 ~9 & 0.60 ~0.44& JCMT ~IRAM\\            
            \noalign{\smallskip}
            \hline
         \end{tabular}
\begin{list}{}{}
\item[${\mathrm{a}}$)] The columns are divided in two sub-columns 
(\emph{left}: OSO, SEST, JCMT, and \emph{right} IRAM) indicating the 
telescope used to obtain the corresponding parameter. 
The IRAM 30m telescope was used to observe NGC~3079. The other sources 
were observed with the OSO 20m, SEST and JCMT telescopes.
\end{list}               
  \end{table}

For the SEST observations we alternated between a 500 MHz and 1 GHz backend, 
depending on weather. Simultaneous observations with the 1 and 3 mm receiver 
were taking place. In addition, the data of the Seyfert galaxy NGC~3079 
obtained during the IRAM survey in 2006 is included in this work. We used 
the software package XS (written by P. Bergman) to reduce the data and 
fit the gaussians. Beamsizes and efficiencies are shown in Table~\ref{tab:beams}.

The sample consists of five Seyfert galaxies of which two are Seyfert 1 type and 
three are considered mainly Seyfert 2. Table~\ref{tab:galaxies} lists the 
coordinates of the center positions observed in these galaxies, their 
sub-classification as Seyfert galaxies and heliocentric radial velocities, 
according to the NASA/IPAC Extragalactic Database (NED) 
(http://nedwww.ipac.caltech.edu/). The distances were calculated assuming 
a Hubble constant of H$_0$ = 74.37 $\pm$ 2.27 \kms~Mpc$^{-1}$ 
~(Ngeow and Kanbur, \cite{ngeow06}).

\section{Results}

\subsection{NGC 1068}

   \begin{table}[!b]
      \caption[]{NGC~1068 line parameters.}
         \label{tab:1068-intensities}
				\centering
         \begin{tabular}{lcccc}
            \hline
            \noalign{\smallskip}
            Transition & Gaussian  & $V$ & $T_{A}^{*}$ & $\Delta V$ \\
                       & Component & [\kms] & [mK] & [\kms] \\
            \noalign{\smallskip}
            \hline
            \noalign{\smallskip}
		CO 1-0 & 1 & 1007$\pm$~~1 & 225.2$\pm$14.9 & ~~58$\pm$~~3\\
		       & 2 & 1100$\pm$~~3 & 265.1$\pm$~~3.8 & 156$\pm$11\\
		       & 3 & 1234$\pm$~~2 & 215.3$\pm$10.8 & ~~96$\pm$~~4\\
            \noalign{}\\
		CO 2-1 & 1 & 1007$\pm$~~2 & 206.8$\pm$32.5 & ~~63$\pm$~~9\\
		       & 2 & 1121$\pm$~~4 & 529.3$\pm$~~8.8 & 179$\pm$16\\
		       & 3 & 1245$\pm$~~3 & 222.7$\pm$37.8 & ~~81$\pm$10\\
            \noalign{}\\
		CN 1-0 & 1 & 1001$\pm$23 & ~~12.8$\pm$13.2 & ~~94$\pm$42\\
		       & 2 & 1091$\pm$68 & ~~19.4$\pm$~~4.7 & 154$^{~\rm a}$\\
		       & 3 & 1213$\pm$32 & ~~15.6$\pm$~~9.5 & 136$\pm$54\\
		       & 4 & 1946$\pm$26 & ~~8.3$^{~\rm b}$ & 154$^{~\rm a}$\\		       		       
            \noalign{}\\
		CN 2-1 & 1 & 1093$\pm$~~9 & ~~~10.4$\pm$~~0.9 & 154$\pm$66\\
		       & 2 & 1289$\pm$14  & ~~~~6.2$\pm$~~0.9 & 154$^{~\rm a}$\\
            \noalign{}\\
		HCN 3-2 & 1 & 1103$\pm$~~7 & ~~48.7$\pm$~~2.4 & 275$\pm$17\\
            \noalign{}\\
		HNC 1-0 & 1 & 1073$\pm$13 & ~~~~9.7$\pm$~~1.8 & 114$\pm$30\\
		        & 2 & 1268$\pm$20 & ~~~~6.7$\pm$~~1.5 & 170$\pm$62\\
            \noalign{}\\
		HNC 3-2 & 1 & 1071$\pm$~~9 & ~~12.9$\pm$~~1.8 & 134$\pm$23\\
		        & 2 & 1250$\pm$24 & ~~~~3.7$\pm$~~2.5 & ~~70$\pm$58\\						
            \noalign{\smallskip}
            \hline
         \end{tabular}
\begin{list}{}{}
\item[$^{\mathrm{a}}$] The line widths were set to the value found in the main component of the CN 2--1 line.
\item[$^{\mathrm{b}}$] In order to avoid the effect of the artefact in the backend, the intensity was locked to the value expected for this spingroup.
\end{list}     
   \end{table}

\begin{figure*}[!ht]
  
  \hfill\includegraphics[width=6.5cm]{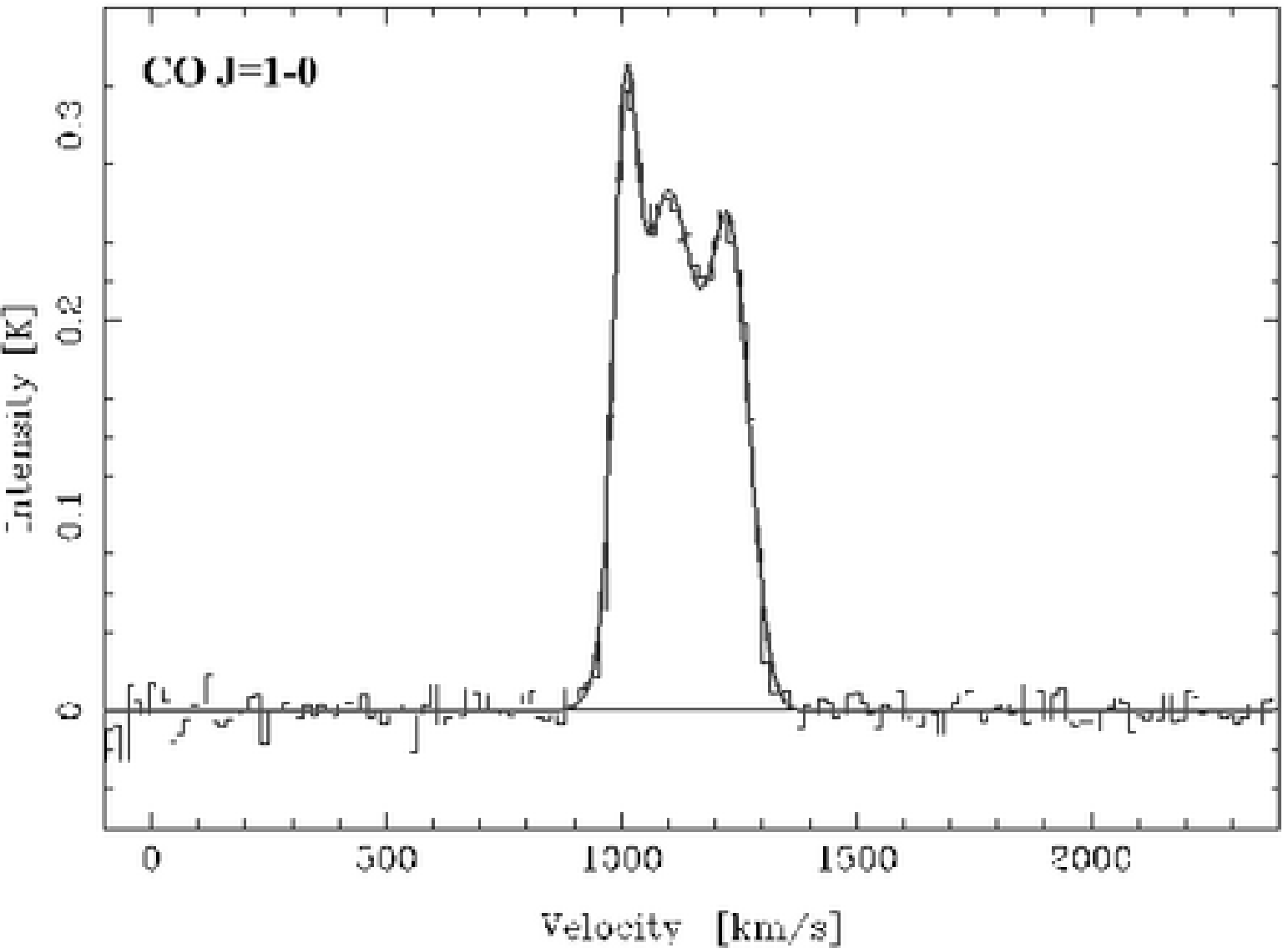}%
  \hfill\includegraphics[width=6.5cm]{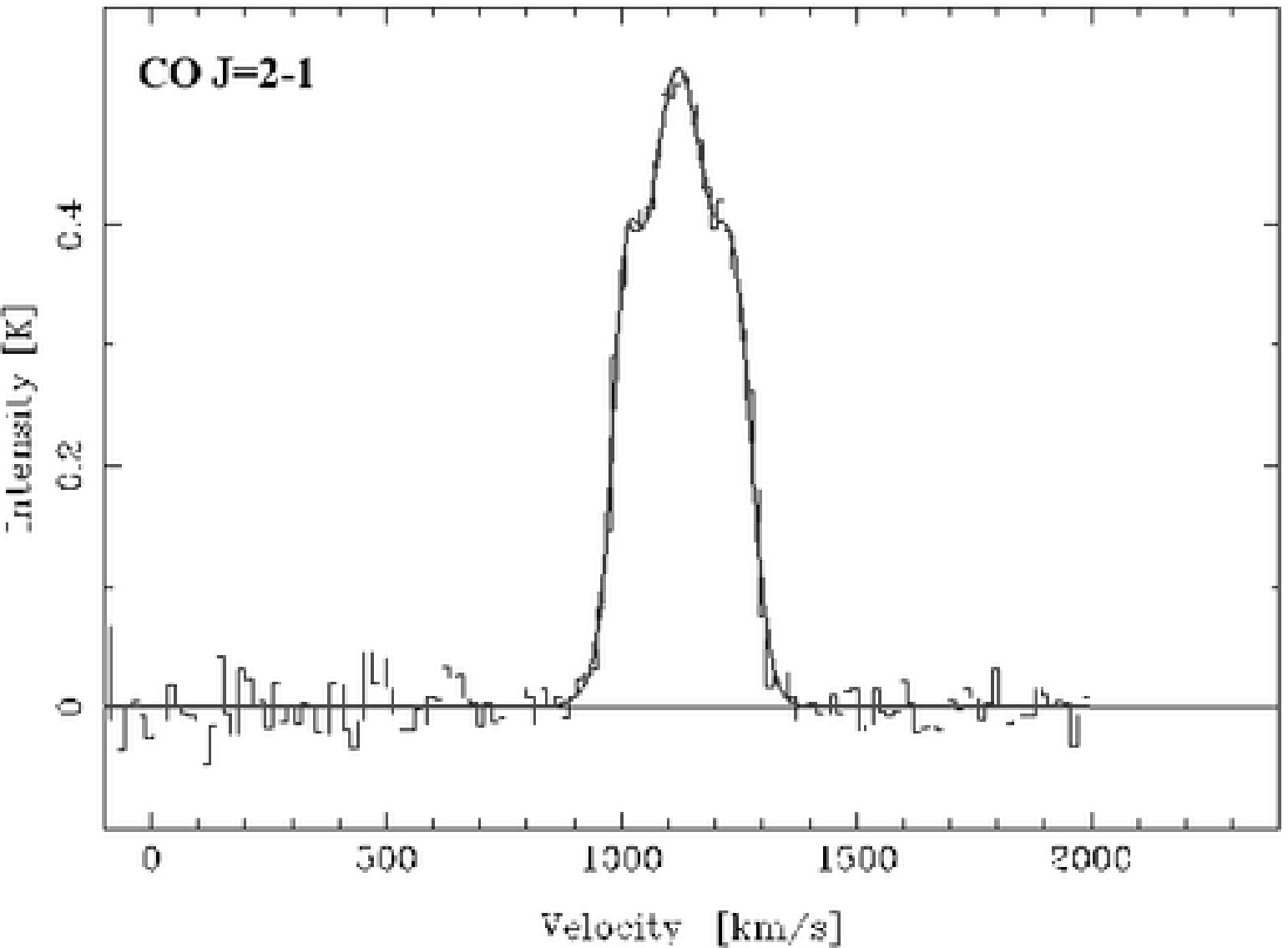}\hspace*{\fill}\\
  
  \hspace*{\fill}\includegraphics[width=6.5cm]{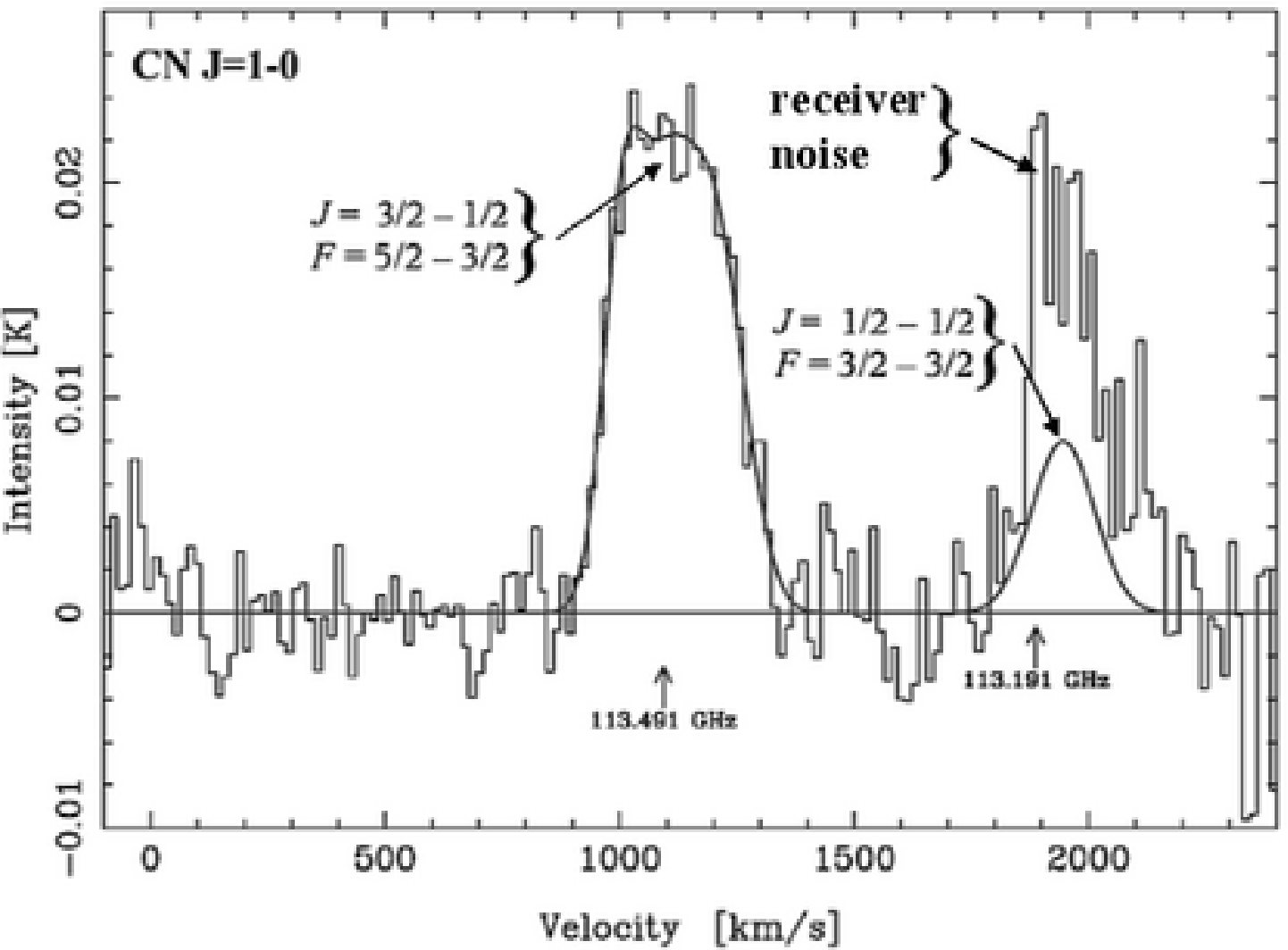}%
  \hfill\includegraphics[width=6.5cm]{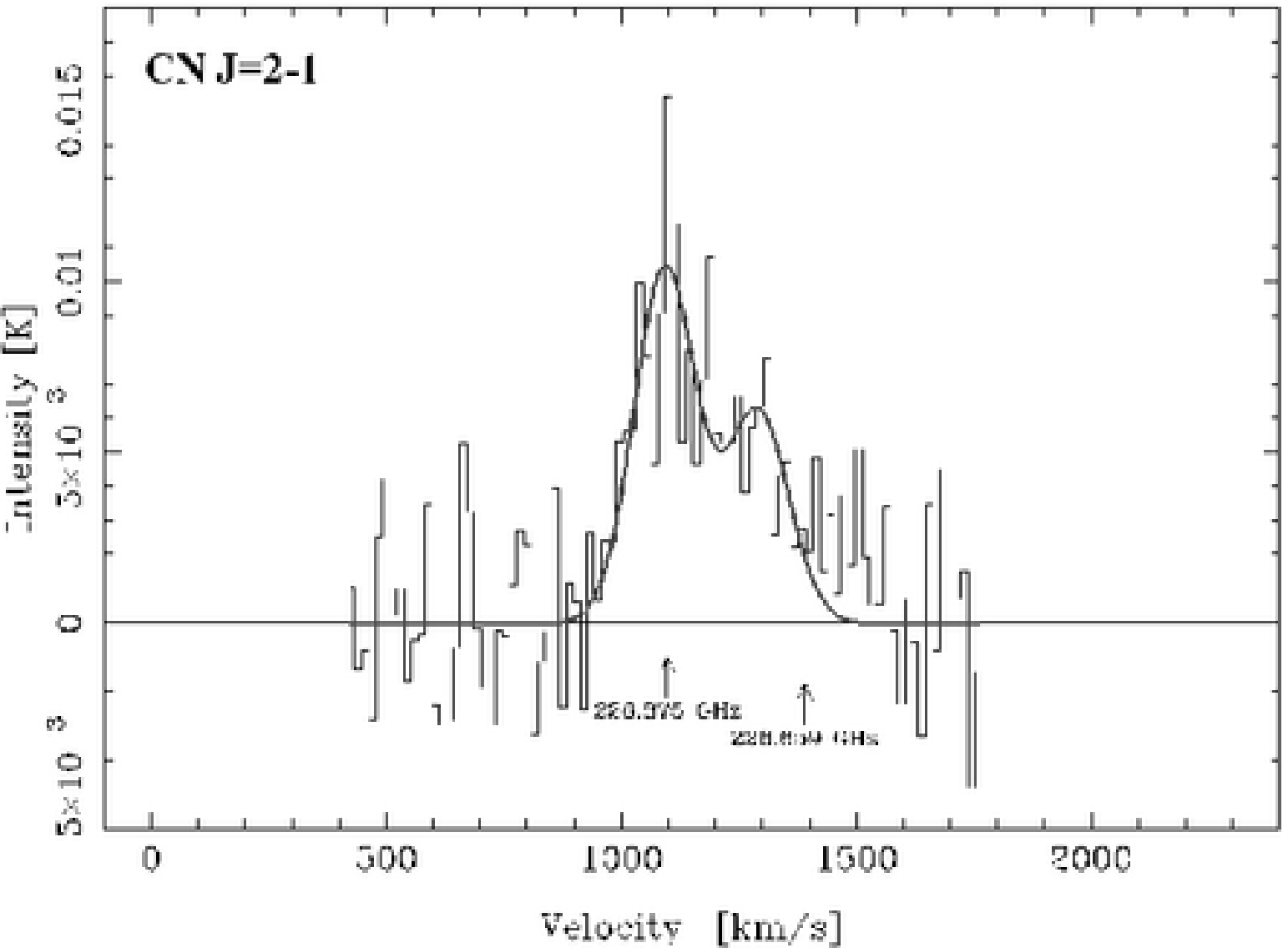}\hspace*{\fill}\\
  
  \hspace*{\fill}\includegraphics[width=6.5cm]{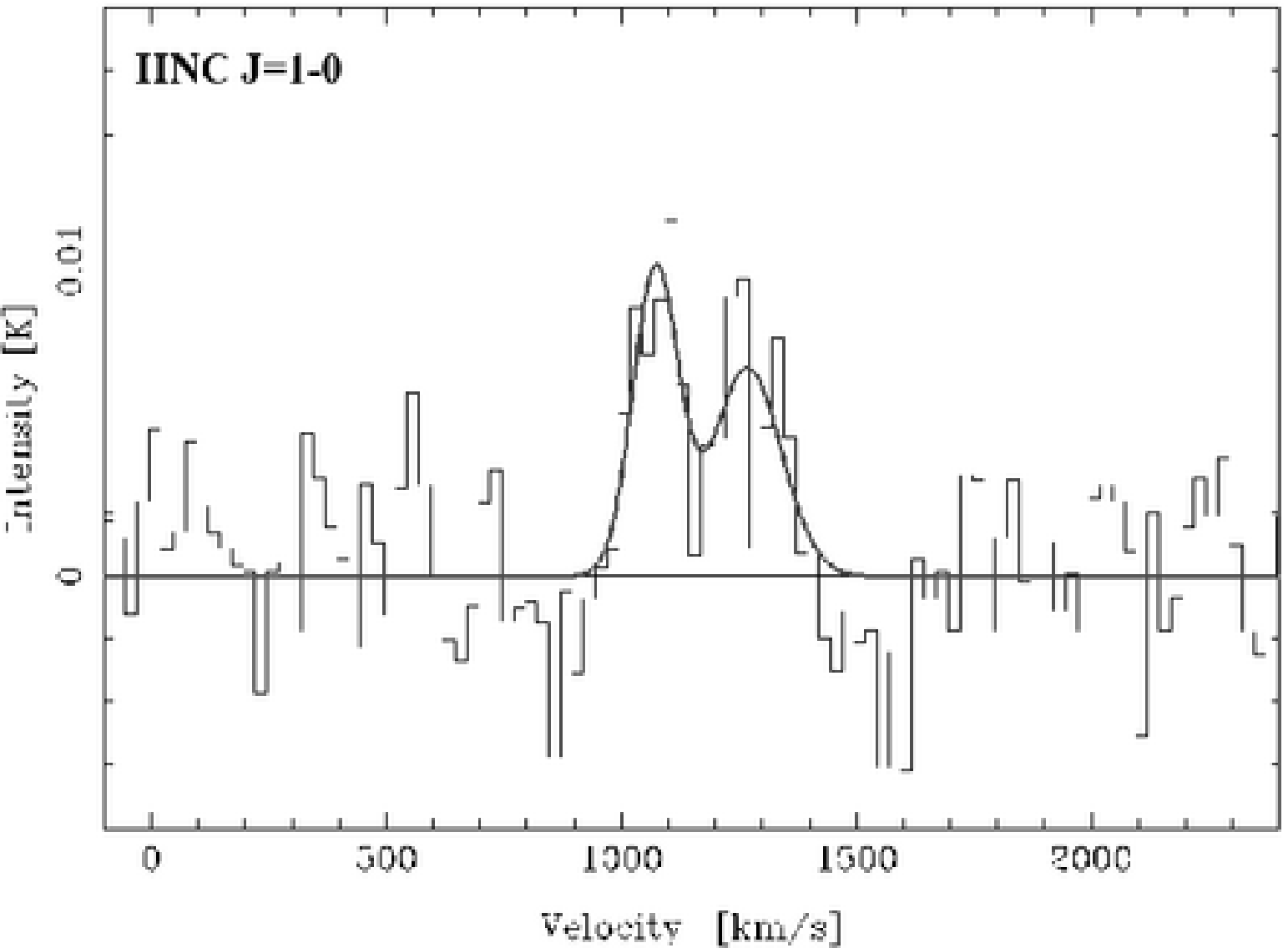}%
  \hfill\includegraphics[width=6.5cm]{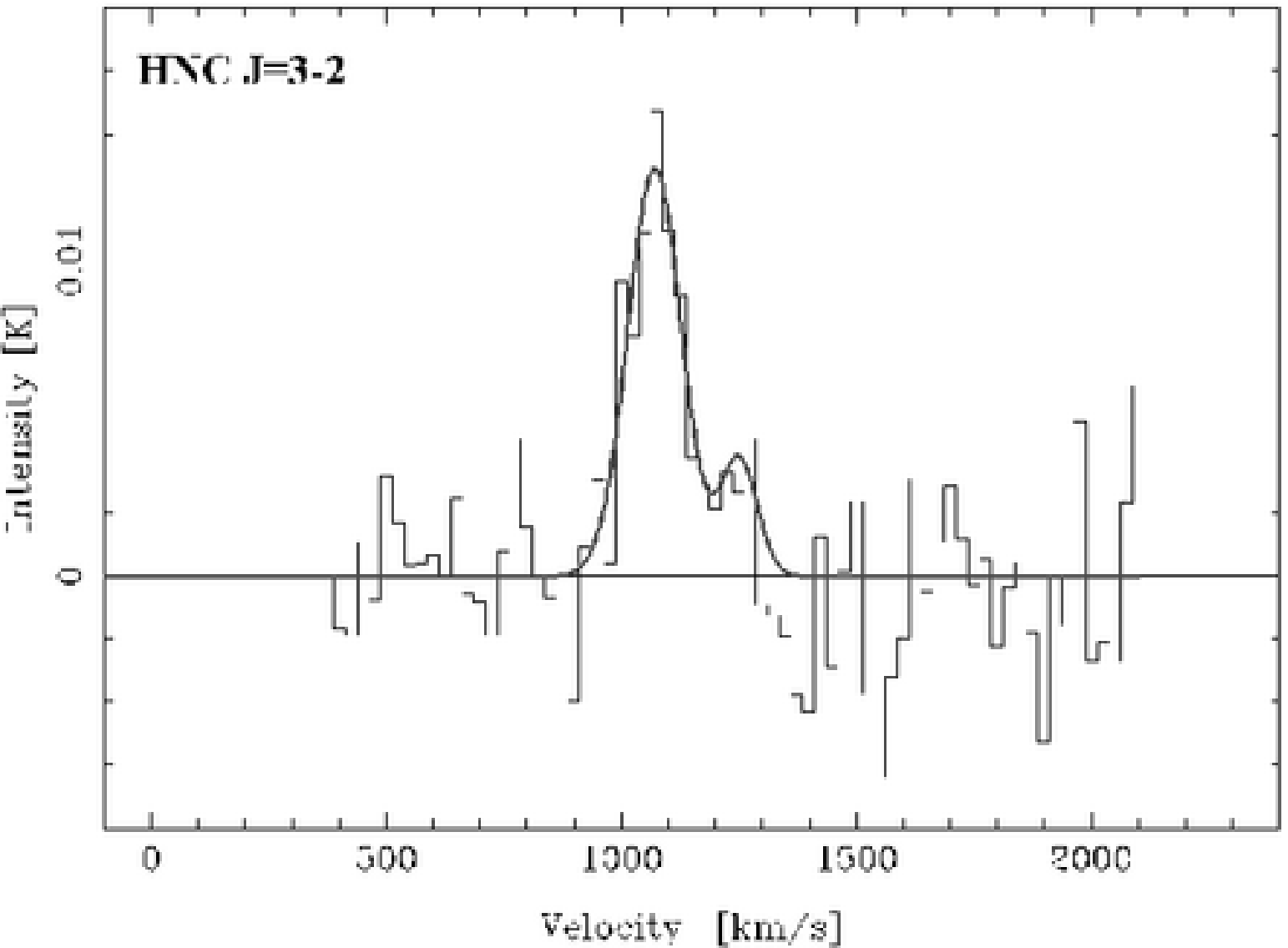}\hspace*{\fill}\\
  
  \centering
  \includegraphics[width=6.5cm]{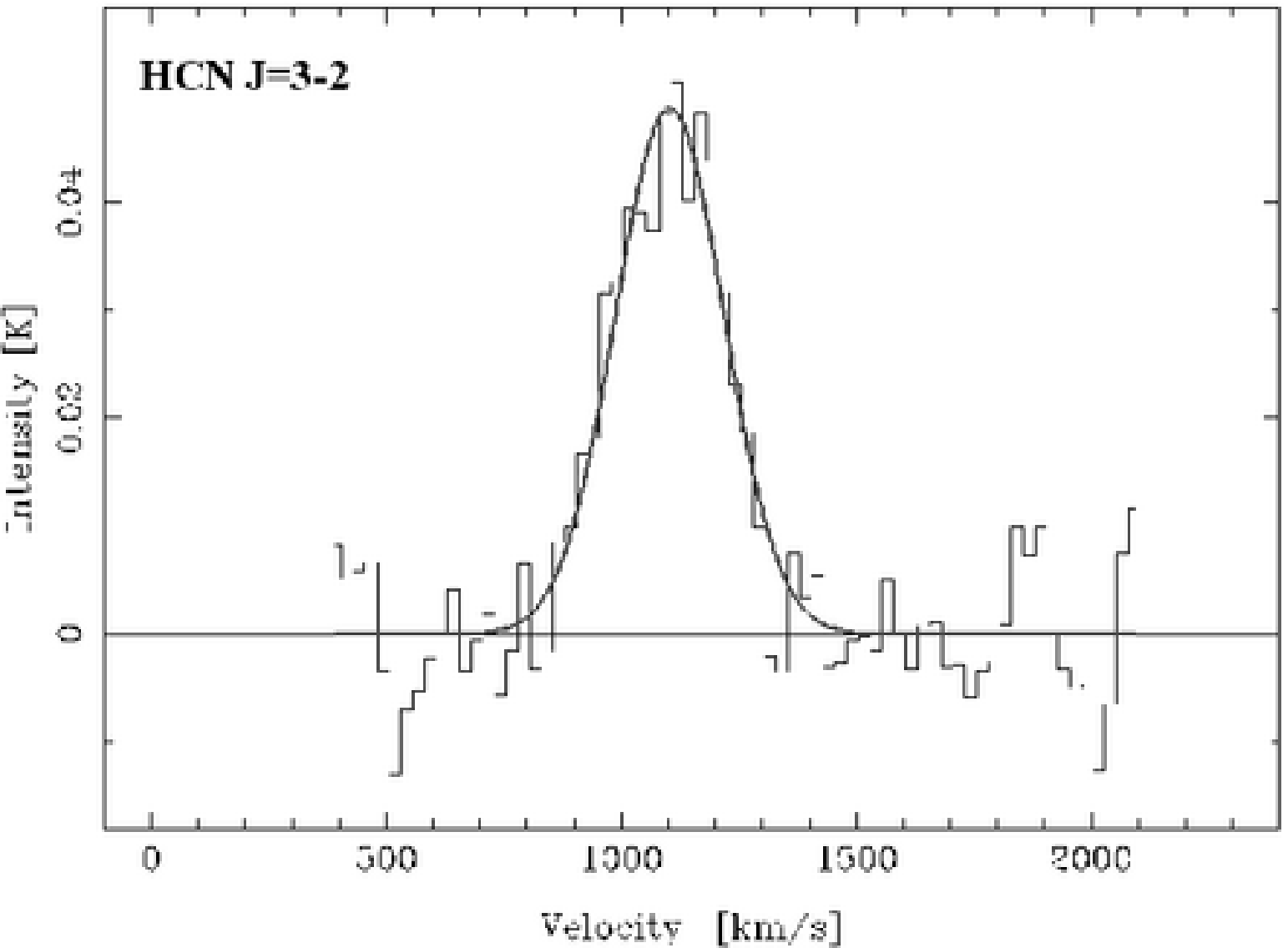}
  
  \caption{\footnotesize{Molecular line emissions in \textbf{NGC~1068}. 
  The velocity resolution was set to 25 \kms for HNC and HCN, 
  and to 15 \kms for CO and CN. The spectra are centered with 
  respect to the heliocentric systemic velocity $\rm v_{sys} = 1137$ \kms. 
  Emission from the spiral arms are detected in the CO 1-0 
  line. The CO 2-1 line is dominated by the emission coming 
  from the CND. The two main spingroups of CN are detected in
  both J=1--0 and J=2--1 transitions. In the CN 1--0 the second
  spingroup is corrupted by noise in the spectrum. The two spingroups
  are blended in the CN 2--1 line.
  The different line shapes (profiles) of the 
  HCN and HNC spectra seem to indicate that their emissions 
  emerge from different regions.}}
  \label{fig:1068-spectra}
\end{figure*}

The molecular line emissions observed in NGC~1068 are shown in 
Figure~\ref{fig:1068-spectra}. A first-order polynomial was used 
in most cases to correct the baselines, with the exception of 
the HCN 3--2 and CN 2--1 spectra, for which a second-order polynomial was 
required. The velocity resolution was set to 15 \kms~for CO and CN, whereas a 
25 \kms resolution was used for HNC and HCN. These velocity resolutions 
represent less than 10\% of the line widths. The spectra are centered 
with respect to the heliocentric systemic velocity 
$\rm v_{sys} = 1137$ \kms~(from NED). 

The $J$=1--0 and $J$=2--1 lines of CO show a triple structure
where the peaks can be attributed mainly to the bars and the
spiral arms, as described by Helfer \& Blitz~\cite{helfer95}.

We detect two of the main spingroups of CN 1--0: the 1--0 
(\textit{J} = 3/2 -- 1/2, \textit{F} = 5/2 -- 3/2) line at the 
center of the spectrum, and the 1-0 (\textit{J} = 1/2 -- 1/2, 
\textit{F} = 3/2 -- 3/2) line shifted 856 \kms~to the right.
This spectrum shows three components as well, however they are 
hard to distinguish due to the blending. 
In fact if we freely fit three gaussian components, the uncertainties 
of the center velocity, amplitude and line width are about 100\% or
larger. Instead, if we set the line width of the central component 
to 154 \kms~, which corresponds to the line width found for the 
CN $J$=2--1 line as described below, we get reasonable values. Only
the first gaussian component shows high uncertainties in the 
amplitude and line width. Since the beam sizes of the CN and CO
molecules are similar at the frequencies of the $J$=1--0 lines,
we think that both beams pick up emission coming from the spiral
arms and bars. Although, the nuclear region seems to be the predominant
component in the case of CN.

Due to the second spingroup is corrupted by noise in the backend,
we set the amplitude of the gaussian to 8.3 mK,
which corresponds to the expected factor of about 0.43 times the 
amplitude of the main spingroup, according to the National Institute 
of Standards and Technology (NIST). 
The resulting central velocity of the second component was 1946 \kms 
($\sim113.169$ GHz), i.e., about 22 MHz shifted from 
the expected frequency for this spingroup.

   \begin{table}[!b]
      \caption[]{NGC~1365 line parameters.}
         \label{tab:1365-intensities}
				\centering
         \begin{tabular}{lcccc}
            \hline
            \noalign{\smallskip}
            Transition & Gaussian  & $V$ & $T_{A}^{*}$ & $\Delta V$ \\
                       & Component & [\kms] & [mK] & [\kms] \\
            \noalign{\smallskip}
            \hline
            \noalign{\smallskip}
		CO 1-0 & 1 & 1532$\pm$~~2 & 322.7$\pm$5.2 & 136$\pm$4\\
		       & 2 & 1709$\pm$~~3 & 260.3$\pm$5.0 & 142$\pm$5\\
            \noalign{}\\
		CN 1-0 & 1 & 1511$\pm$13   & ~~16.8$\pm$3.5 & 130$\pm$25\\
		       & 2 & 1685$\pm$33   & ~~~~8.9$\pm$1.4  & 212$\pm$92\\
		       & 3 & 2356$\pm$18   & ~~~~4.7$\pm$1.2  & 130$^{~\rm a}$\\		       
            \noalign{}\\
		CN 2-1 & 1 & 1526$\pm$11 & ~~10.4$\pm$1.5  & 130$^{~\rm a}$\\
		       & 2 & 1727$\pm$35     & ~~~~8.4$\pm$1.9 & 212$^{~\rm b}$\\
		       & 3 & 1846$\pm$35     & ~~~~4.11$\pm$3.5 & 130$^{~\rm a}$\\						 
            \noalign{}\\            
		HCN 3-2 & 1 & 1534$\pm$~~5 & ~~51.4$\pm$3.4 & 136$\pm$13\\
		        & 2 & 1706$\pm$25  & ~~11.4$\pm$3.3 & 142$^{~\rm c}$\\
            \noalign{}\\
		HNC 1-0 & 1 & 1539$\pm$11 & ~~16.0$\pm$1.7 & 142$\pm$23\\
		        & 2 & 1750$\pm$34 & ~~~~5.1$\pm$1.4 & 195$\pm$98\\
            \noalign{\smallskip}
            \hline
         \end{tabular}
\begin{list}{}{}
\item[$^{\mathrm{a}}$] The two spingroups are supposed to have the same line width, so these were set according to the value found for the main spingroup of the CN 1--0 line.
\item[$^{\mathrm{b}}$] The line width of the second component of the double peak structure was set to the corresponding value found in the CN 1--0 line.
\item[$^{\mathrm{c}}$] The second component of HCN 3--2 seems to be affected by noise, so its line width was locked at $\Delta V = 142$ \kms, i.e., the corresponding line width of the CO 1--0 line.
\end{list}     
   \end{table}

In the CN 2--1 line, the two spingroups at 226.8746 GHz (2--1, 
\textit{J} = 5/2 -- 3/2, \textit{F} = 7/2 -- 5/2) and 226.6596 GHz 
(2--1, \textit{J} = 3/2 -- 1/2, \textit{F} = 5/2 -- 3/2) are 
severely blended since the shift is $\sim$300 \kms. 
In order to identify the two spingroups we first fit two
gaussian components to get the line width of the main spingroup,
which is $154\pm66$. We then set this value to both gaussian components and
fit again the other parameters.
The second spingroup is, in the optically thin limit, a factor 
0.54 weaker than the main spingroup. We get a factor $\sim0.6$ between the 
intensities obtained from the gaussian fit.
On the other hand, the resulting center velocity of the second 
component is 1289 \kms, which corresponds to a shift of about 
68 MHz with respect to the expected frequency of the second 
spingroup. Besides the noise in the data, this shift may also 
be produced by the influence of the unresolved spingroups, 
(2--1, \textit{J} = 3/2 -- 1/2, \textit{F} = 1/2 -- 1/2) and 
(2--1, \textit{J} = 3/2 -- 1/2, \textit{F} = 3/2 -- 1/2), 
located in between the two main spingroups. Together, these 
inter-spingroups would produce an intensity comparable to 
that of the second spingroup.

The line shape and intensity of the CN 2--1 spectrum differs from 
the one obtained by Usero \etal~\cite{usero04}
(thereafter U04). Besides the lack of baseline coverage observed 
in the U04 spectrum, there is a substantial discrepancy between the 
estimated source sizes. In U04 the emission of most of the molecules was
assumed to emerge from a $6''\times$4$''$ region, which corresponds to 
their estimate of the size of the circumnuclear disk (CND), 
based on the CO 2--1 high resolution
map presented by Schinnerer \etal~(\cite{schinnerer00}).

   \begin{table}[tp]
      \caption[]{NGC~3079 line parameters.}
         \label{tab:3079-intensities}
				\centering
         \begin{tabular}{lcccc}
            \hline
            \noalign{\smallskip}
            Transition & Gaussian  & $V$ & $T_{A}^{*}$ & $\Delta V$ \\
                       & Component & [\kms] & [mK] & [\kms] \\
            \noalign{\smallskip}
            \hline
            \noalign{\smallskip}
		CO 1-0 & 1 & ~932$\pm$~~4 & 129.4$\pm$11.8 & ~143$\pm$~~9\\
		       & 2 & 1146$\pm$~~3 & 374.3$\pm$~~3.3 & 269$\pm$13\\
		       & 3 & 1312$\pm$~~6 & 145.1$\pm$13.1 & ~~74$\pm$~~9\\
		       & 4 & 1389$\pm$~~5 & 186.3$\pm$~~9.1 & ~~72$\pm$~~5\\						       
            \noalign{}\\
		CO 2-1 & 1 & ~~964$\pm$15 & 233.3$\pm$18.9 & 175$\pm$18\\
		       & 2 & 1062$\pm$18 & 224.3$\pm$84.9 & ~~98$\pm$18\\
		       & 3 & 1160$\pm$20 & 321.7$\pm$35.3 & 126$\pm$20\\
		       & 4 & 1311$\pm$14 & 428.3$\pm$~~9.4 & 190$\pm$~~7\\	
		       					       
            \noalign{}\\
		CN 1-0 & 1 & ~~967$\pm$~~9  & ~~~5.6$\pm$~~0.6 & 162$\pm$23\\
		       & 2 & 1287$\pm$15 & ~~~4.5$\pm$~~0.6 & 215$\pm$33\\
			       
            \noalign{}\\
		CN 2-1 & 1 & ~~997$\pm$14 & ~~~~8.1$\pm$~~1.5 & 137$\pm$34\\
		       & 2 & 1251$\pm$25 & ~~~~5.3$\pm$~~1.3 & 203$\pm$69\\
						       
            \noalign{}\\
		HCN 1-0 & 1 & 1005$\pm$30 & ~~~~9.7$\pm$~~1.4 & 249$\pm$73\\
		        & 2 & 1274$\pm$21 & ~~10.7$\pm$~~1.7 & 188$\pm$49\\						
						
            \noalign{}\\
		HCN 3-2 & 1 & 1033$\pm$42 & ~~~~6.4$\pm$~~2.1 & 176$\pm$93\\
		        & 2 & 1287$\pm$32 & ~~10.4$\pm$~~1.8 & 219$\pm$70\\						

            \noalign{}\\
		HNC 1-0 & 1 & ~~983$\pm$17 & ~~~~6.5$\pm$~~1.7 & 129$\pm$41\\
		        & 2 & 1236$\pm$21 & ~~~~7.8$\pm$~~1.5 & 157$\pm$38\\
            \noalign{}\\
		HNC 3-2 & 1 & ~~996$\pm$48 & ~~~~7.4$\pm$~~3.2 & 178$\pm$124\\
		        & 2 & 1197$\pm$43 & ~~~~7.5$\pm$~~3.8 & 120$\pm$90\\	
						        					
            \noalign{\smallskip}
            \hline
         \end{tabular}

   \end{table}

If we correct the peak antenna temperature ($T_A^* \approx9.7$ mK) of our 
CN 2--1 spectrum, in order to obtain the main beam brightness temperature 
$T_{\rm mb}$ in the same way done in U04 
($T_{mb}=T_A^* \times \eta_{mb}^{-1}\times \Omega_{mb} \times \Omega_S^{-1}$), 
we get a peak $T_{\rm mb}$ of about 0.42 K, which is about 0.18 K (or a 
factor of $\sim$1.8) larger than the peak temperature obtained by U04. 

Due to the chemical link between HCN and CN, here we rather estimate 
the source size based on the high resolution map of HCN 1--0, published 
in Helfer \& Blitz (\cite{helfer95}). We estimate that the emissions of
HCN, CN (and very likely HNC as well) emerge from a more extended region 
of about 10$''\times$10$''$. This source size can also be inferred from 
the HCN 1--0 map presented by Kohno \etal~(\cite{kohno01}). 

By correcting the peak antenna temperature (as described in \S3.5) we get 
a peak $T_{\rm mb}$ of about 0.12 K, which is a factor 2 
smaller than what was found by U04.  

The HCN 3--2 line has its center velocity at $V \approx 1103$ \kms~which 
coincides with the main HCN emission shown in the position-velocity 
(\emph{p-v}) map by Tacconi \etal~(\cite{tacconi94}) (this is discussed in \S4).
The line width and line shape of this spectrum are consistent with HCN 
1--0 spectra published in previous work (Nguyen-Q-Rieu \etal~\cite{nguyen92}, 
Curran \etal~\cite{curran00}).

   \begin{table}[!t]
      \caption[]{Line parameters of NGC~2623 and NGC~7469. The third column 
      indicate the gaussian components (G.C.) used to fit the spectral lines.}
         \label{tab:2623-7469-intensities}
				\centering
         \begin{tabular}{llcccc}
            \hline
            \noalign{\smallskip}
            Source & Transition & G.C. & $V$ & $T_{A}^{*}$ & $\Delta V$ \\
                   &            &      & [\kms] & [mK] & [\kms] \\
            \noalign{\smallskip}
            \hline
            \noalign{\smallskip}
					       
		N2623 & HNC 3-2 & 1 & 5465$\pm$34 & ~~~~3.4$\pm$~~0.7 & 337$\pm$88\\
		
		\noalign{\smallskip}		

		N7469 & CO 2-1 & 1 & 4773$\pm$0 & ~~~~0.15$\pm$~~0 & 126$\pm$0\\
		         &        & 2 & 4927$\pm$0 & ~~~~0.16$\pm$~~0 & 123$\pm$0\\		
						        					
            \noalign{\smallskip}
            \hline
         \end{tabular}

   \end{table}

The HNC spectra present a double peak profile in which the main 
peak has its center velocity at $V \approx 1070$ \kms~in both the $J$=1--0 
and $J$=3--2 transitions. However, the line shape and intensity of the HNC 1--0 spectrum 
differs from the one obtained by H\"uttemeister \etal~(\cite{hutte95}). 
The HNC 3--2 spectrum seems to be affected by noise, since the amplitude and line 
width of the secondary component have uncertainties larger than 50\%.
The different line shapes observed between the HNC and HCN spectra would 
indicate that they emerge from regions of different kinematics. The line 
parameters are summarized in Table~\ref{tab:1068-intensities}.

\begin{figure*}[!ht]
  
  \hfill\includegraphics[width=6.5cm]{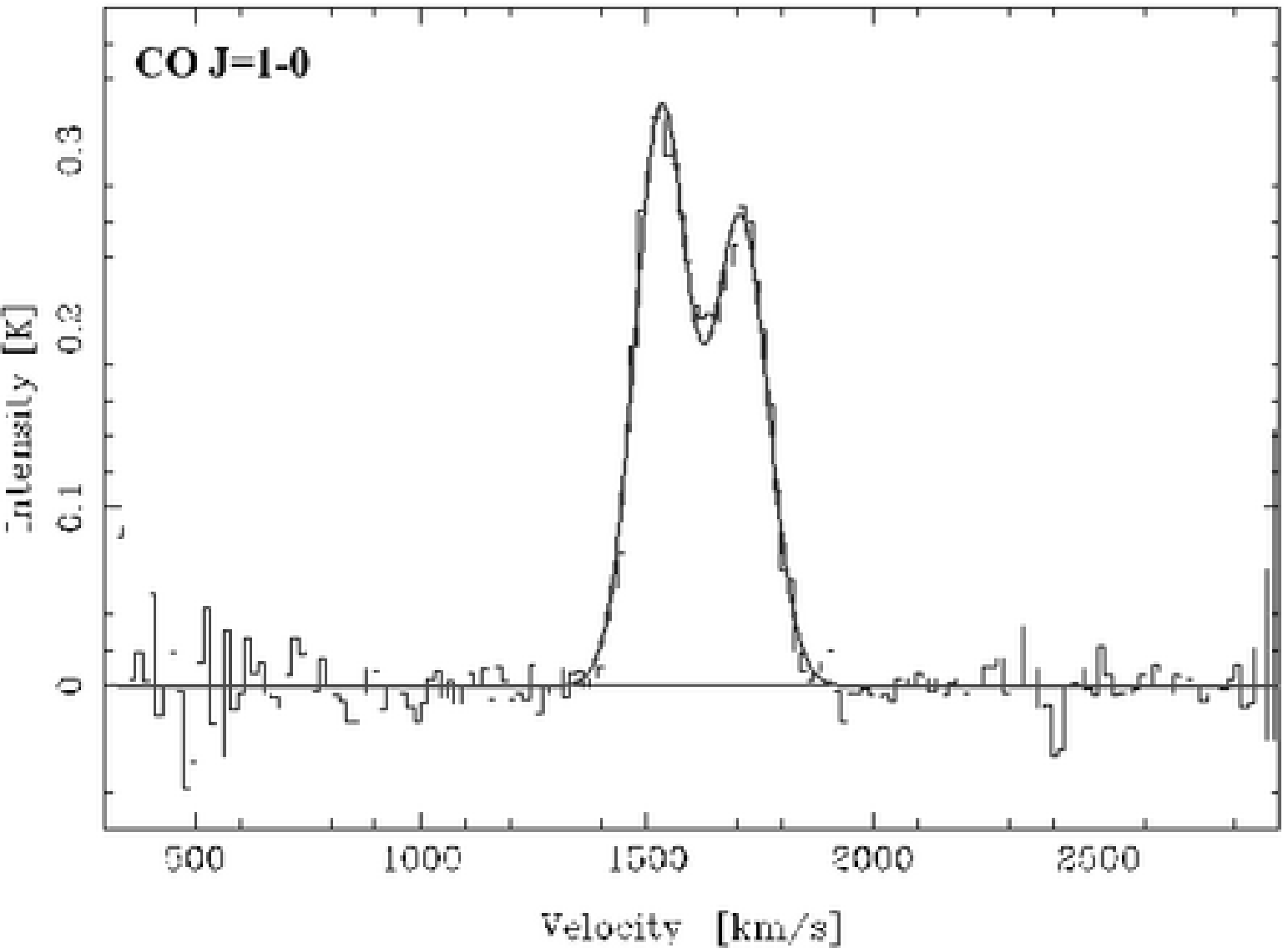}%
  \hfill\includegraphics[width=6.5cm]{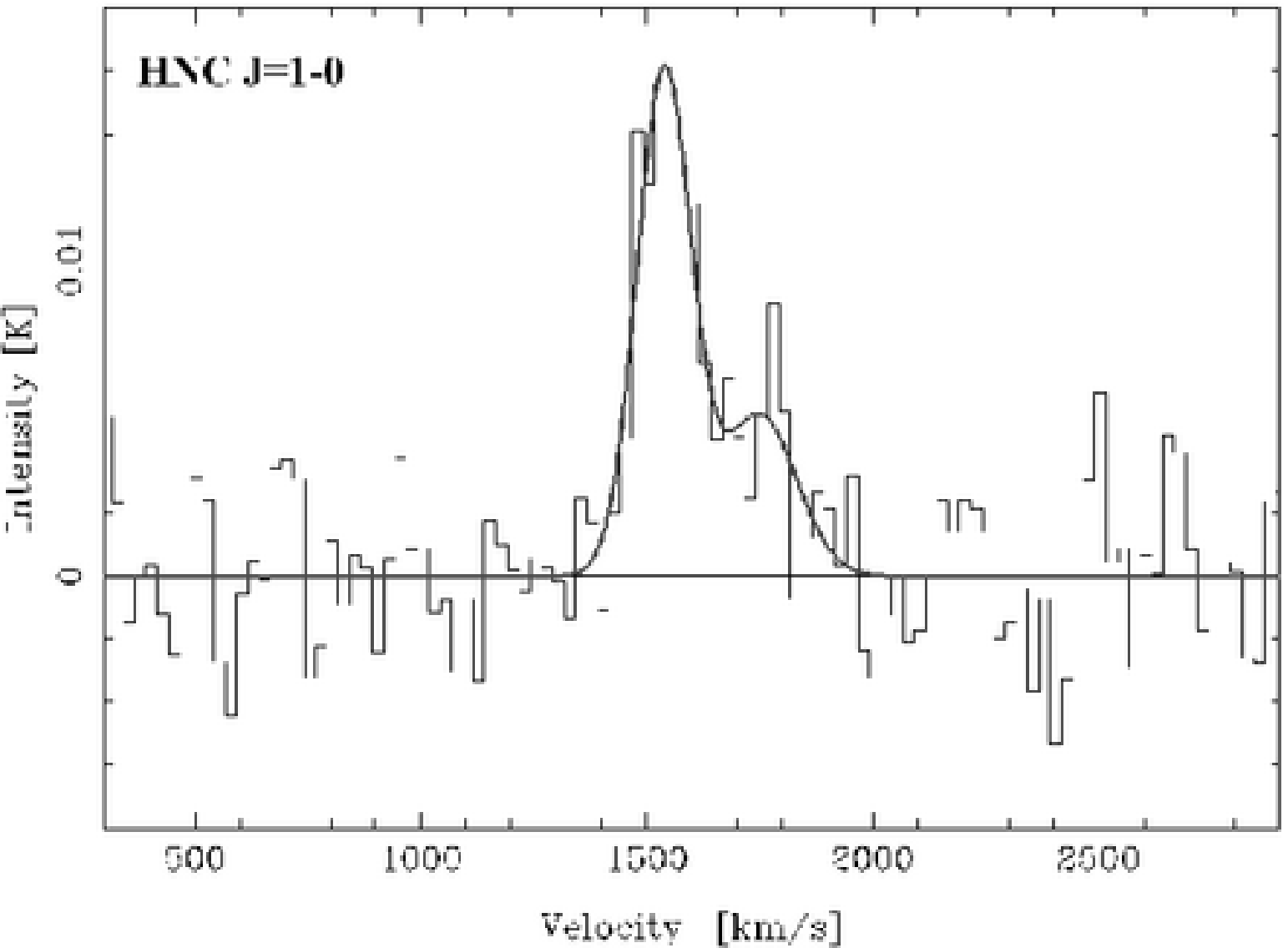}\hspace*{\fill}\\\\
  
  \hspace*{\fill}\includegraphics[width=6.5cm]{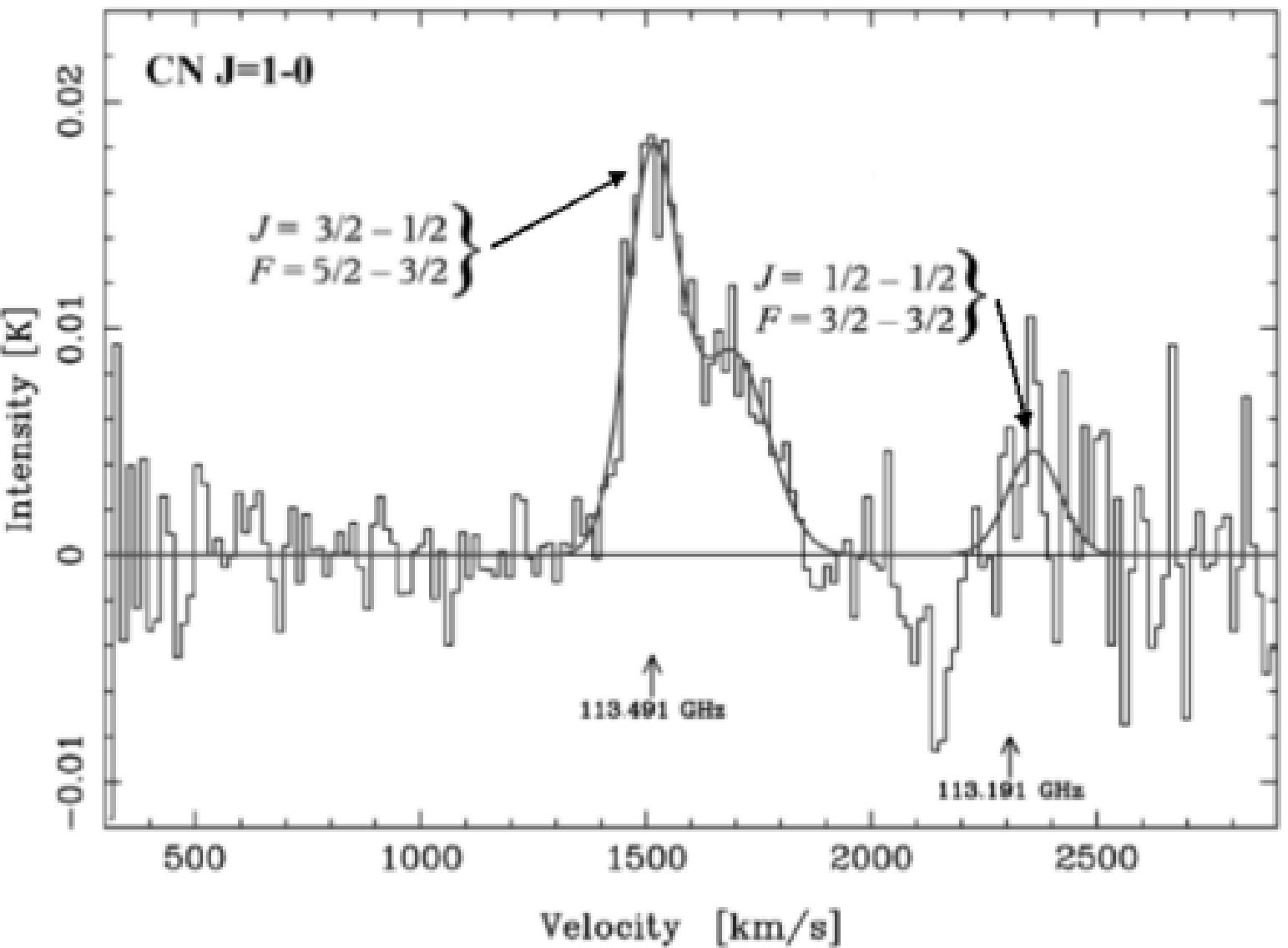}%
  \hfill\includegraphics[width=6.5cm]{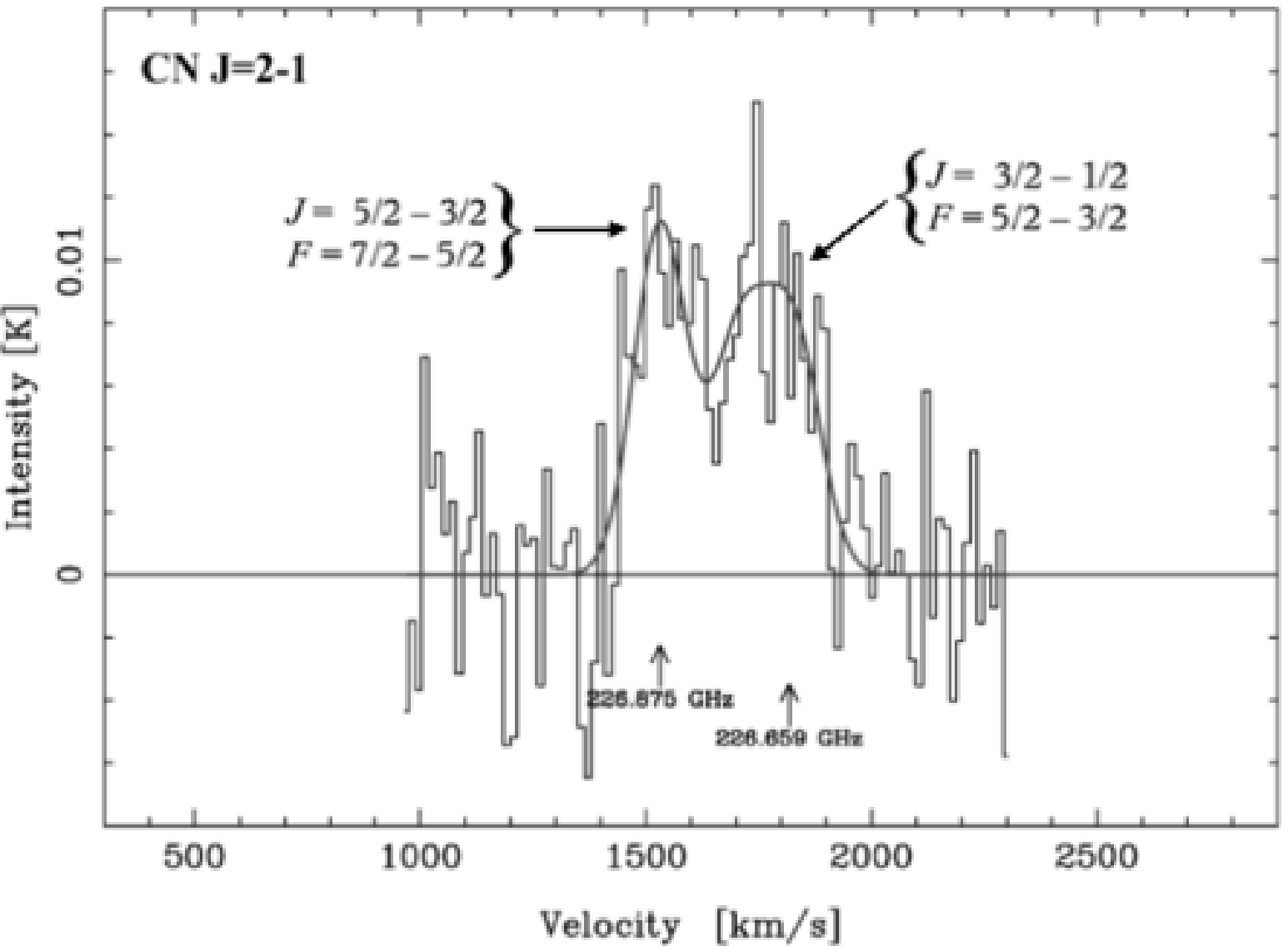}\hspace*{\fill}\\

  \centering  
  \includegraphics[width=6.5cm]{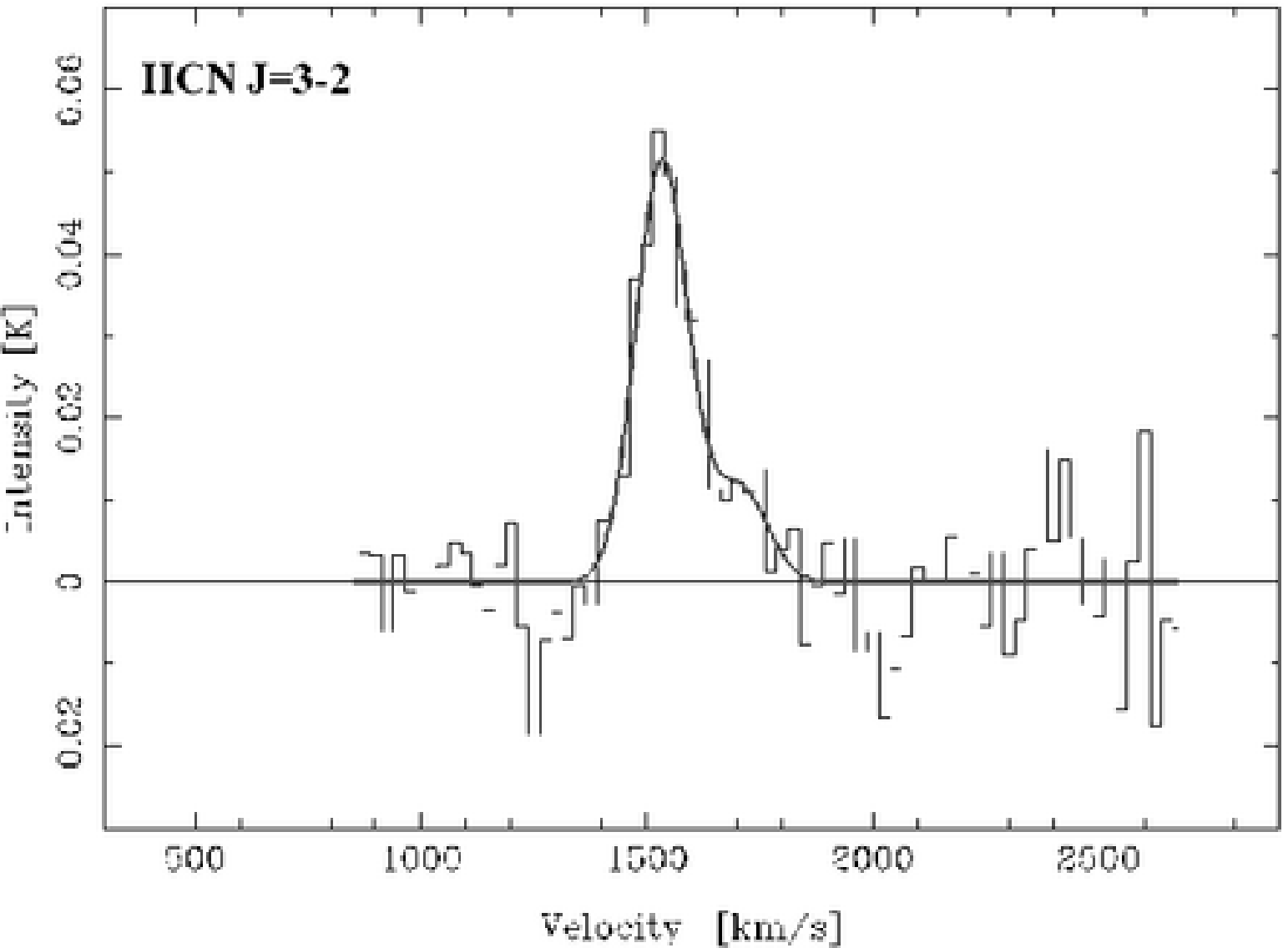}
  
  \caption{\footnotesize{Molecular line emissions in \textbf{NGC~1365}. The spectra 
  are centered with respect to the heliocentric systemic velocity 
  $\rm v_{sys} = 1636$ \kms. The velocity resolution was set to 25 \kms 
  for HNC and HCN,  and to 15 \kms for CO and CN. A double peak line shape 
  is observed in all the spectra. This structure is related with the double 
  peak emission coming from the center of the galaxy. In the CN spectra both 
  spingroups are detected. The second spingroup of the CN 2--1 line overlaps
  the double structure itself.}}
  \label{fig:1365-spectra}  
\end{figure*}

\subsection{NGC~1365}

Figure~\ref{fig:1365-spectra} shows the molecular line emissions 
observed in NGC~1365. Without considering the second spingroups of 
CN, all the molecular spectra of NGC~1365 present a double 
peak structure, irrespective of the beam size or line transition.

This double 
peak structure can also be seen in the CO 2--1 and CO 3--2 spectra 
showed in Sandqvist et al. (\cite{sandqvist95}) and in Sandqvist 
(\cite{sandqvist99}), respectively.

\begin{figure*}[!htp]
  
  \hfill\includegraphics[width=6.5cm]{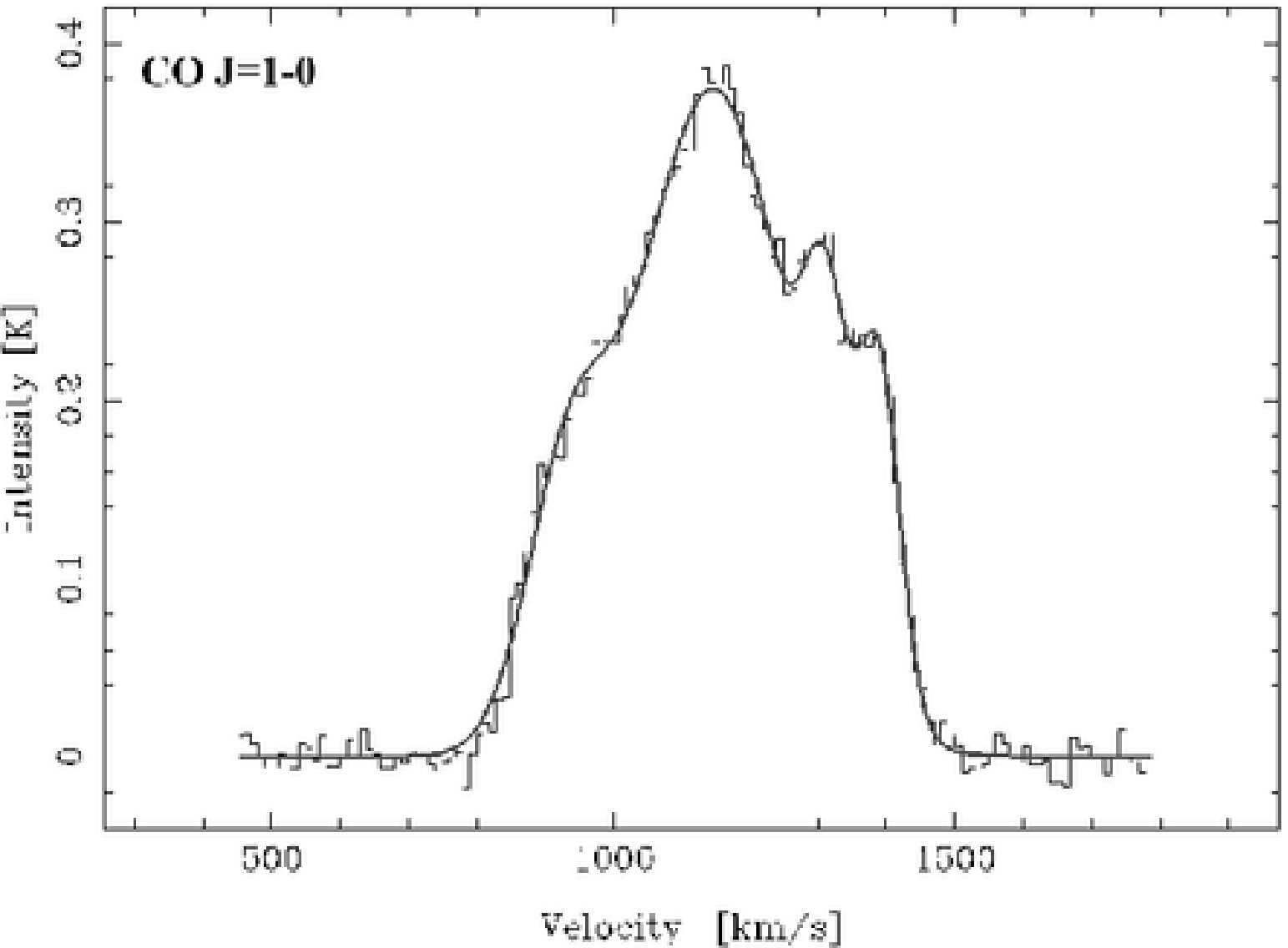}%
  \hfill\includegraphics[width=6.5cm]{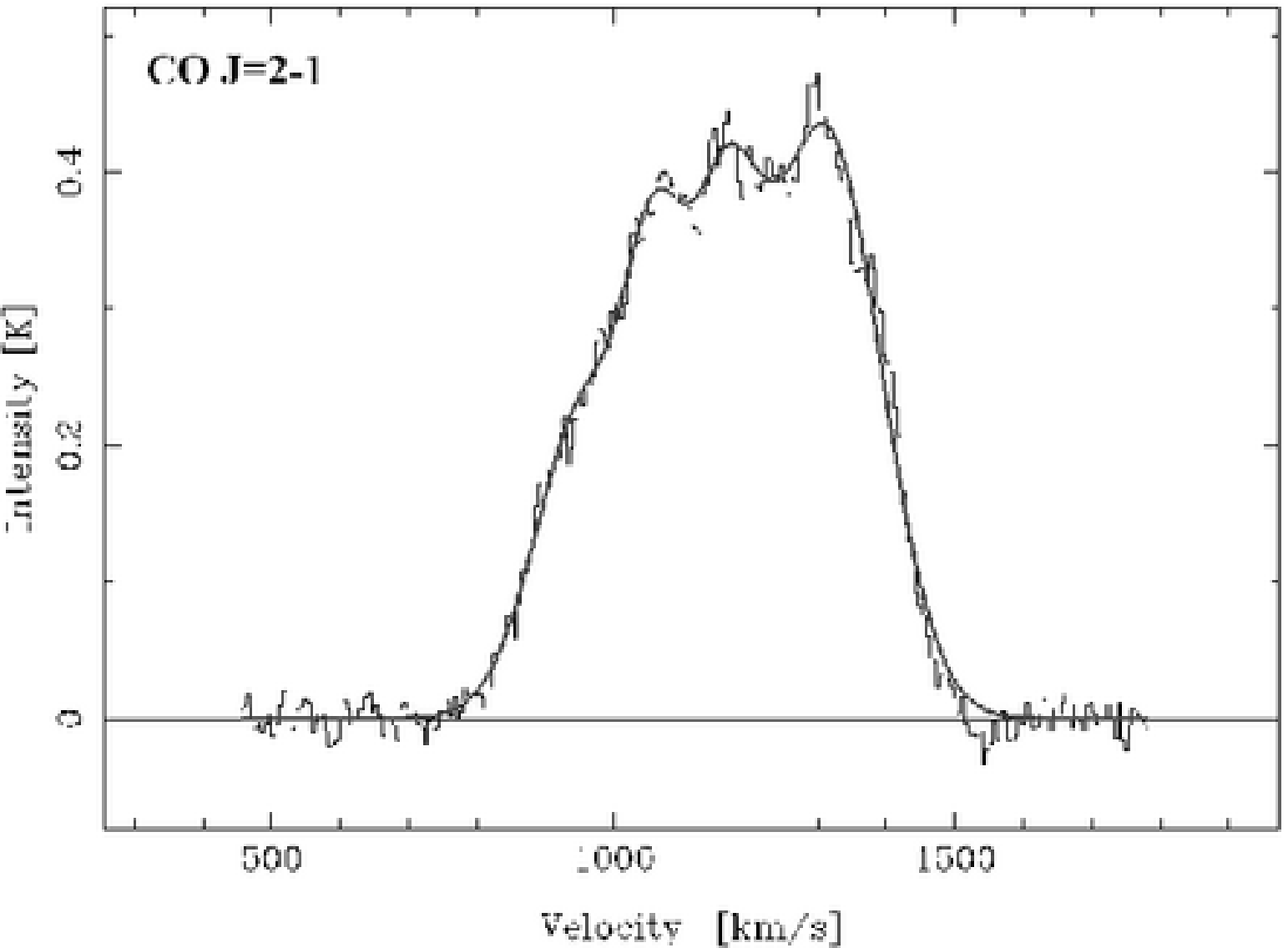}\hspace*{\fill}\\
  
  \hspace*{\fill}\includegraphics[width=6.5cm]{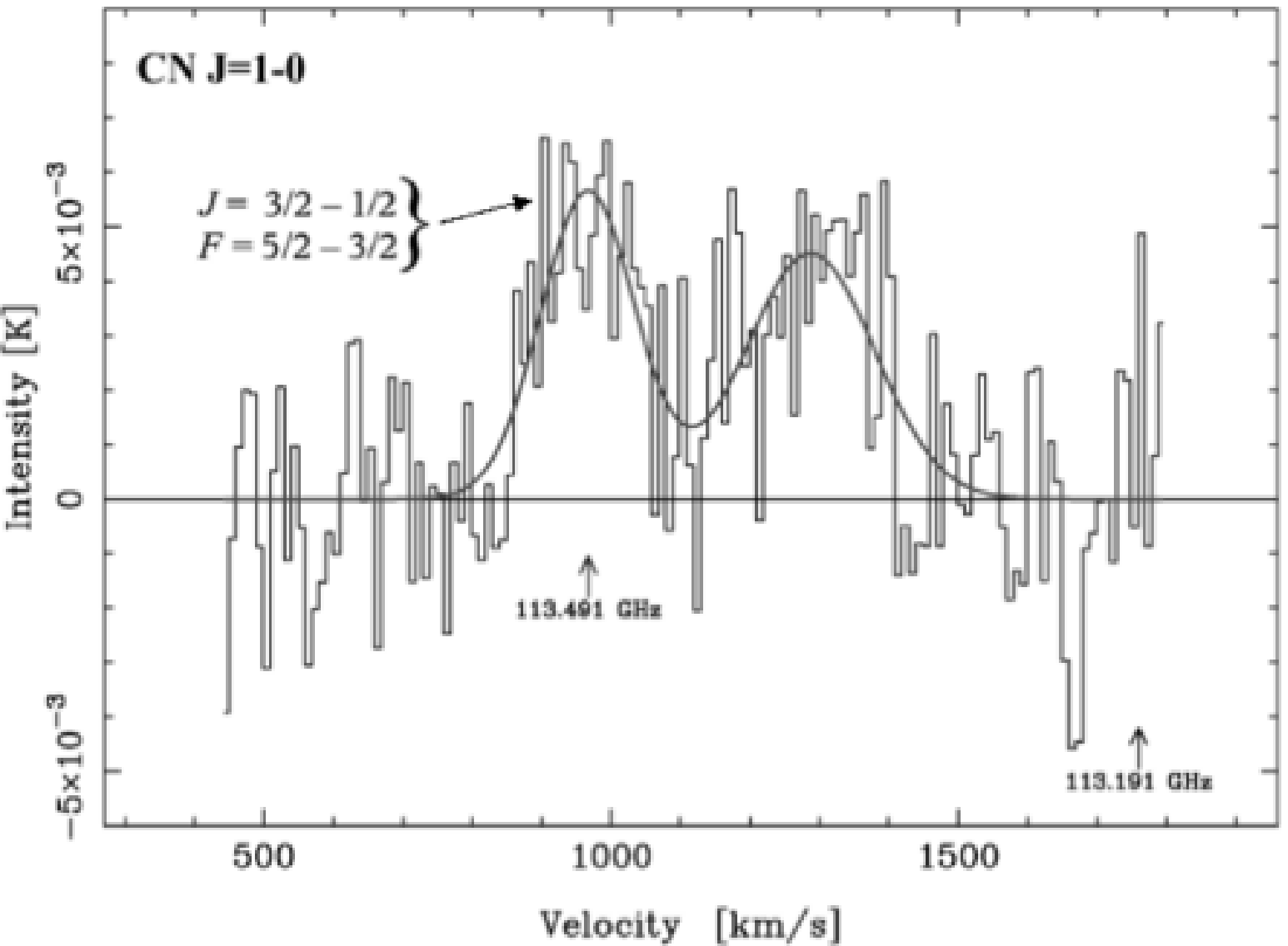}%
  \hfill\includegraphics[width=6.5cm]{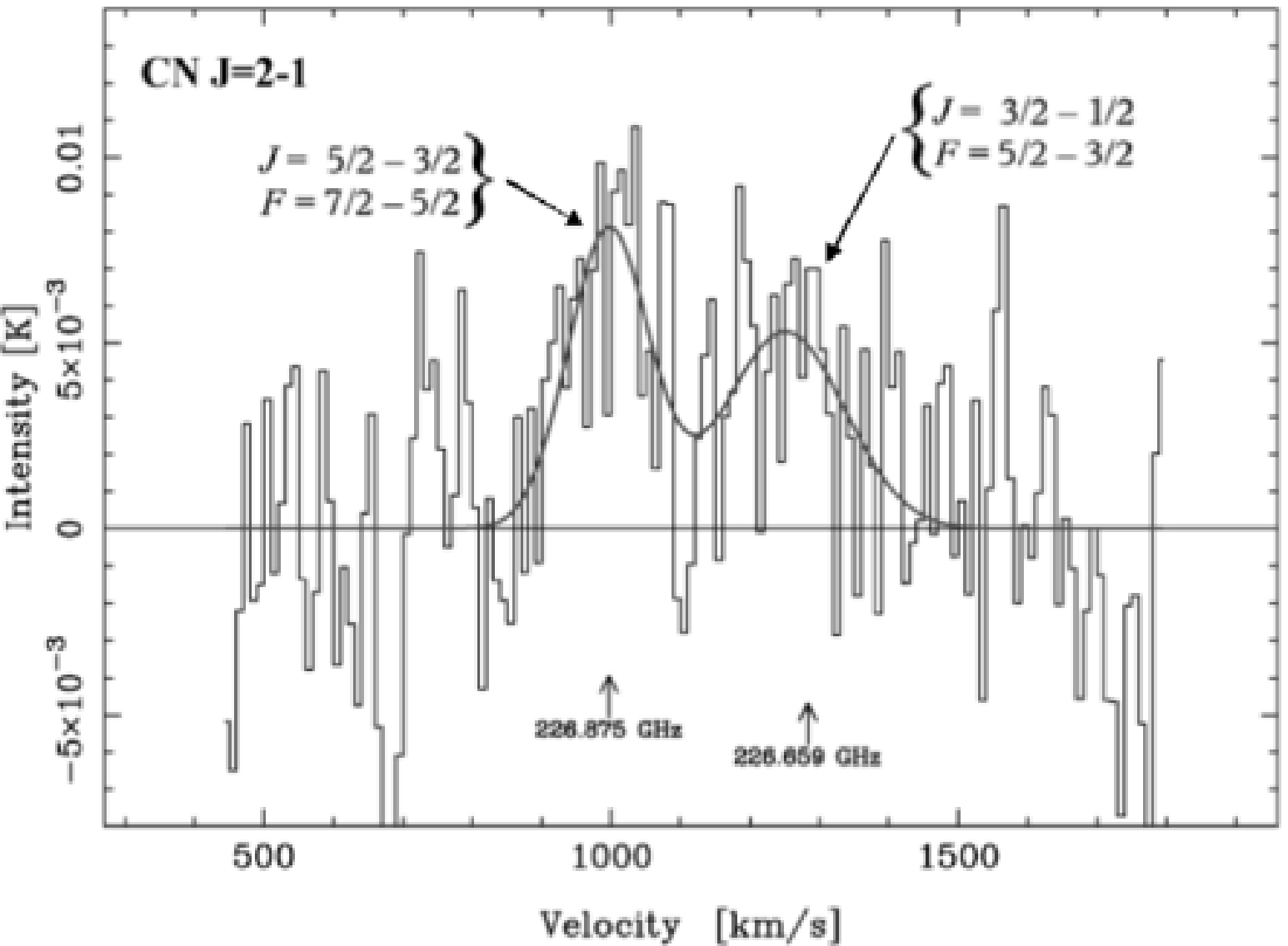}\hspace*{\fill}\\
  
  \hspace*{\fill}\includegraphics[width=6.5cm]{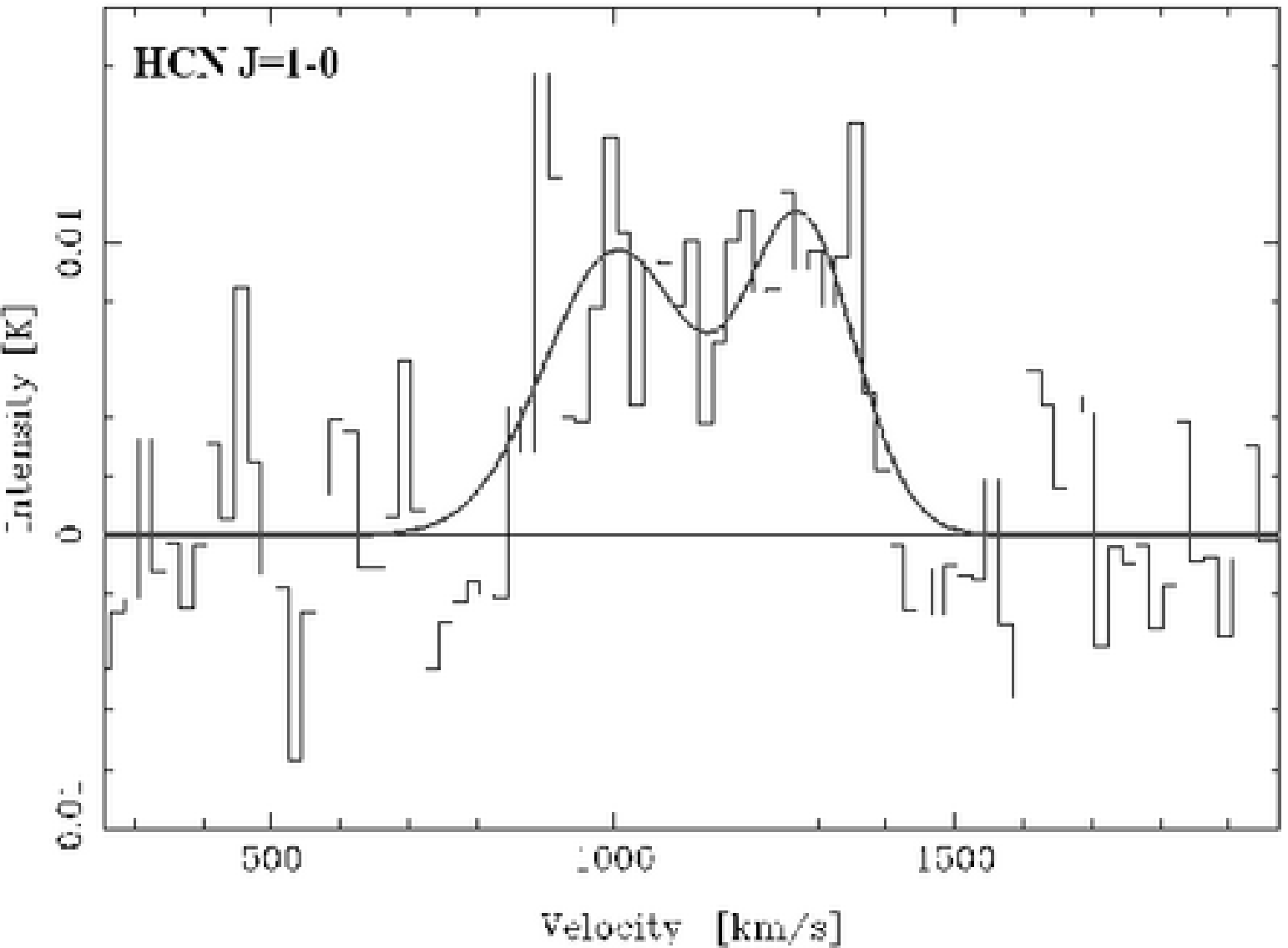}%
  \hfill\includegraphics[width=6.5cm]{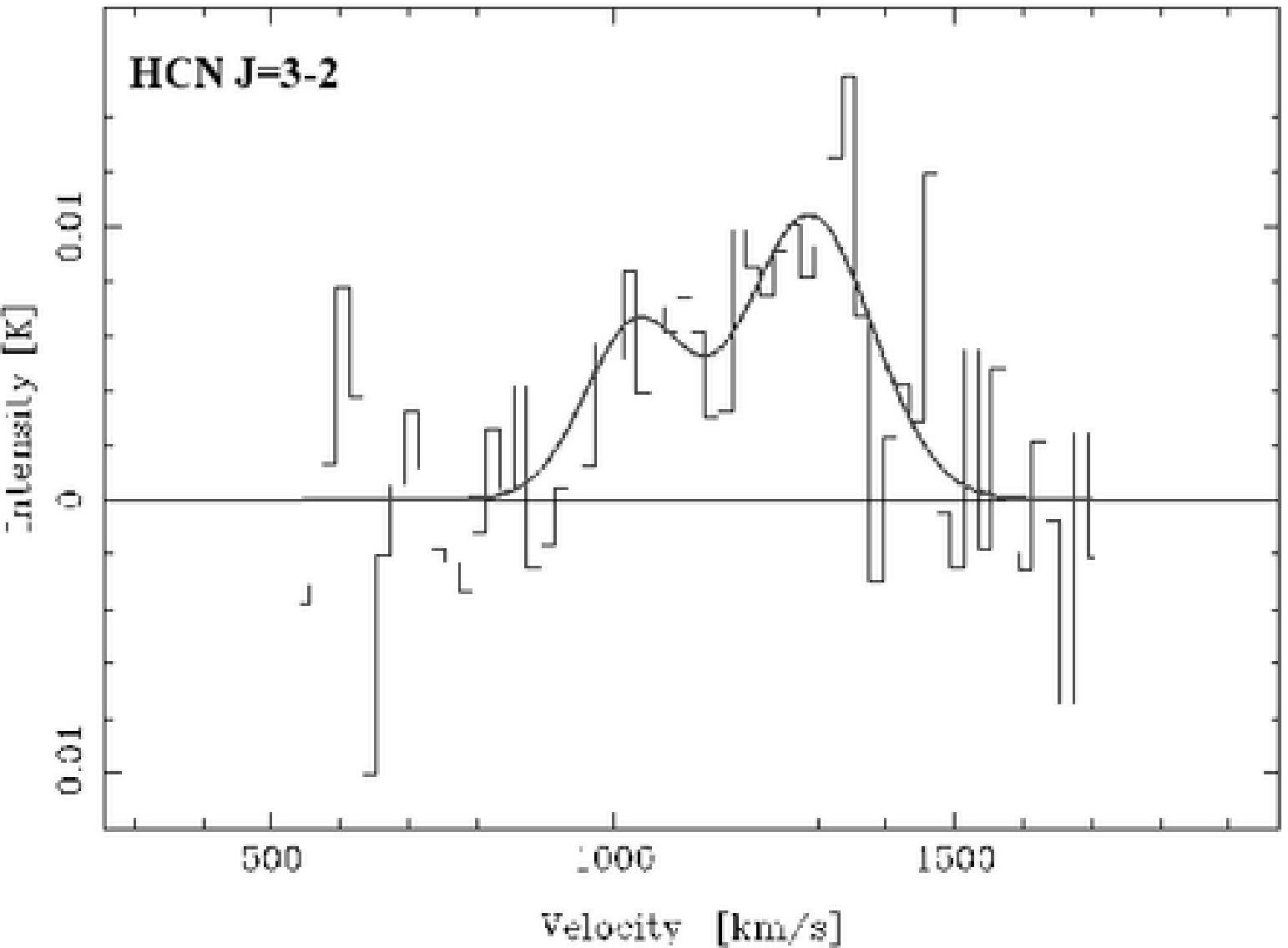}\hspace*{\fill}\\
  
  \hspace*{\fill}\includegraphics[width=6.5cm]{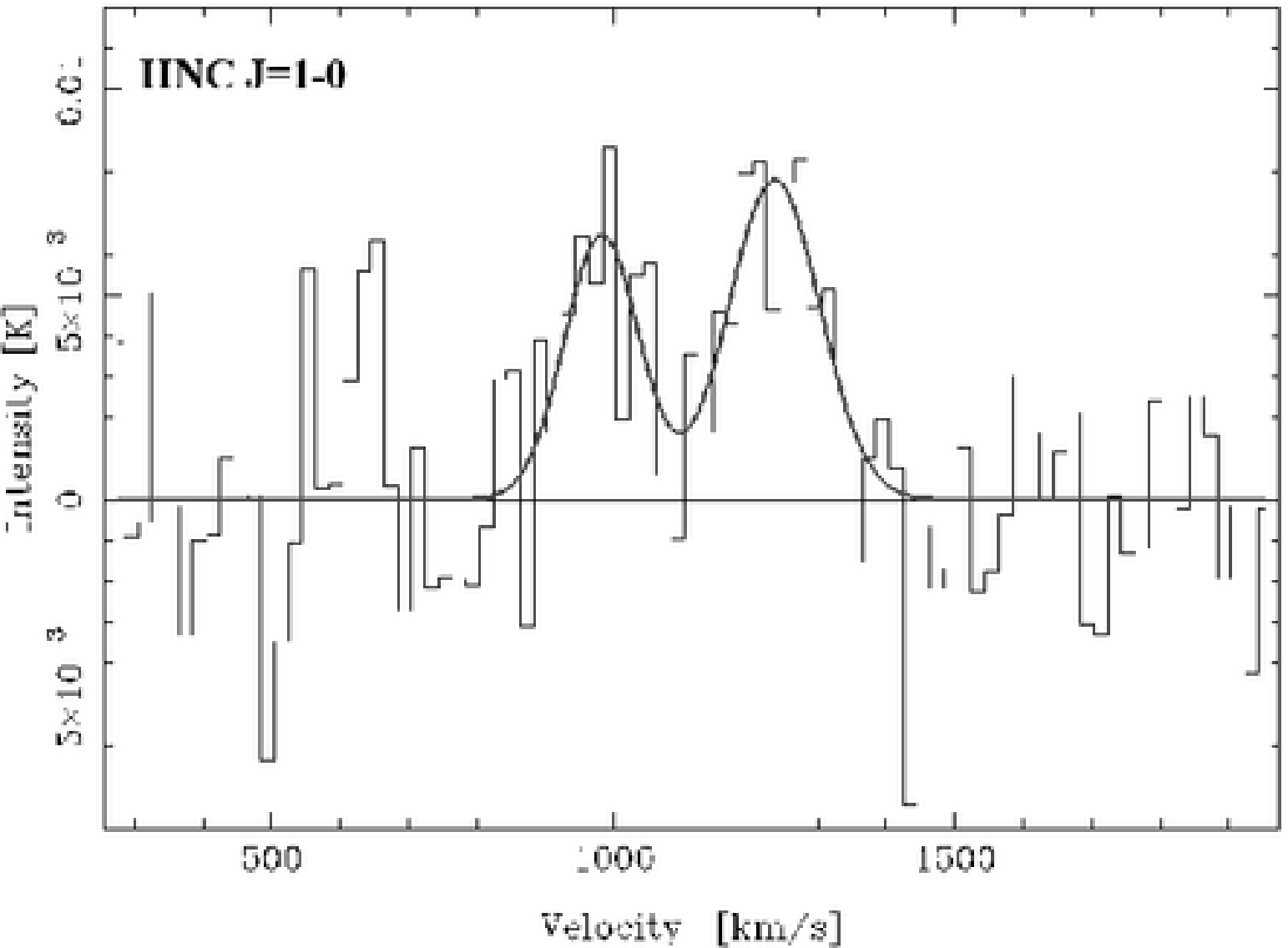}%
  \hfill\includegraphics[width=6.5cm]{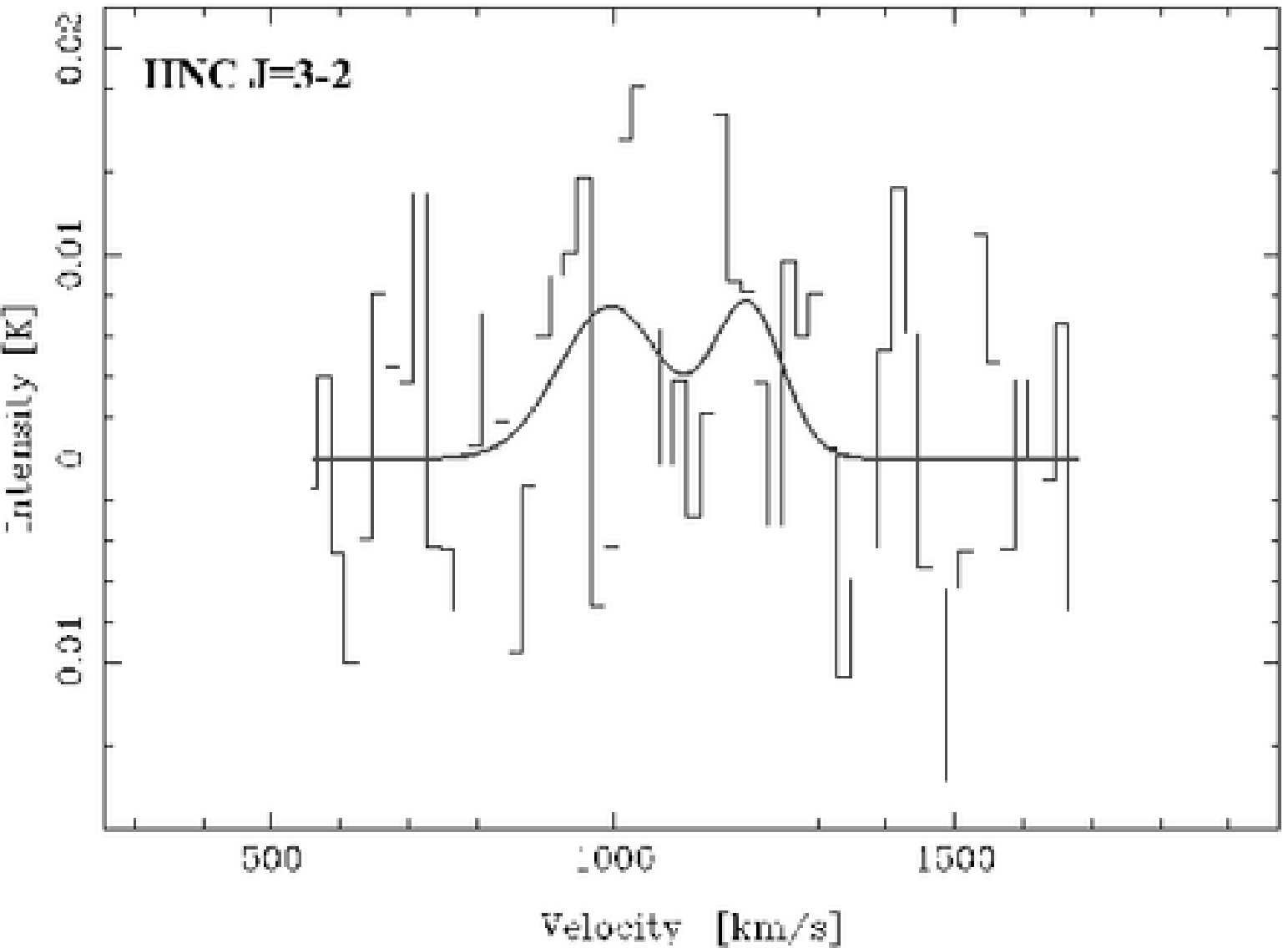}\hspace*{\fill}\\

  \caption{\footnotesize{Molecular line emissions in \textbf{NGC~3079}. The 
  velocity resolution was set to 20 \kms for HNC and HCN, and to 10 \kms 
  for CO and CN. The spectra are centered with respect to the heliocentric 
  systemic velocity $\rm v_{sys} = 1116$ \kms. Emission from four structures 
  are observed in the CO lines. Only the main spingroup is observed in the CN 1--0
  line. The second spingroup is on the right edge of the spectrum, out of the bandwidth.
  Instead, both spingroups are observed in the CN 2--1 line, although the second spingroup
  overlaps the double structure of the CN emission. The both transitions of HCN 
  and HNC have similar line shapes, which indicates that their emissions emerge from the
  same region.}}
  \label{fig:3079-spectra}
\end{figure*}

The two strongest spingroups of the CN 1--0 line are also 
detected in this galaxy, with a double peak structure at the
nuclear region. The strongest peak was used as reference for the main 
spingroup. We first fit two gaussian components to the main group and
then we set the line width of the second spingroup, observed
at around 2400 \kms, to the value found for the main
spingroup. The proportion found between the amplitude of the spingroups is 
about 0.3 instead of the expected value of 0.43, according to NIST. 
The data around the second spingroup are strongly affected 
by noise, so its central velocity is shifted by about 50 \kms~from the 
expected value of about 800 \kms.

\begin{figure*}[!htp]
 
  \hspace*{\fill}\includegraphics[width=6.5cm]{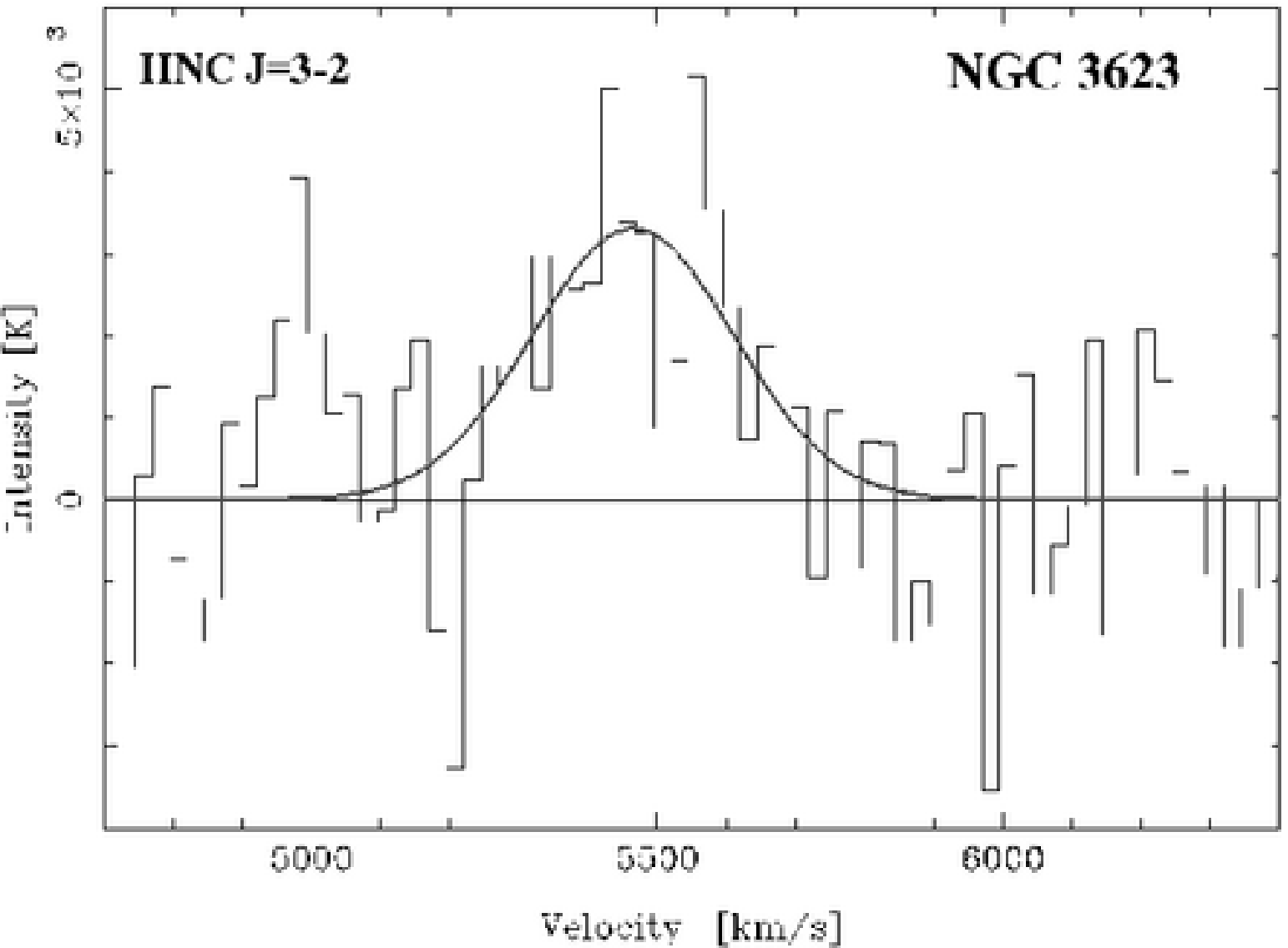}%
  \hfill\includegraphics[width=6.5cm]{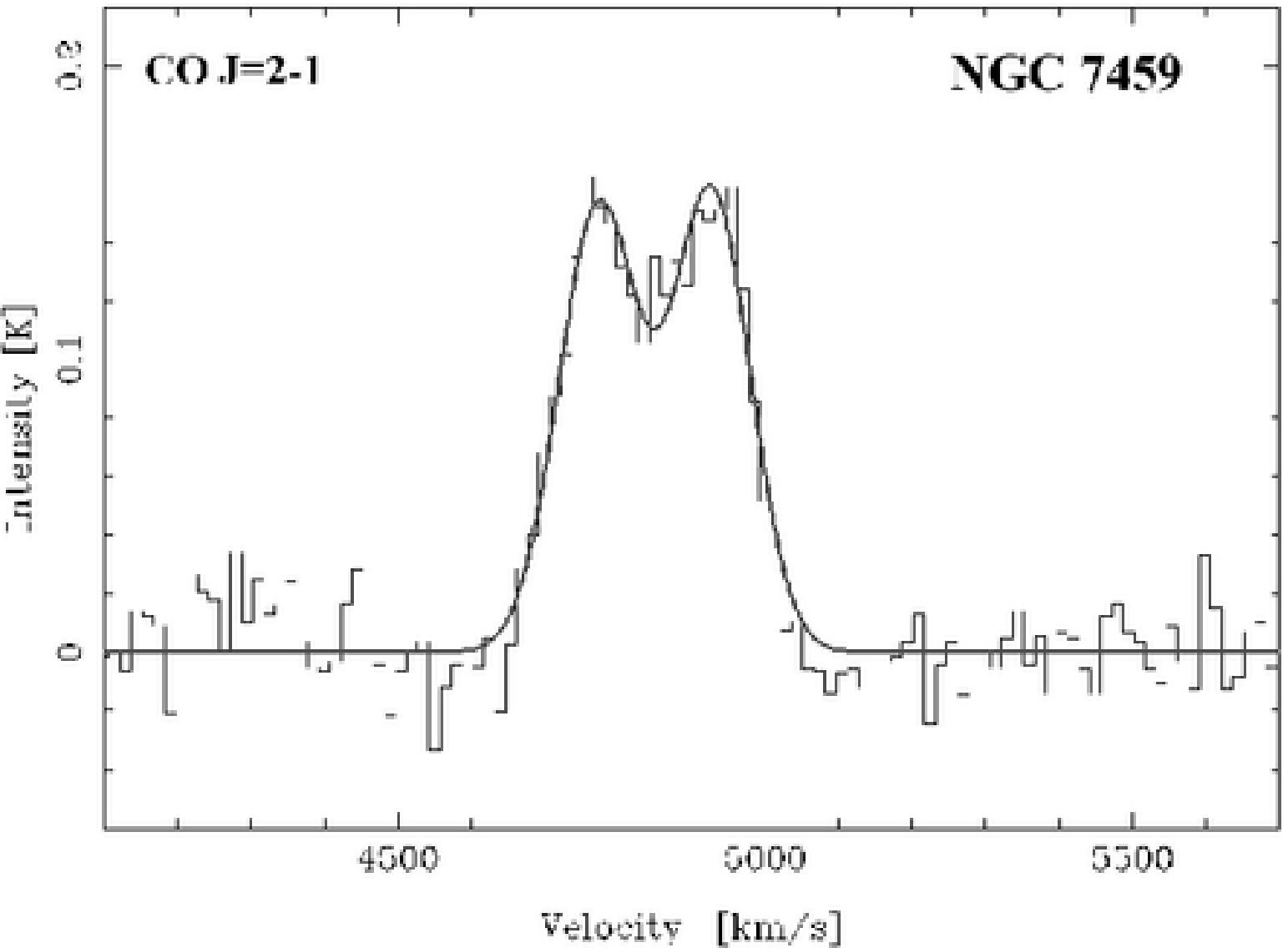}\hspace*{\fill}\\
  
  \hspace*{\fill}\includegraphics[width=6.5cm]{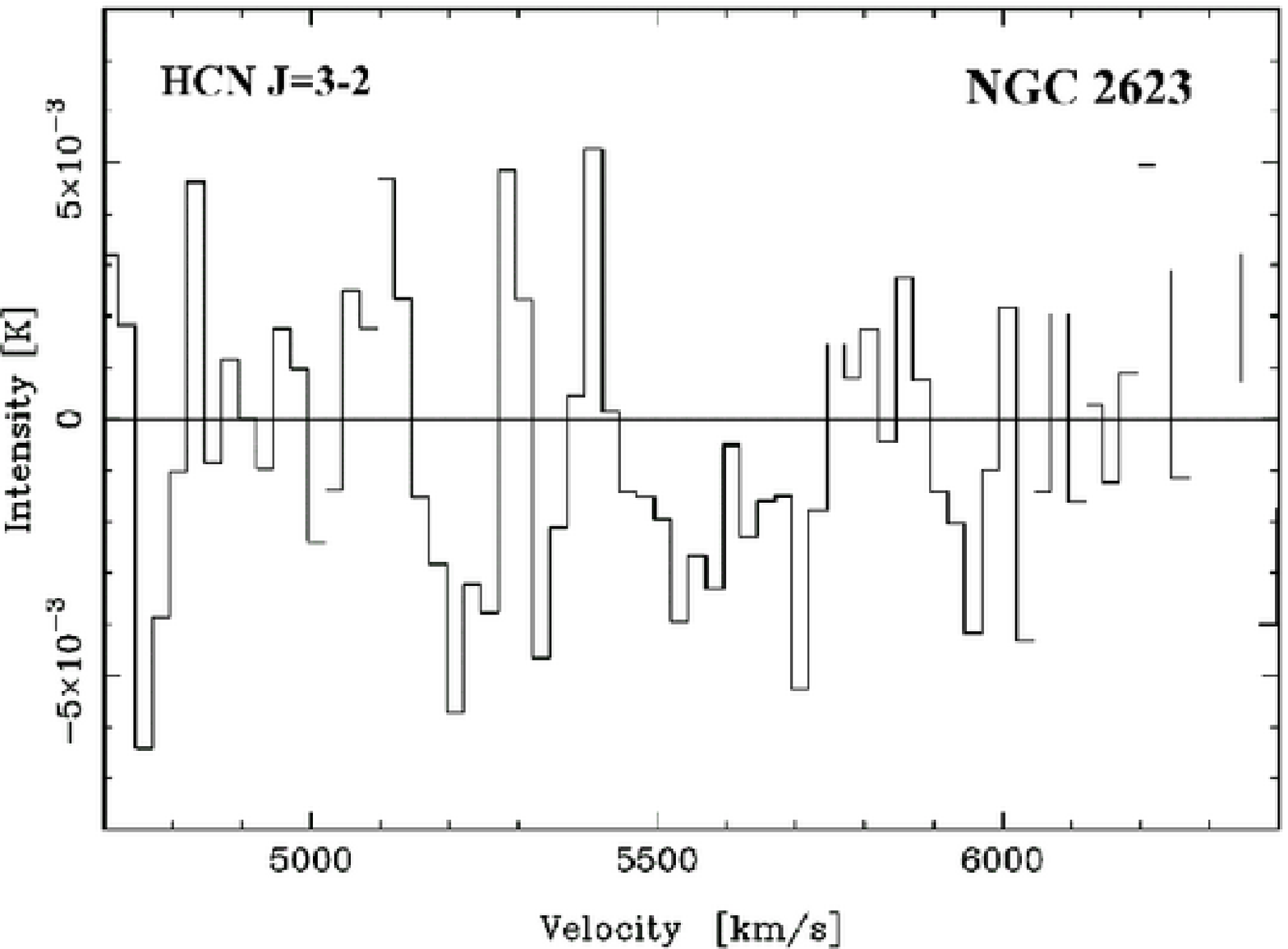}%
  \hfill\includegraphics[width=6.5cm]{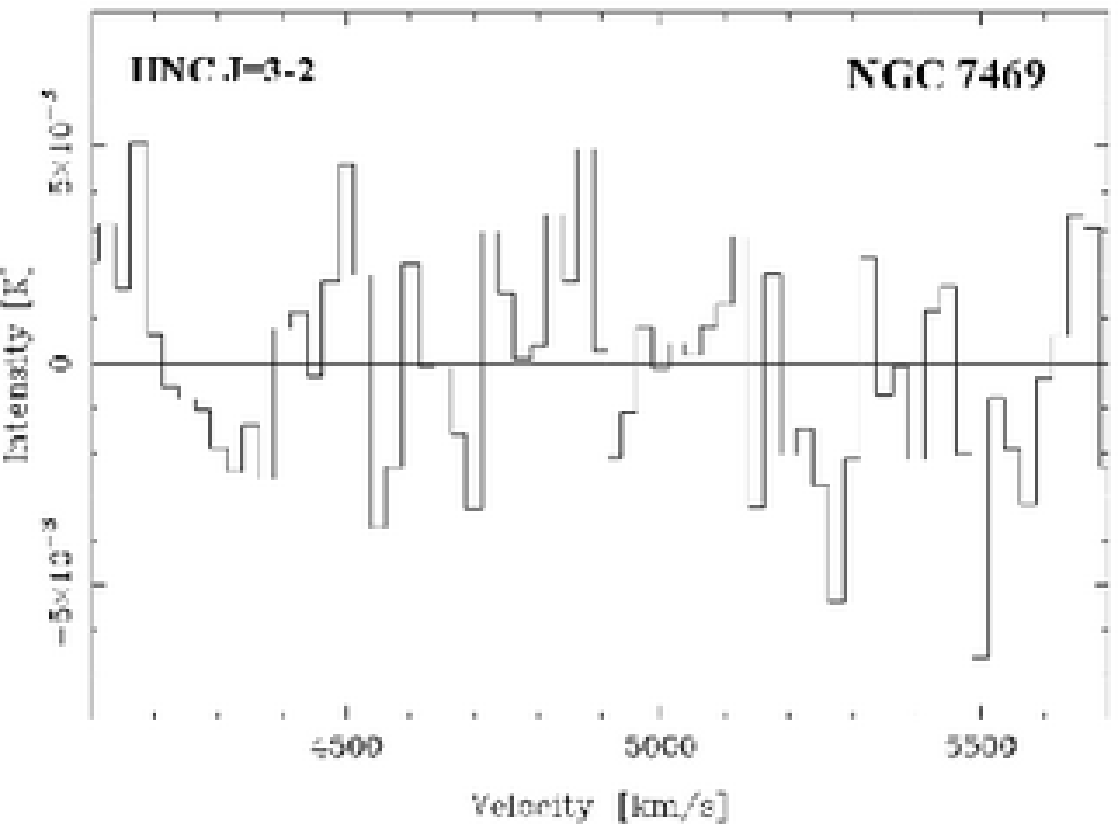}\hspace*{\fill}\\  
  
  \caption{{\footnotesize Molecular line emissions in \textbf{NGC~2623} (\emph{left}) 
  and \textbf{NGC~7469} (\emph{right}). The velocity resolution was set to 25 \kms for 
  HNC and to 15 \kms for CO. The spectra are centered with respect to the 
  heliocentric systemic velocities $\rm v_{sys} = 5535$ and $\rm v_{sys} = 4892$ 
  \kms for NGC~2623 and NGC~7469, respectively. We do not detect HCN 3-2 emission 
  in NGC~2623 nor HNC 3-2 in NGC~7469. Their observed intensities were less than 
  $2\sigma$.}}
  \label{fig:non-detection}
\end{figure*}

In the CN 2--1 line, the second spingroup overlaps with the 
double structure itself. The line widths of the two spingroups 
are set according to the value found for the CN 1--0 line.
Given their combined intrinsic line 
strengths, the integrated line intensity ratio of the two 
spingroups is expected to be about 1.85 (assuming optically 
thin emission). If we also set the line width of the second peak
of the double structure at 212 \kms, as found for CN 1--0, we get a 
proportion of about 0.4 between the two spingroups. Instead, if we
set the amplitude of the second spingroup to 5.7 mK (the expected proportion)
and we fit the line width of the double peak structure, we get a value
of 181 \kms, which is about 30 \kms less than the width found in the CN 1--0
line. In any case, the center velocity of the second spingroup is shifted by 
about 30 \kms~more than the expected velocity.

The HNC 1--0 and HCN 3--2 spectra have a structure similar to that of
the CN 1--0 spectrum. Note that in CO 1--0 the double-peak is more
pronounced than in the CN, HCN or HNC spectra, where the low velocity
peak dominates. The line parameters are summarized in 
Table~\ref{tab:1365-intensities}.

\subsection{NGC~3079}

The molecular line emissions observed in NGC~3079 are shown in 
Figure~\ref{fig:3079-spectra}. The spectra are centered with respect 
to the heliocentric systemic velocity $\rm v_{sys} = 1116$ \kms. The 
velocity resolution was set to 20 \kms~for HNC and HCN, and to 10 \kms~for 
CO and CN. Four structures are observed in the CO lines, which are 
not present in the other spectra. The CO lines are picking up extended
lower density gas, compared to the CN lines (which have similar beam
size). Hence the difference in line shape.

The CN 1-0 line present a double peak structure. 
Only the main spingroup (J = 3/2 - 1/2, F = 5/2 - 3/2) of the CN 1-0 line is 
observed since the second spingroup falls beyond the bandwidth of the backend. 
In the CN 2-1 line, the two main spingroups, (J = 5/2 - 3/2, F = 7/2 - 5/2) 
and (J = 3/2 - 1/2, F = 5/2 - 3/2), are detected. But they are severely blended 
and merged with the double peak structure. In this case, only two gaussians 
were fitted because a third (middle) component presented large uncertainties in 
both the amplitude and line width, due to the noise in the spectrum. The two 
components are separated by only 254 \kms, and the proportion between their 
amplitudes is about 0.65. The difference between these and the expected values 
is attributed to the noise and the blending of the emission of the second 
spingroup.

The HCN spectrum of the $J$=1--0 and $J$=3--2 transitions agree with the HCN 1--0 
spectrum obtained by Nguyen-Q-Rieu \etal~(\cite{nguyen92}).

The HNC 1--0 spectrum is different from the tentative detection presented by 
H\"uttemeister \etal~(1995), which is broader and more intense than the spectrum 
showed here. Besides the similar line shapes, our HCN and HNC spectra
extends from 900 \kms~to 1300 \kms. Instead, the HNC spectrum showed in H\"uttemeister 
\etal~(1995) peaks at around 1300 \kms. Hence, we believe that our HNC detections are 
correct. In this case, because of the similar line shapes and 
widths (although the HNC 3--2 line is strongly affected by noise, and its peak 
intensities have uncertainties between 40\% and 50\%.), the HCN and HNC emissions 
are likely emerging from the same gas. The line parameters are summarized in 
Table~\ref{tab:3079-intensities}.

\subsection{NGC~2623 and NGC~7469}

The HNC 3--2 line emission observed in NGC~2623 is shown in the 
\emph{left top panel} of Figure~\ref{fig:non-detection}. The 
velocity resolution was set to 25 \kms. The spectrum is centered 
with respect to the heliocentric systemic velocity $\rm v_{sys} = 5535$ 
\kms (NED). After a total integration time of 2 hours we do not 
detect HCN 3--2 emission in this galaxy.

The \emph{right top panel} of Figure~\ref{fig:non-detection} shows the 
CO 2-1 spectrum observed in NGC~7469. The velocity resolution was set 
to 15 \kms. The spectrum is centered with respect to the systemic 
velocity $\rm v_{sys} = 4892$ \kms. The CO spectrum shows a double 
peak structure, like the one observed in NGC~1365. We observed the HNC 3--2
line for about 1 hour of integration time, but we do not detect any 
emission. The \emph{bottom panels} show the spectra of the not detected lines.
The line parameters of the detected transition lines are 
summarized in Table~\ref{tab:2623-7469-intensities}.


\subsection{Line intensities and ratios}

The velocity integrated intensities are showed in Table~\ref{tab:intensities}.
In order to compute the line intensity ratios 
were corrected for the different beam sizes obtained with different 
frequencies and telescopes, according to the correction factors 
defined for compact sources in Rohlfs and Wilson (\cite{rohlfs03}). 
In the case of extended sources ($\Omega_S>\Omega_{mb}$) the ratios 
were corrected for the main beam filling factor, which was approximated 
as $f_{mb}=\Omega_S/(\Omega_S+\Omega_{mb})$. The beams, as well as 
the source structures, were assumed to be gaussians.

\begin{table*}[!ht]
    \begin{minipage}{18cm}   
    \centering
      \caption[]{Velocity-integrated intensities$^{~\rm a}$.}
         \label{tab:intensities}
     		\centering
         \begin{tabular}{lcccccccc}
            \hline
            \noalign{\smallskip}
            Galaxy & $I$(CO) 1-0 & $I$(CO) 2-1 & $I$(CN) 1-0 & $I$(CN) 2-1 & $I$(HCN) 1-0$^{~\rm b}$ & $I$(HCN) 3-2 & $I$(HNC) 1-0 & $I$(HNC) 3-2 \\
            \noalign{\smallskip}
            \hline
            \noalign{\smallskip}
		NGC~3079 & 205.8$\pm$20.6 & 375.4$\pm$37.6 & 2.7$\pm$0.4 & 4.5$\pm$0.7 & 5.7$\pm$0.8 & 8.0$\pm$1.4 & 2.9$\pm$0.5 & 5.0$\pm$1.9\\
		NGC~1068 & 113.7$\pm$11.4 & 203.6$\pm$20.5 & 9.4$\pm$1.0 & 6.1$\pm$0.7 & 10.0$\pm$1.7 & 22.7$\pm$2.5 & 3.2$\pm$0.5 & 3.5$\pm$0.6\\
		NGC~2623 & 30.2$\pm$3.2$^{~\rm c}$ & - & $\lesssim$1.3$^{~\rm c}$ & - & - & $\lesssim$0.5 & 0.9$\pm$0.3$^{~\rm c}$ & 1.8$\pm$0.4\\
		NGC~1365 & 122.3$\pm$12.3 & 181.0$\pm$18.1$^{~\rm d}$ & 5.9$\pm$0.7 & 7.5$\pm$0.9 & 6.0$\pm$0.1 & 15.0$\pm$2.0 & 4.7$\pm$0.6 & -\\
		NGC~7469 & 25.6$\pm$3.8$^{~\rm e}$ & 61.5$\pm$6.4 & 1.8$\pm$0.3$^{~\rm c}$ & - & 1.5$\pm$0.3 & - & 1.1$\pm$0.2$^{~\rm c}$ & $\lesssim$0.8\\
            \noalign{\smallskip}
            \hline
         \end{tabular}
\begin{list}{}{}
\footnotesize{
\item[${\mathrm{a}}$)] The values refer to the main-beam brightness temperature, $I_{\rm mb}$, 
in [K \kms]. The errors and upper limits correspond to $1\sigma$ (defined by the r.m.s. in the spectra)
and added in quadrature to the 10\% of error considered for the main beam efficiencies, $\eta_{\rm mb}$, 
reported in Table~\ref{tab:beams} and by the Onsala Space Observatory (OSO).
\item[${\mathrm{b}}$)] HCN 1--0 integrated intensities reported by Curran et al. (\cite{curran00}) 
corrected for a main-beam efficiency of 0.65, according to the on-line values reported by OSO. Our own data are reported for NGC~3079.
\item[${\mathrm{c}}$)] Integrated intensities reported by Aalto et al. (\cite{aalto02}). The values were 
rounded to one decimal figure and corrected by the main beam efficiencies reported by OSO, 0.64, 0.45 and 0.43, of the $J$=1--0 transition of HNC, CN and CO, respectively.
\item[${\mathrm{d}}$)] CO 2--1 integrated intensity obtained by Sandqvist \etal~(\cite{sandqvist95}). A 10\% of error was assumed for the reported intensity.
\item[${\mathrm{e}}$)] CO 1--0 integrated intensity derived from Curran \etal~(\cite{curran00}) considering a main beam efficiency of 0.43 according to the values reported by OSO.
}
\end{list}

\end{minipage}

\end{table*}

The source sizes reported in Table~\ref{tab:galaxies} were estimated from 
high resolution maps available in the literature. For NGC~3079 a source size of $15''\times7.5''$ 
was estimated for the CO 1--0 emission, considering only intensities above 
15\% of the peak integrated intensity of the contour map presented by Koda 
et al. (\cite{koda02}). In NGC~1068 most of the CO 1--0 emission emerges from
the two spiral arms, with the largest extension of $\sim$40'' (e.g. Helfer \& Blitz~\cite{helfer95}, 
Schinnerer et al.~\cite{schinnerer00}). Considering intensities above 20\% of 
the peak emission of the high resolution map by Schinnerer et al. (\cite{schinnerer00}) the
source size of the CO 1-0 emission in NGC~1068 was estimated as $30''\times30''$. 
For NGC~2623 a source size of $8''\times 8''$
was estimated from the CO 1--0 map presented in Bryant et al. (\cite{bryant99}), 
which agrees well with the estimate made by Casoli et al. (\cite{casoli88}).

In the case of NGC~1365 the CO emission is concentrated in the nuclear and bar regions 
(Sandqvist et al. \cite{sandqvist95}). Hence the source size of the CO emission was
estimated as $50''\times50''$, corresponding to intensities above 20\% of the peak
emission in the CO 3--2 map by Sandqvist (\cite{sandqvist99}).



From Papadopoulos \& Allen (\cite{papadopoulos00}) and  Davies et al. (\cite{davies04})
the source size of the CO emission in NGC~7469 was estimated as $8''\times8''$, which
correspond to intensities above 40\% of the peak emission in the high resolution map
by Davies et al. (\cite{davies04}). The criteria of selection of the source size of CO 
varies depending on the gradient of the emission observed in the different sources.


\begin{table}[b]
      \caption[]{Line intensity ratios.}
         \label{tab:line-ratios}
         \begin{tabular}{lcccc}
            \hline
            \noalign{\smallskip}
            Galaxy & CO~$\frac{2-1}{1-0}$ & CN~$\frac{2-1}{1-0}$ & HCN~$\frac{3-2}{1-0}$ &
            HNC~$\frac{3-2}{1-0}$\\
            \noalign{\smallskip}
            \hline
            \noalign{\smallskip}
		NGC~3079 & 0.77$\pm$0.20 & 0.48$\pm$0.15 & 0.18$\pm$0.06 & 0.25$\pm$0.12 \\
		
		NGC~1068 & 0.81$\pm$0.21$^{\mathrm a}$ & 0.19$\pm$0.05 & 0.47$\pm$0.14 & 0.15$\pm$0.05 \\
		
		NGC~2623 & $^{\mathrm b}$ & - & - & 0.36$\pm$0.15 \\
				
		NGC~1365 & 0.84$\pm$0.12$^{\mathrm c}$ & 0.44$\pm$0.12 & 0.42$\pm$0.12 & - \\
		
		NGC~7469 & 0.97$\pm$0.28 & - & - & $\lesssim0.20$ \\

	    \noalign{\smallskip}
            \hline
         \end{tabular}
\begin{list}{}{}
\item[${\mathrm{a}}$)] For NGC~1068 we corrected the JCMT CO $J$=2--1 observations to the beam size
of the SEST CO $J$=1--0 line, applying the factor 20/44, the ratio between the respective beams. With this, the
beam dilution effect is cancelled in both transition lines.
\item[${\mathrm{b}}$)] See Casoli et al. (\cite{casoli88}).
\item[${\mathrm{c}}$)] For NGC~1365 we also corrected the CO $J$=2--1 observations to the beam size
of the CO $J$=1--0 line, applying the factor 25/44 between the respective SEST beams. 
The ratio given above is slightly higher than the one reported by Sandqvist \etal~(\cite{sandqvist95}).
\end{list}
\end{table}
%


The source size of the high density tracers CN, HCN and HNC were estimated 
through HCN maps available in the literature. In all these maps, 
we observed that the HCN 1--0 emissions emerge mainly from the nuclear region 
of the galaxies. Since there are no published maps of CN nor HNC emissions 
for the galaxies studied here, the corresponding source sizes 
were considered equal to that of the HCN emission, due to their chemical link. 
Since there are no HCN maps for NGC~1365 and NGC~2623, the factor 
$\sqrt{\Omega_S(\rm HCN)}/\sqrt{\Omega_S(\rm CO)}\approx0.33$ found in NGC~1068 
was used to estimate the source size of the HCN 1--0 emission, 
based on their corresponding $\Omega_S(\rm CO)$. The estimated
$\Omega_S(\rm HCN)$ are shown in Table~\ref{tab:galaxies}.


The line ratios were computed assuming an error of 10\% in the 
reported beam efficiencies $\eta_{mb}$, 5\% of error in the main 
beam $\theta_{mb}$ (Table~\ref{tab:beams}), and a 10\% error in 
the estimated source sizes $\theta_S$ (Table~\ref{tab:galaxies}). 
The obtained ratios are shown in Table~\ref{tab:line-ratios}. From 
these ratios we can conclude that CO, as well as the high density
tracers - CN, HCN and HNC - are subthermally excited.

\section{Discussion}

\subsection{The distribution of dense gas}

We compare the spectral shape of the high density tracers with HCN and CO 
position-velocity maps available in the literature in order to address the 
location of the dense gas.

\subsubsection{NGC~1068}

\begin{figure*}[!ht]

  \hspace*{\fill}\hspace{0.15cm}\includegraphics[width=4.43cm]{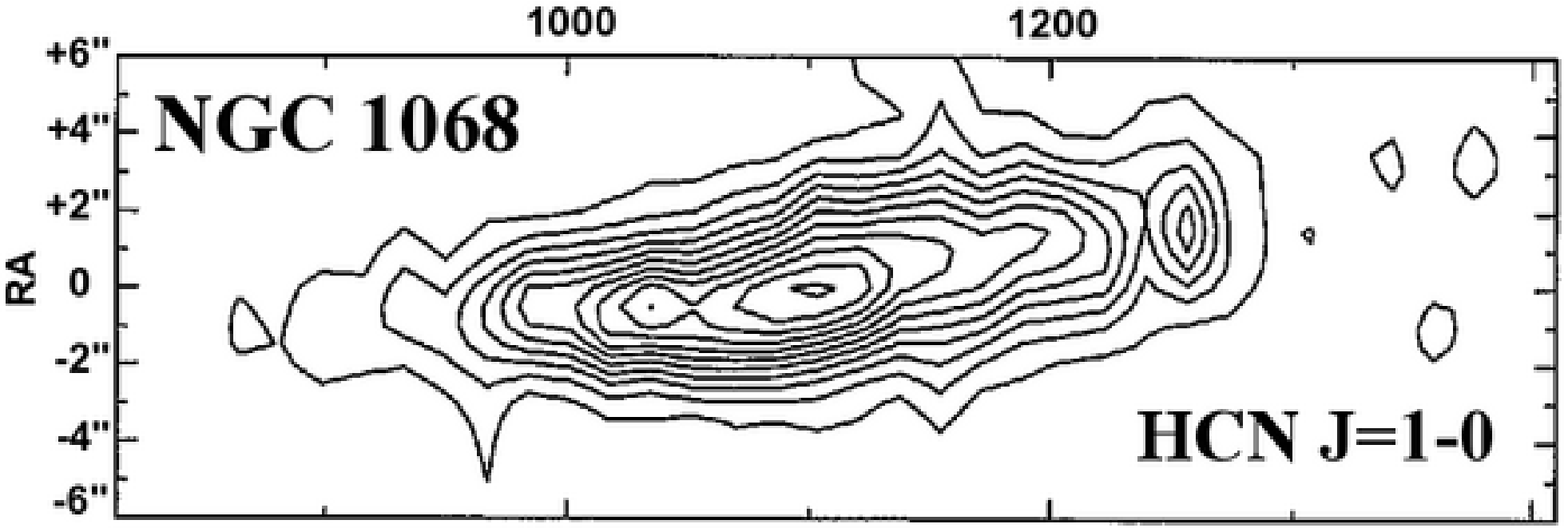}%
  \hfill\hspace{-0.06cm}\includegraphics[width=4.615cm]{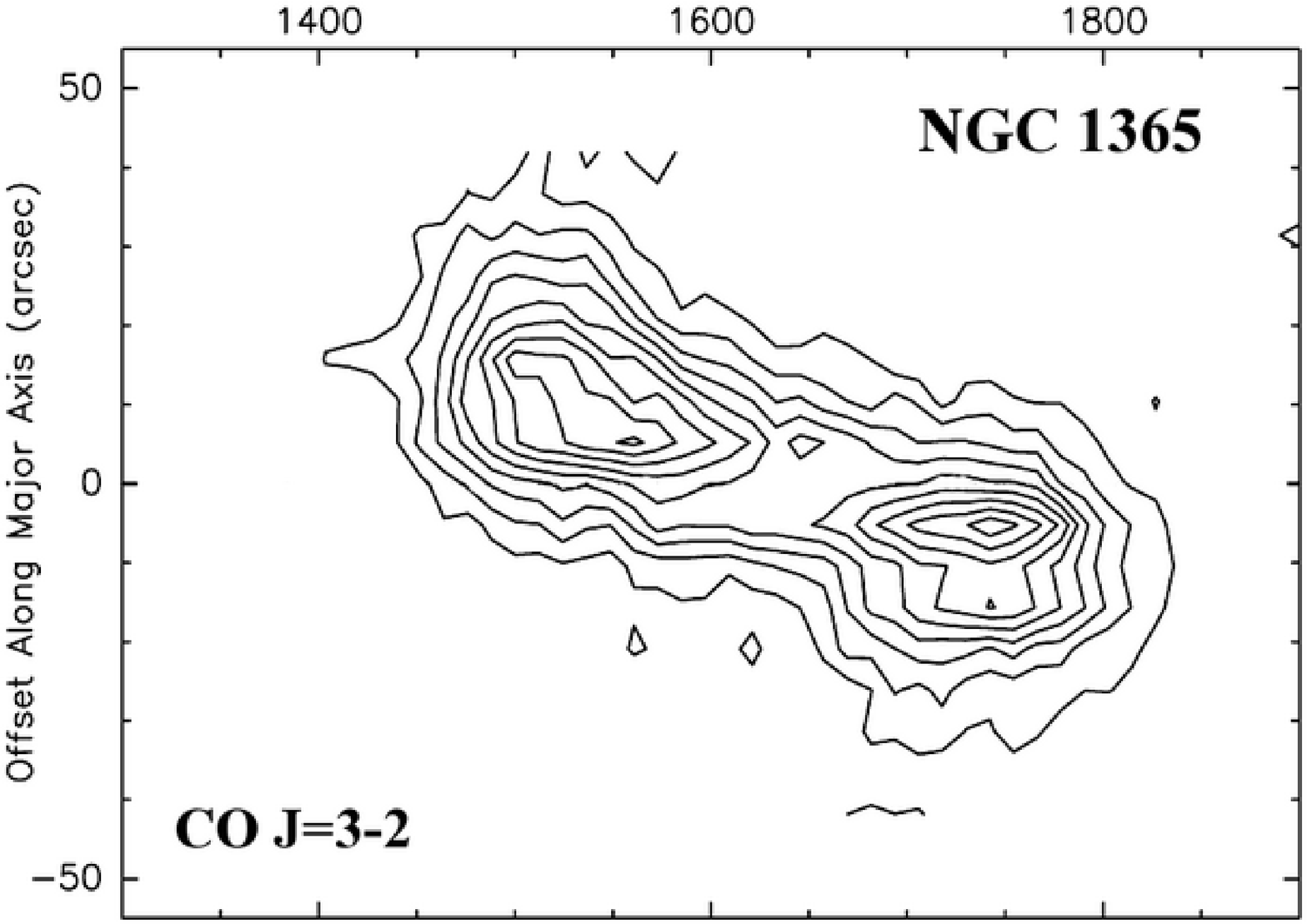}%
  \hfill\hspace{0.06cm}\includegraphics[width=4.4505cm]{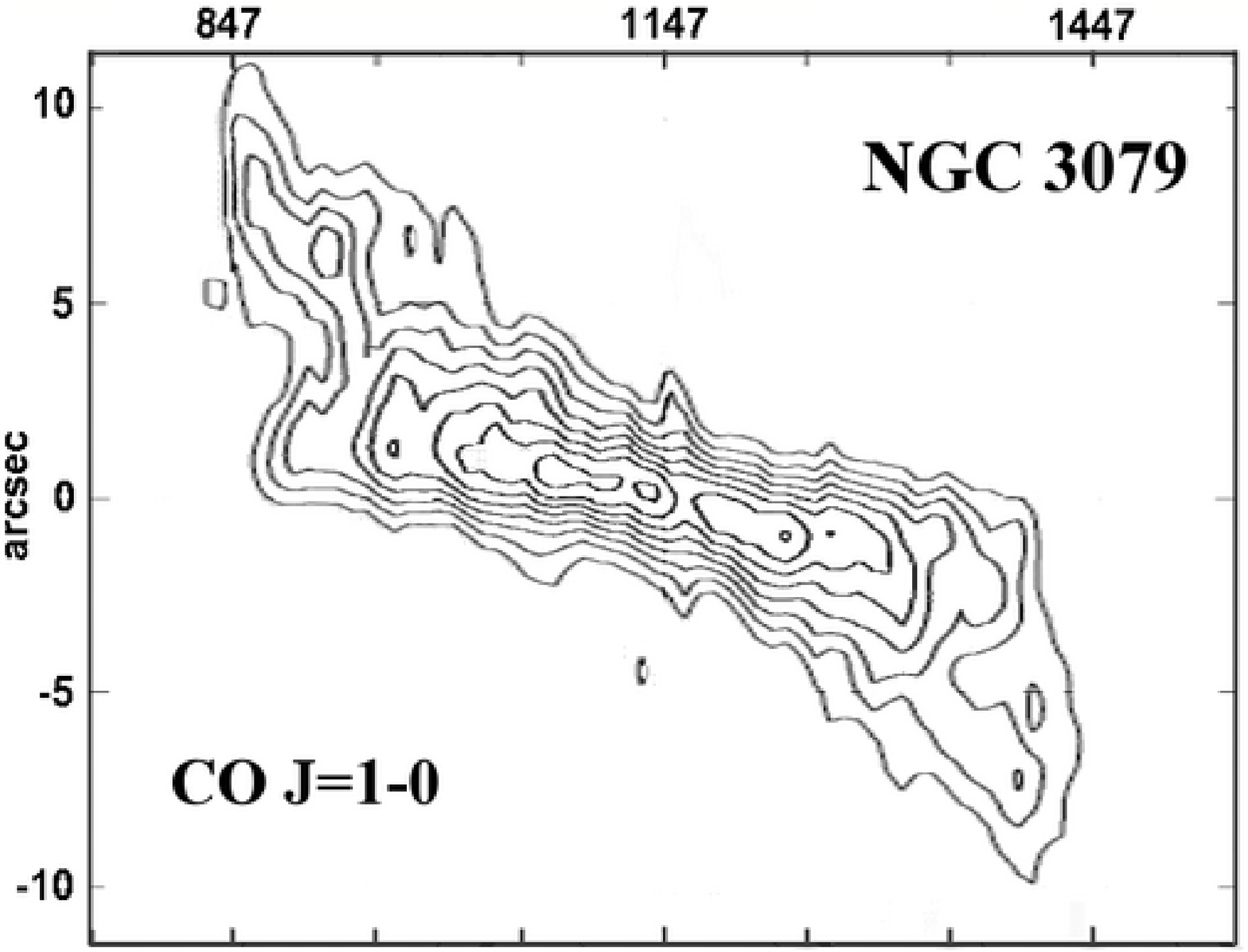}\hspace*{\fill}\\
  \vspace{-0.3cm}
  \hspace*{\fill}\includegraphics[width=4.5cm]{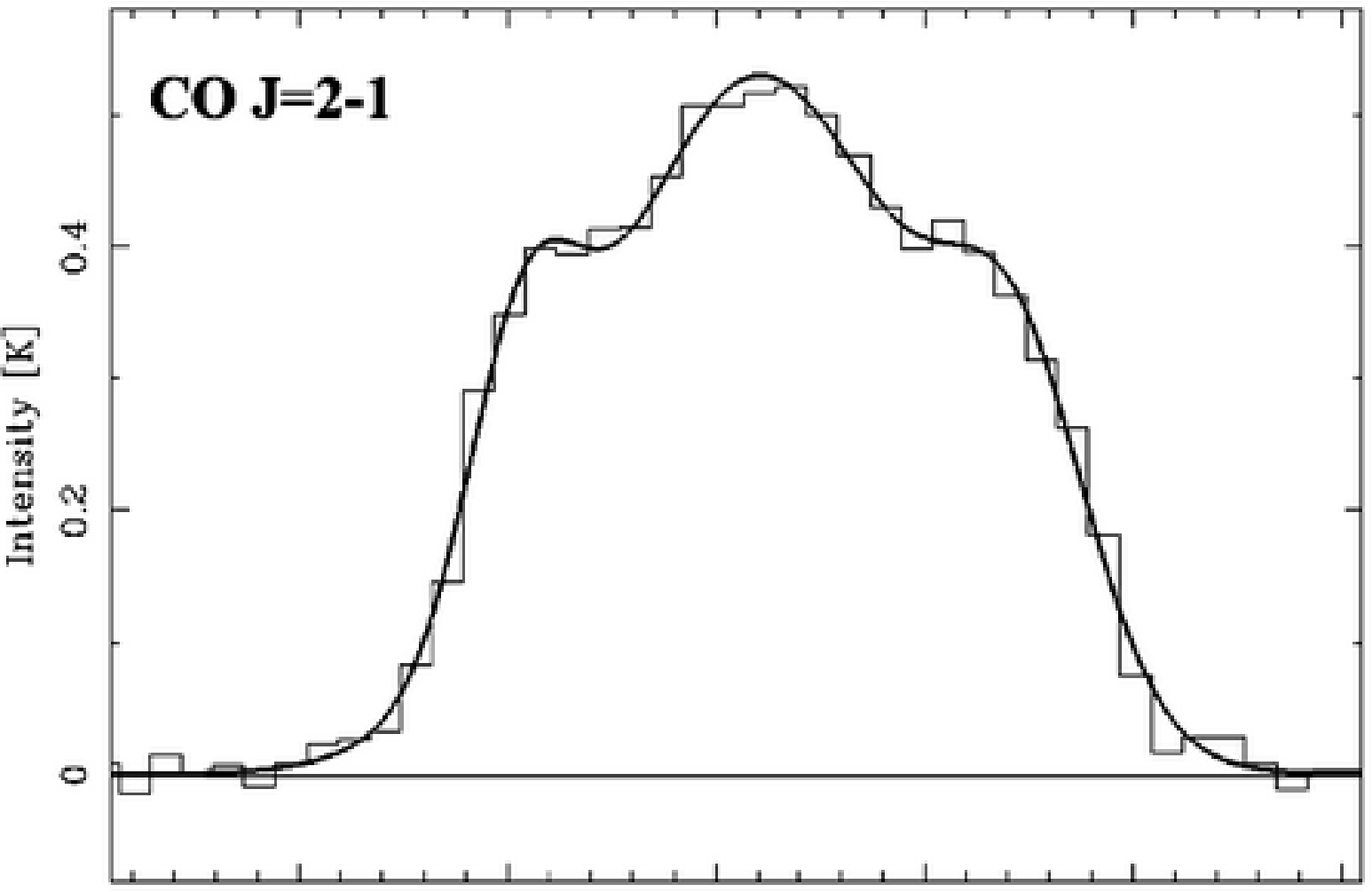}%
  \hfill\includegraphics[width=4.55cm]{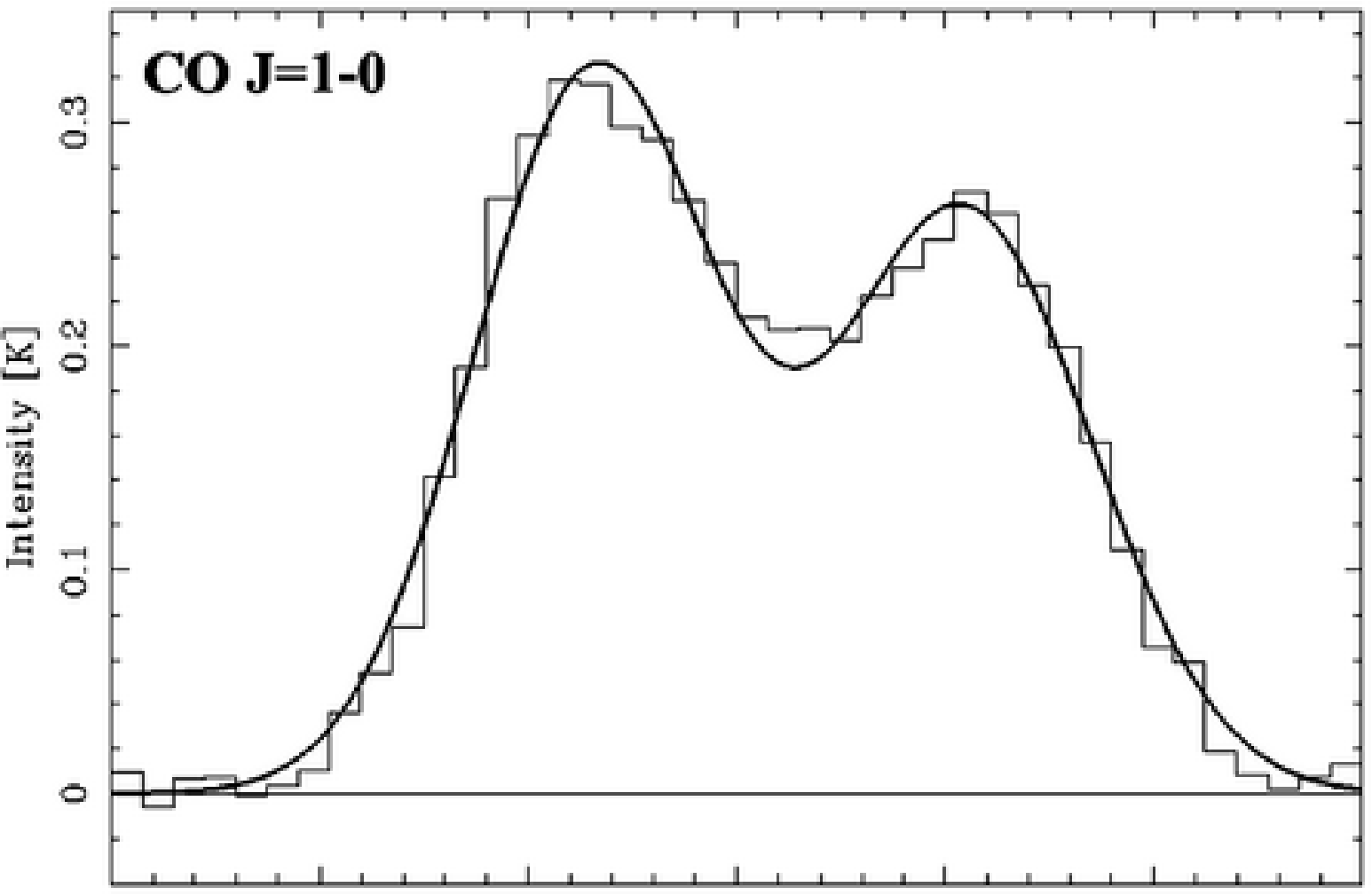}%
  \hfill\includegraphics[width=4.5cm]{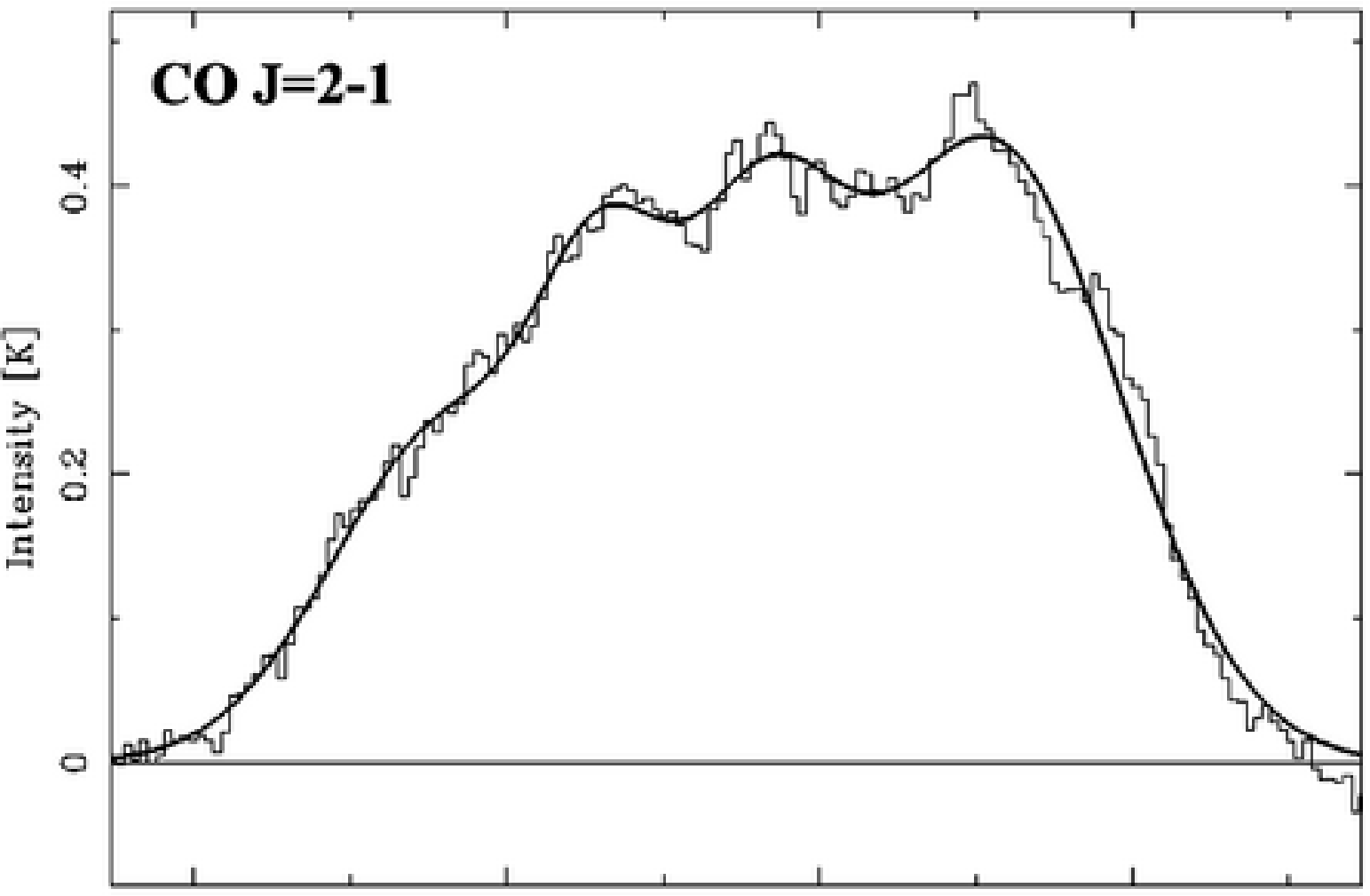}\hspace*{\fill}\\  
  \vspace{-0.3cm}
  \hspace*{\fill}\includegraphics[width=4.5cm]{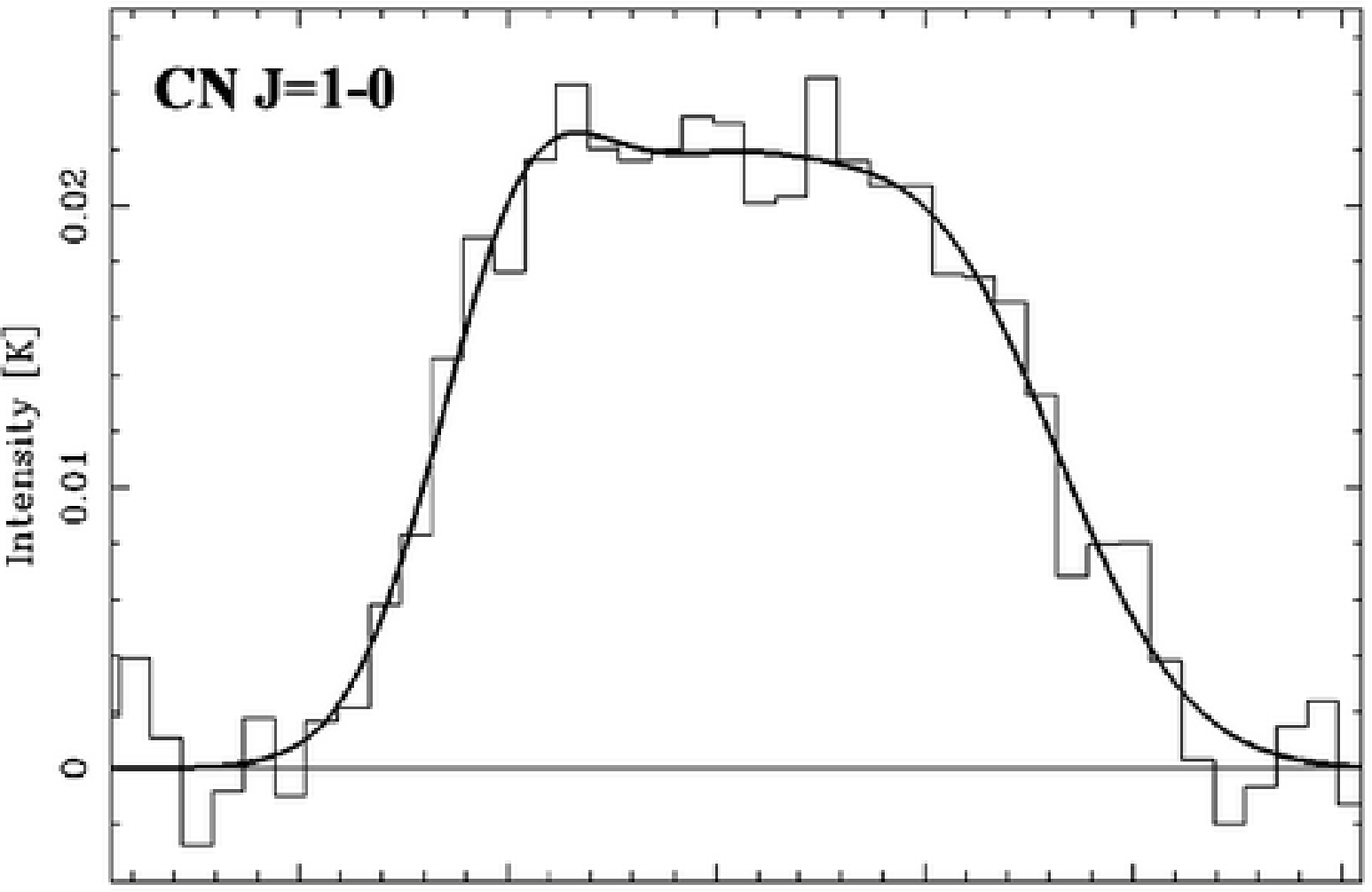}%
  \hfill\includegraphics[width=4.55cm]{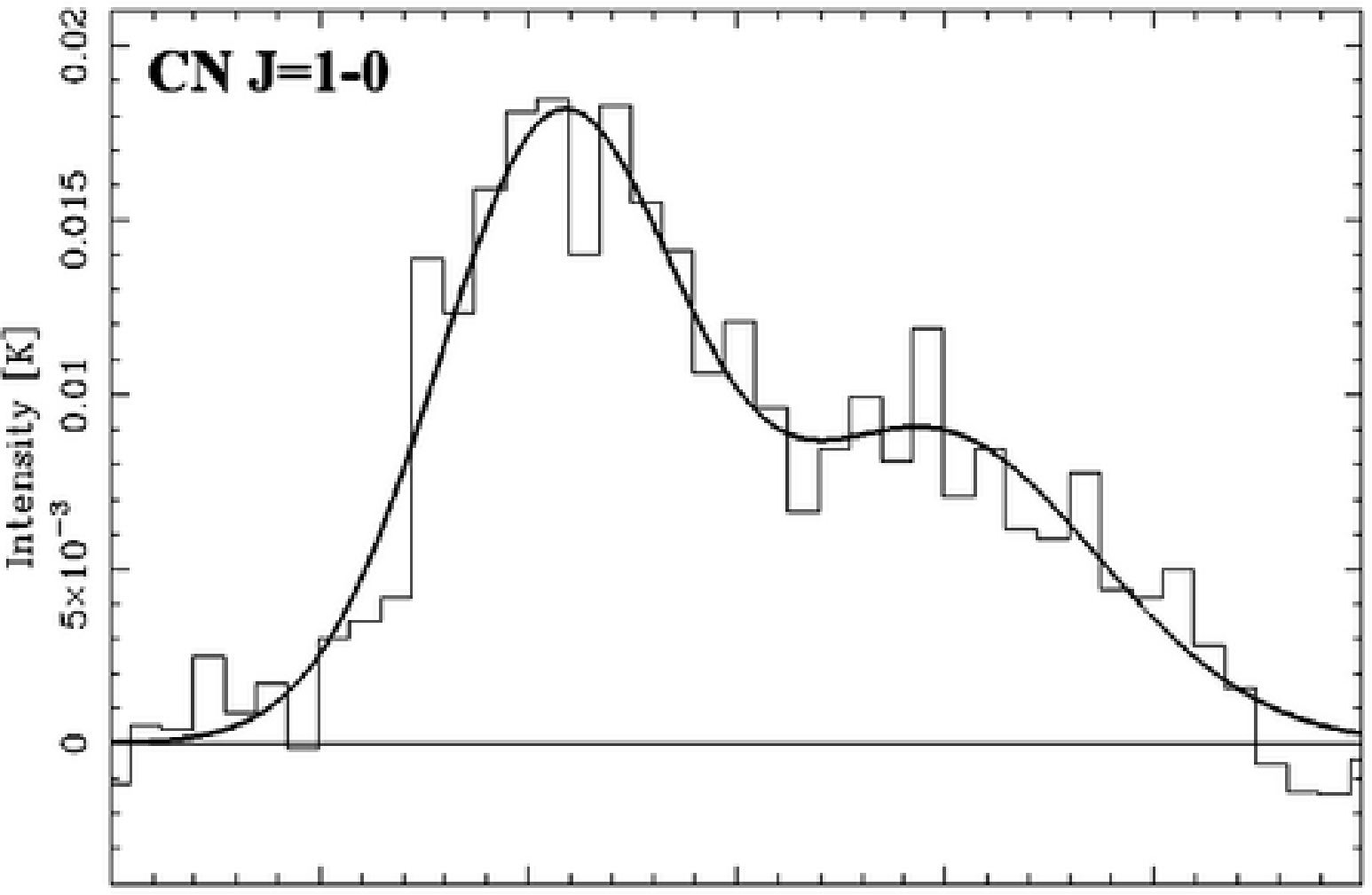}%
  \hfill\includegraphics[width=4.5cm]{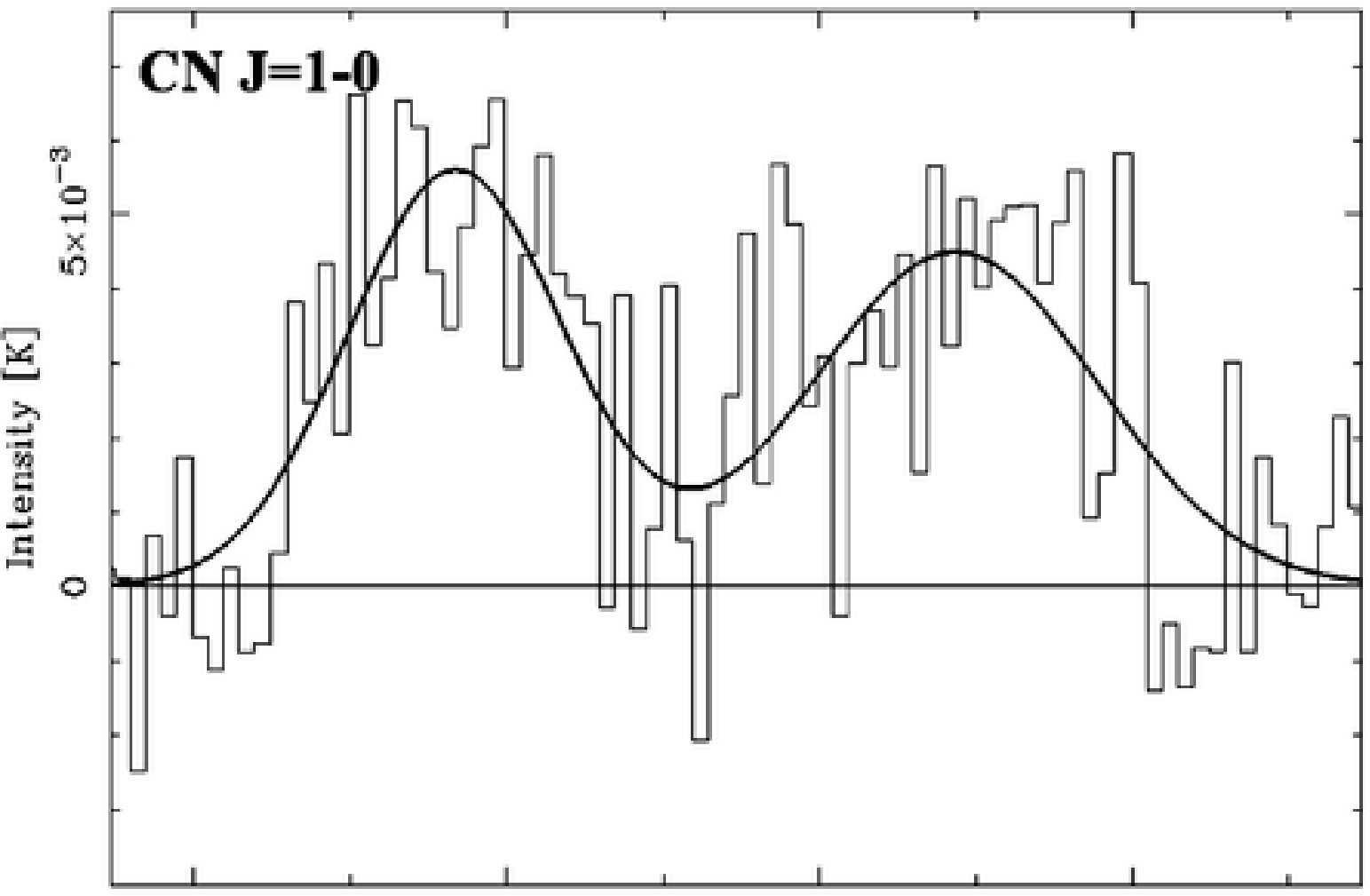}\hspace*{\fill}\\
  \vspace{-0.3cm}
  \hspace*{\fill}\includegraphics[width=4.5cm]{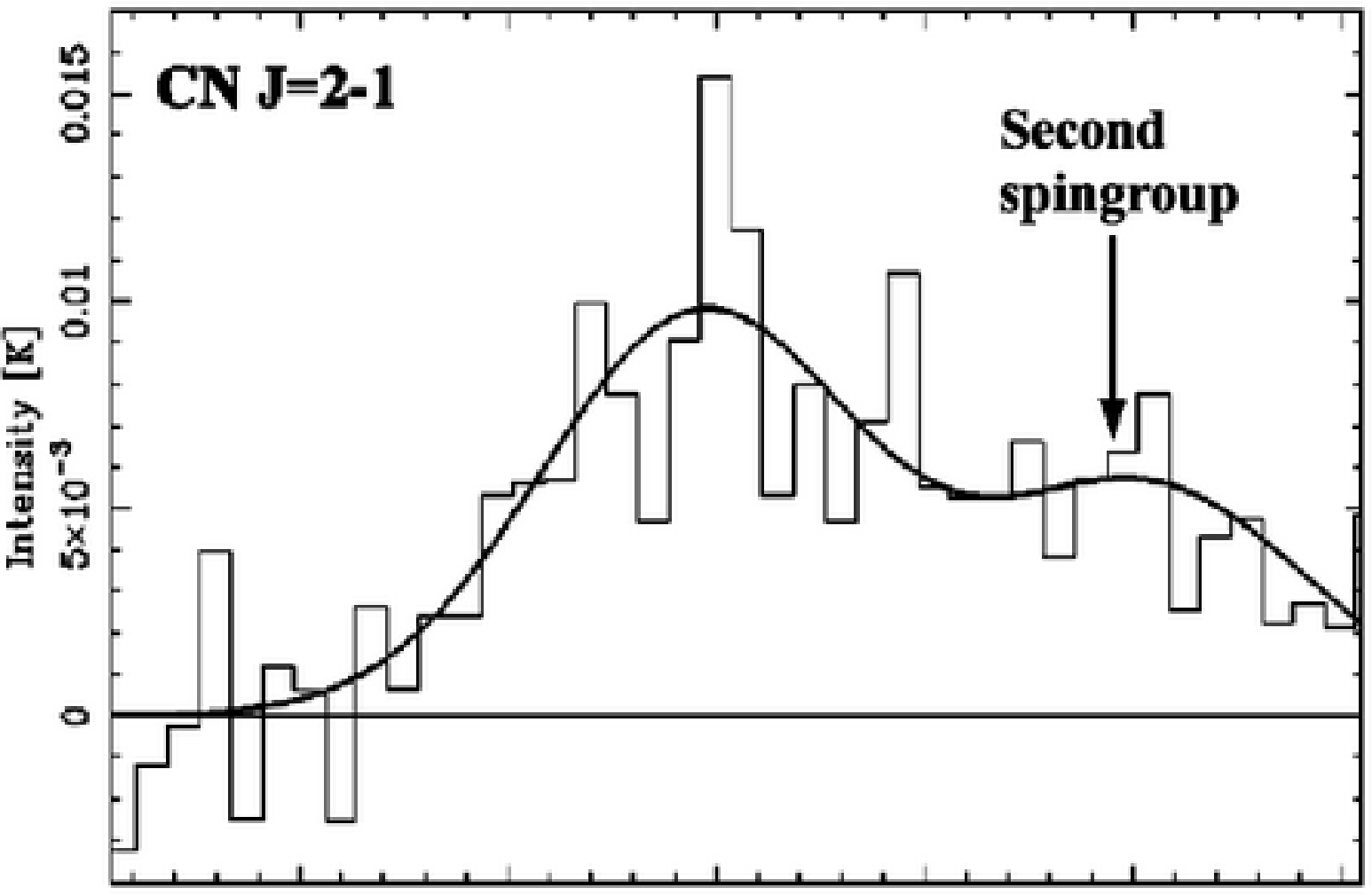}%
  \hfill\includegraphics[width=4.55cm]{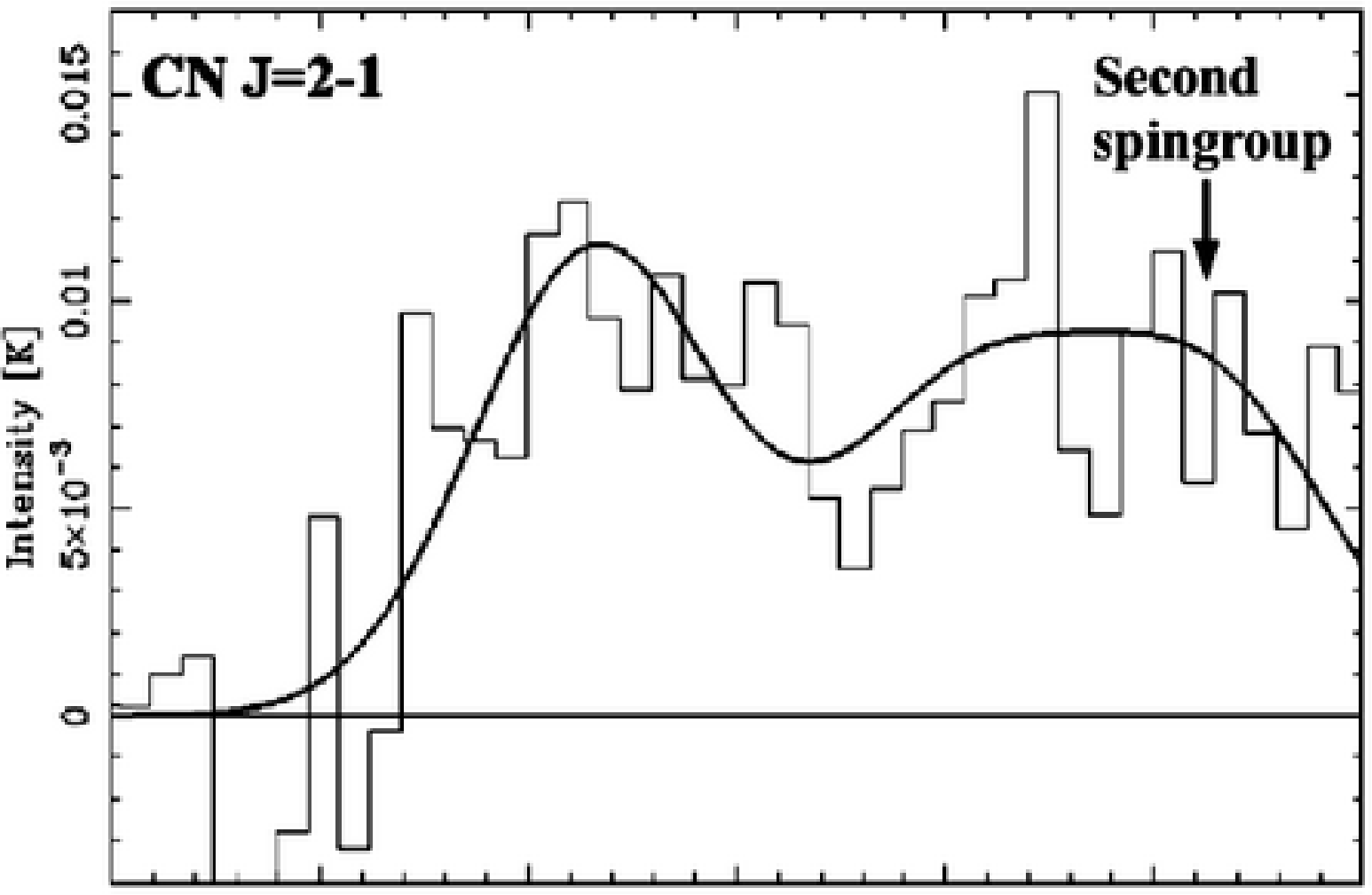}%
  \hfill\includegraphics[width=4.5cm]{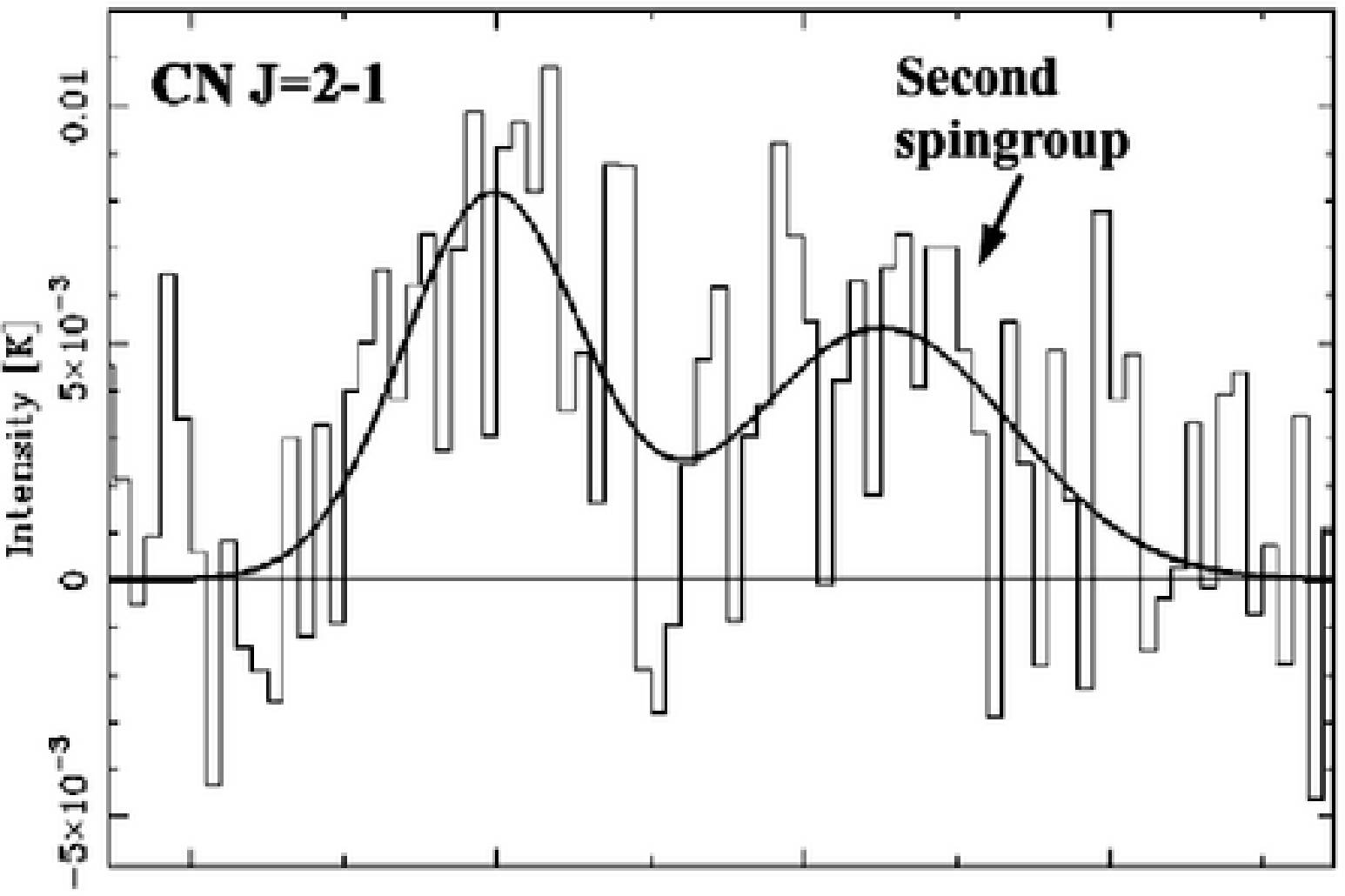}\hspace*{\fill}\\  
  \vspace{-0.3cm}
  \hspace*{\fill}\includegraphics[width=4.5cm]{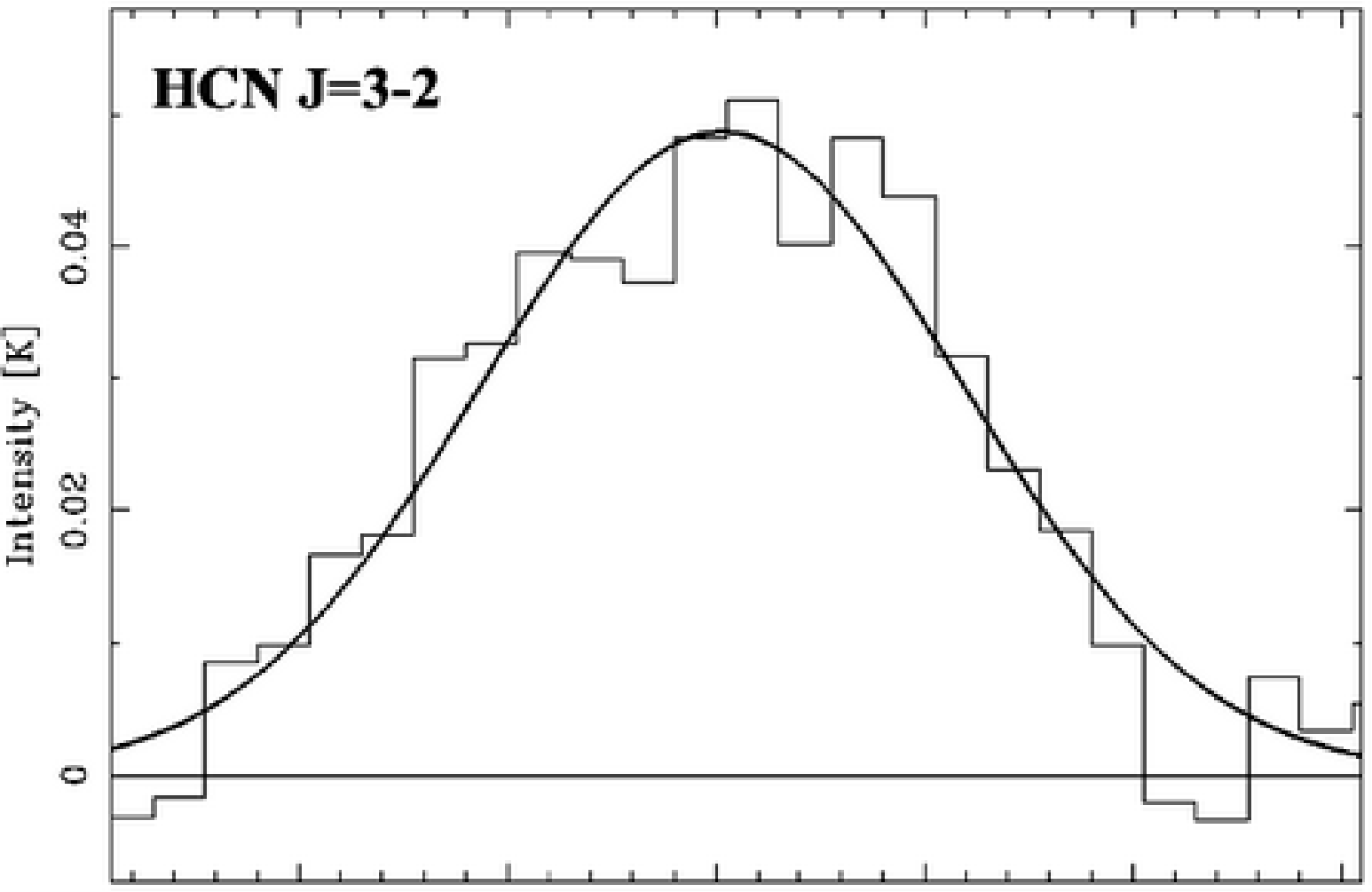}%
  \hfill\includegraphics[width=4.55cm]{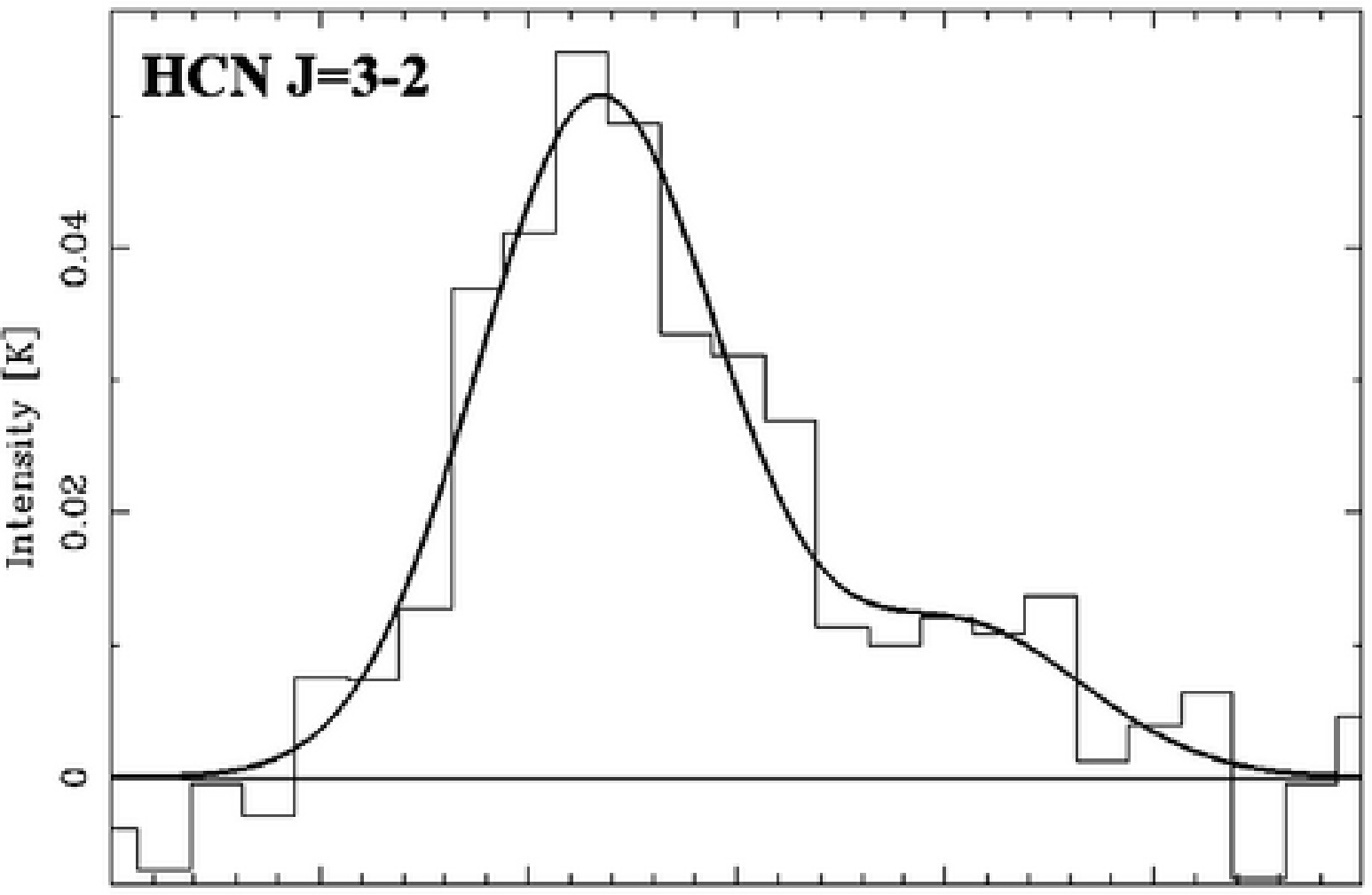}%
  \hfill\includegraphics[width=4.5cm]{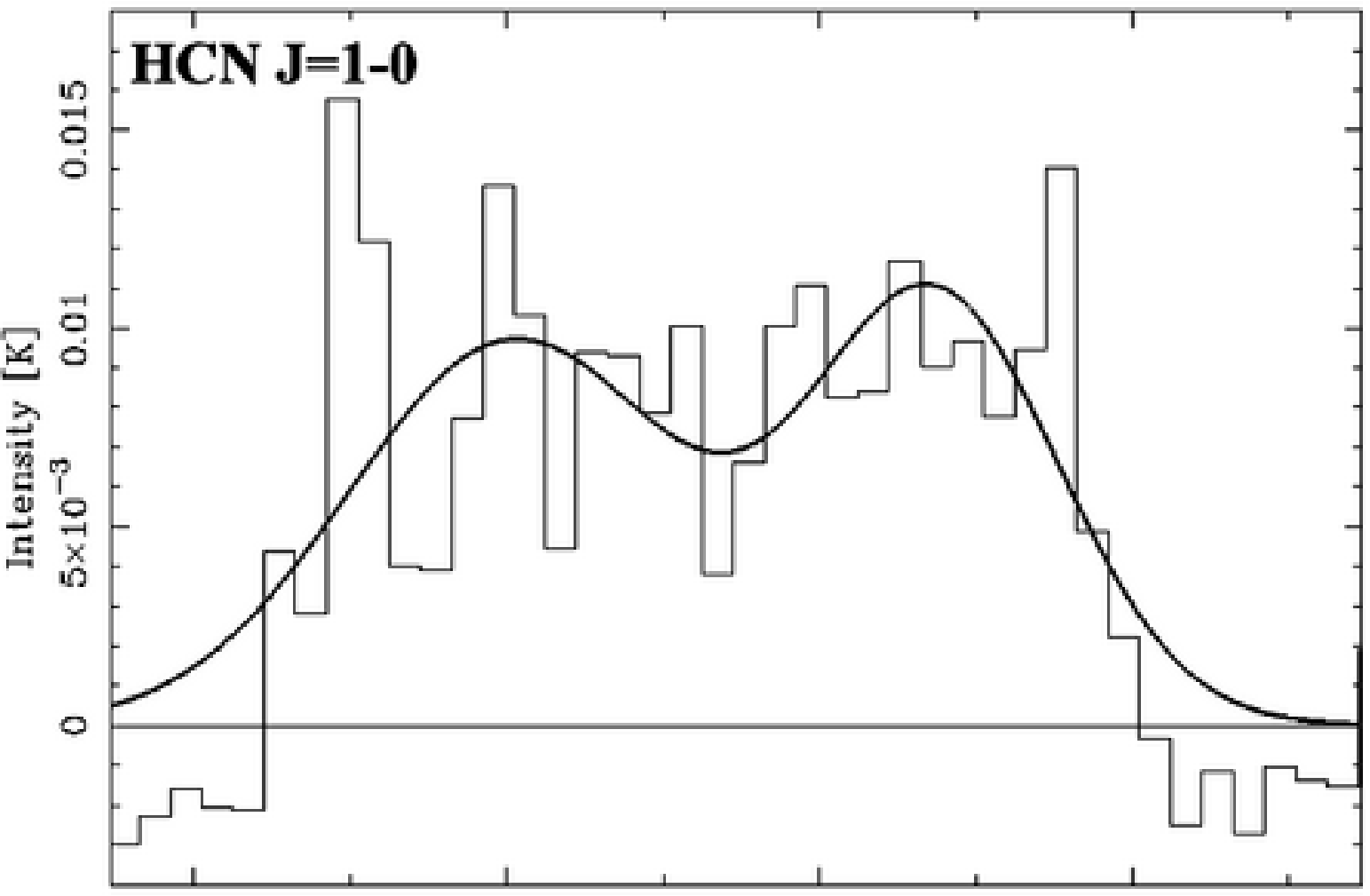}\hspace*{\fill}\\  
  \vspace{-0.3cm}
  \hspace*{\fill}\includegraphics[width=4.5cm]{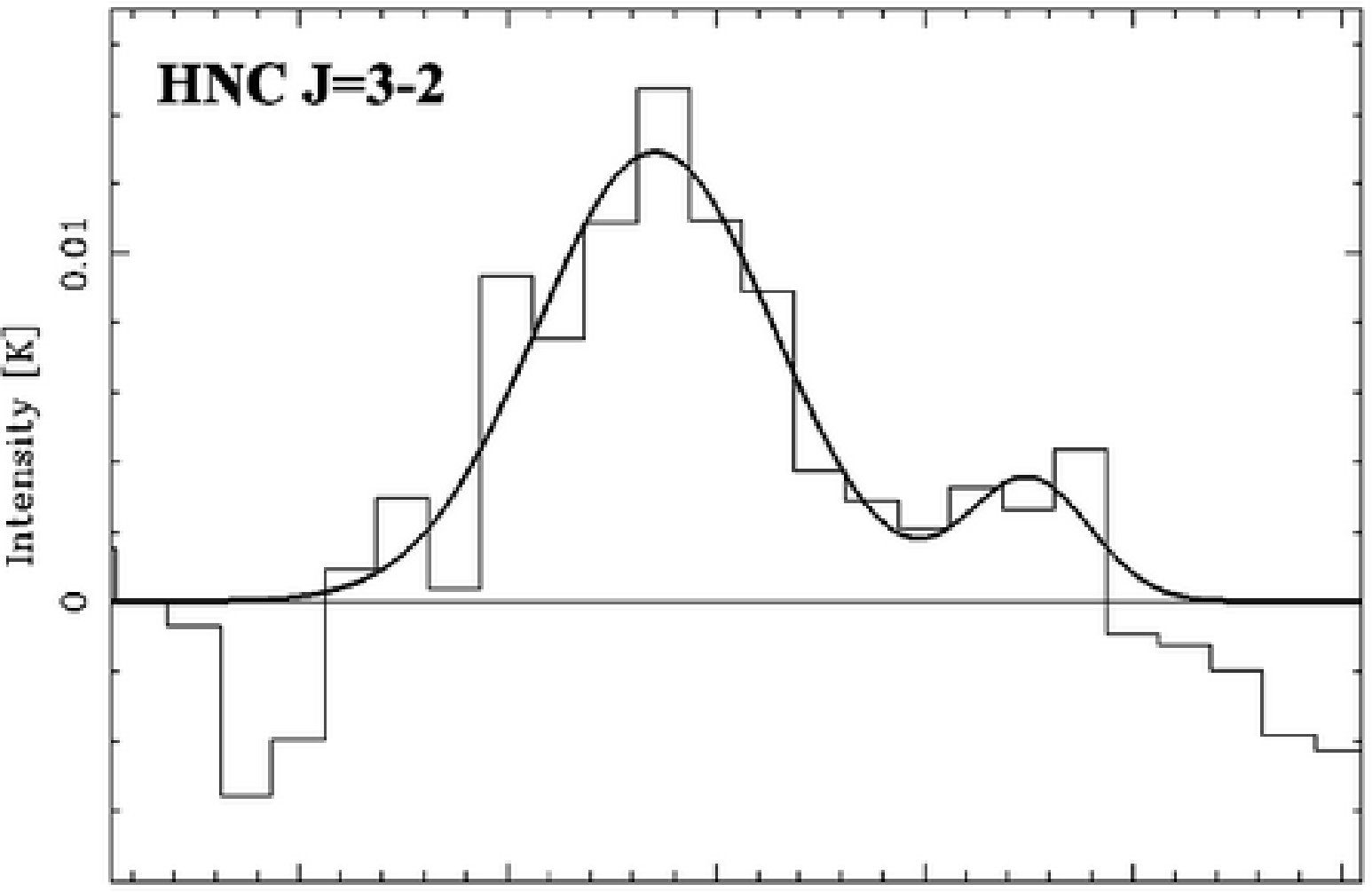}%
  \hfill\includegraphics[width=4.55cm]{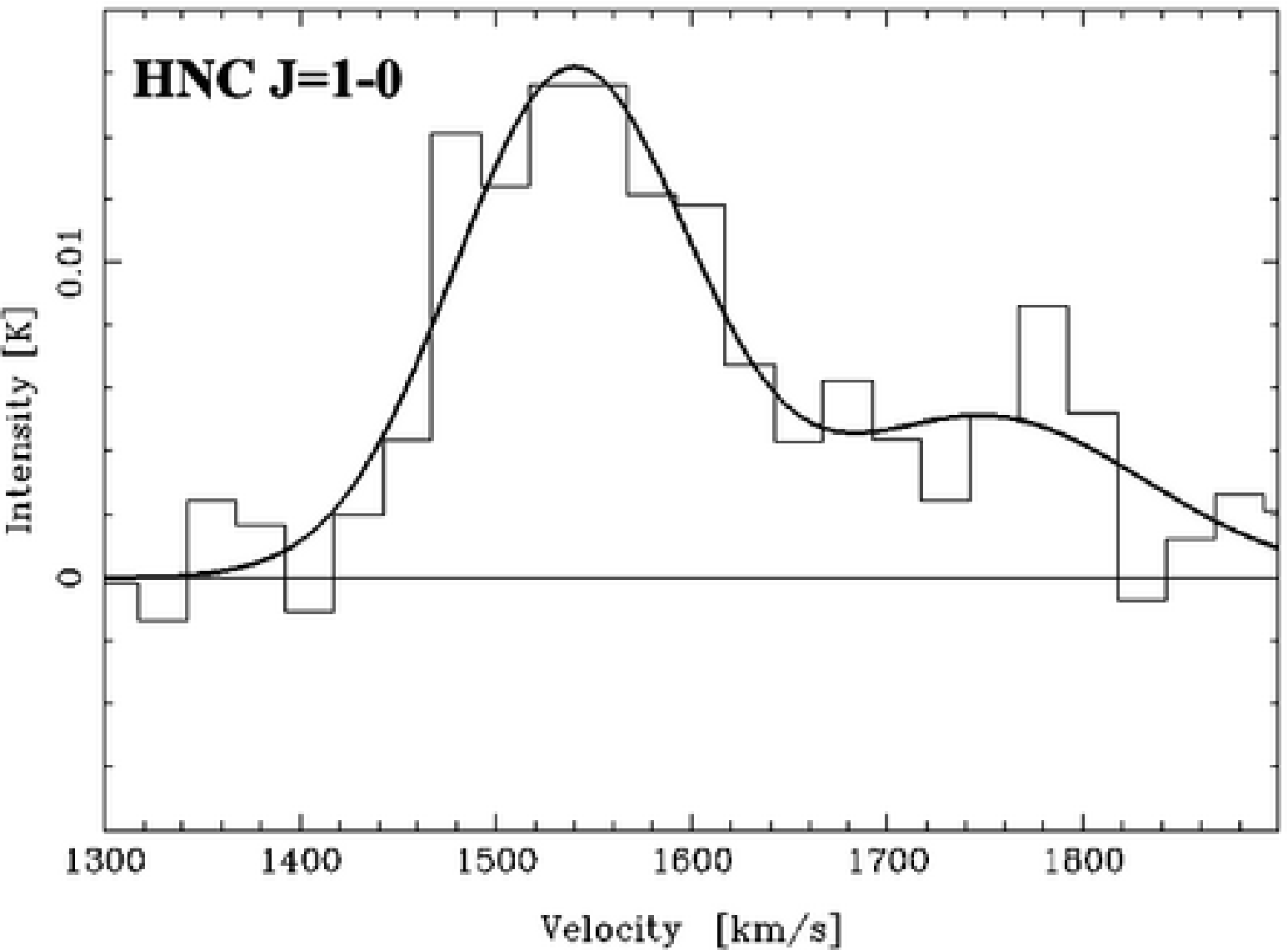}%
  \hfill\includegraphics[width=4.5cm]{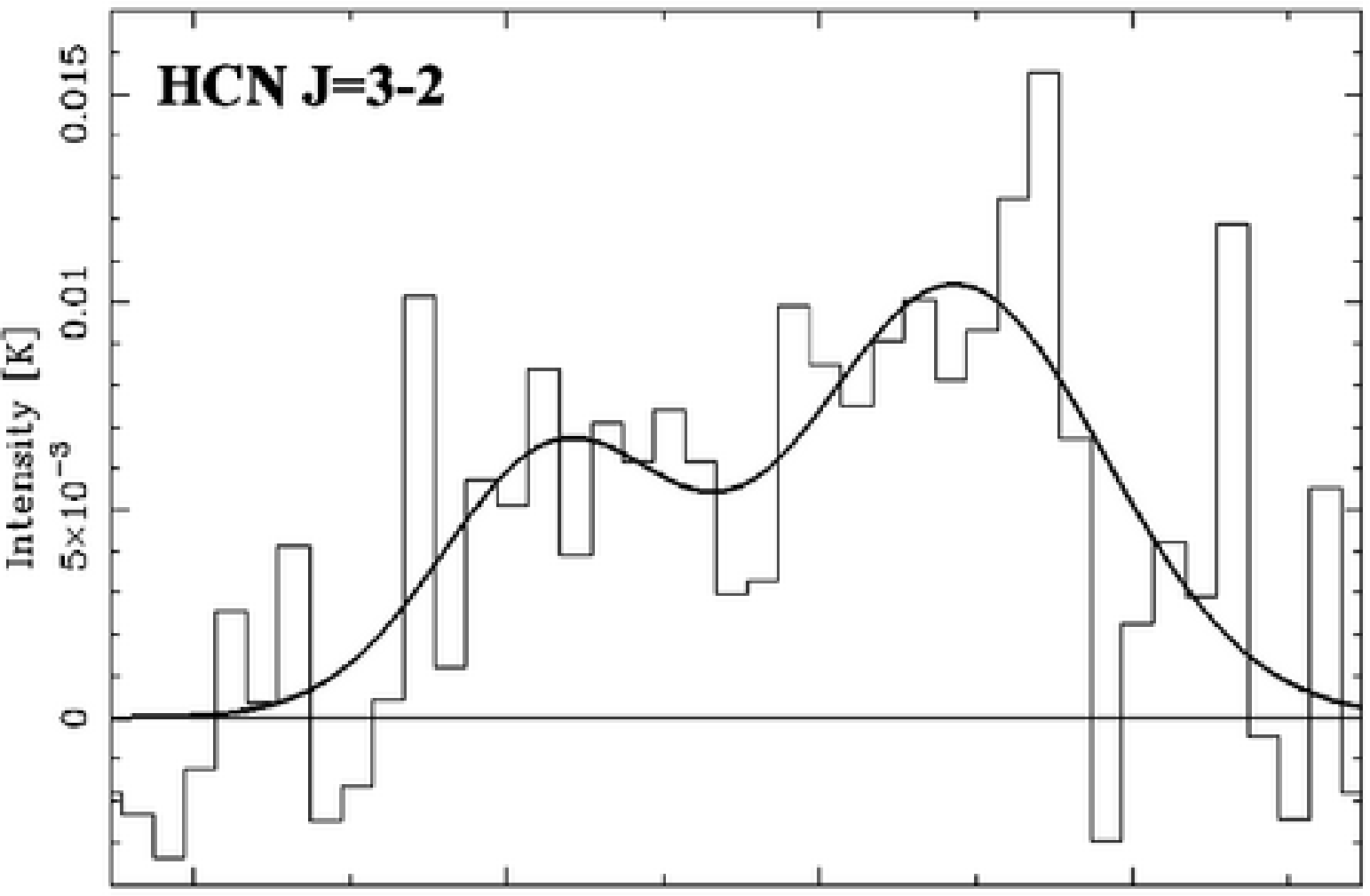}\hspace*{\fill}\\
  \vspace{-0.3cm}
  \hspace*{\fill}\includegraphics[width=4.5cm]{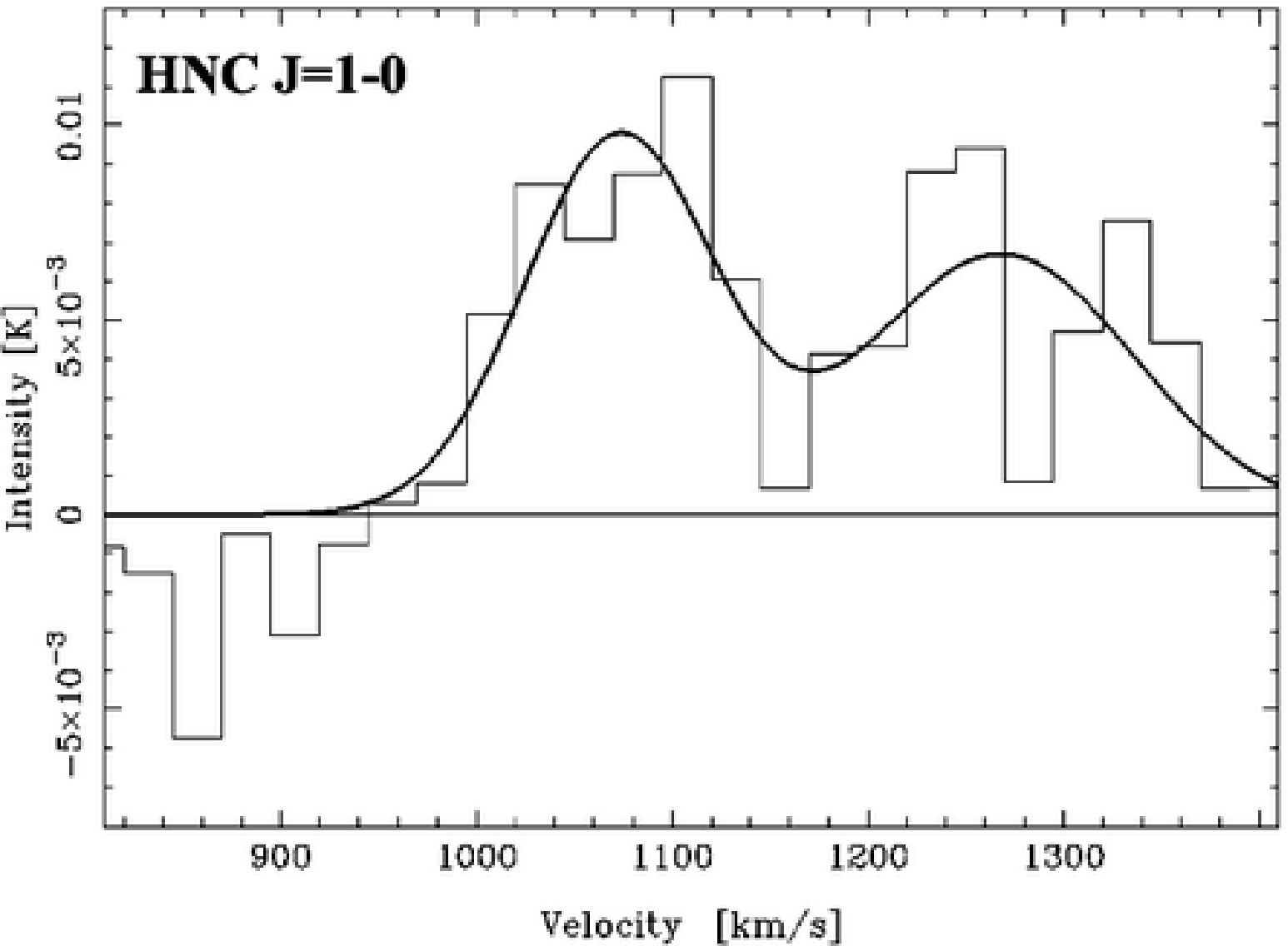}%
  \hfill\hspace{4.5cm}%
  \hfill\includegraphics[width=4.5cm]{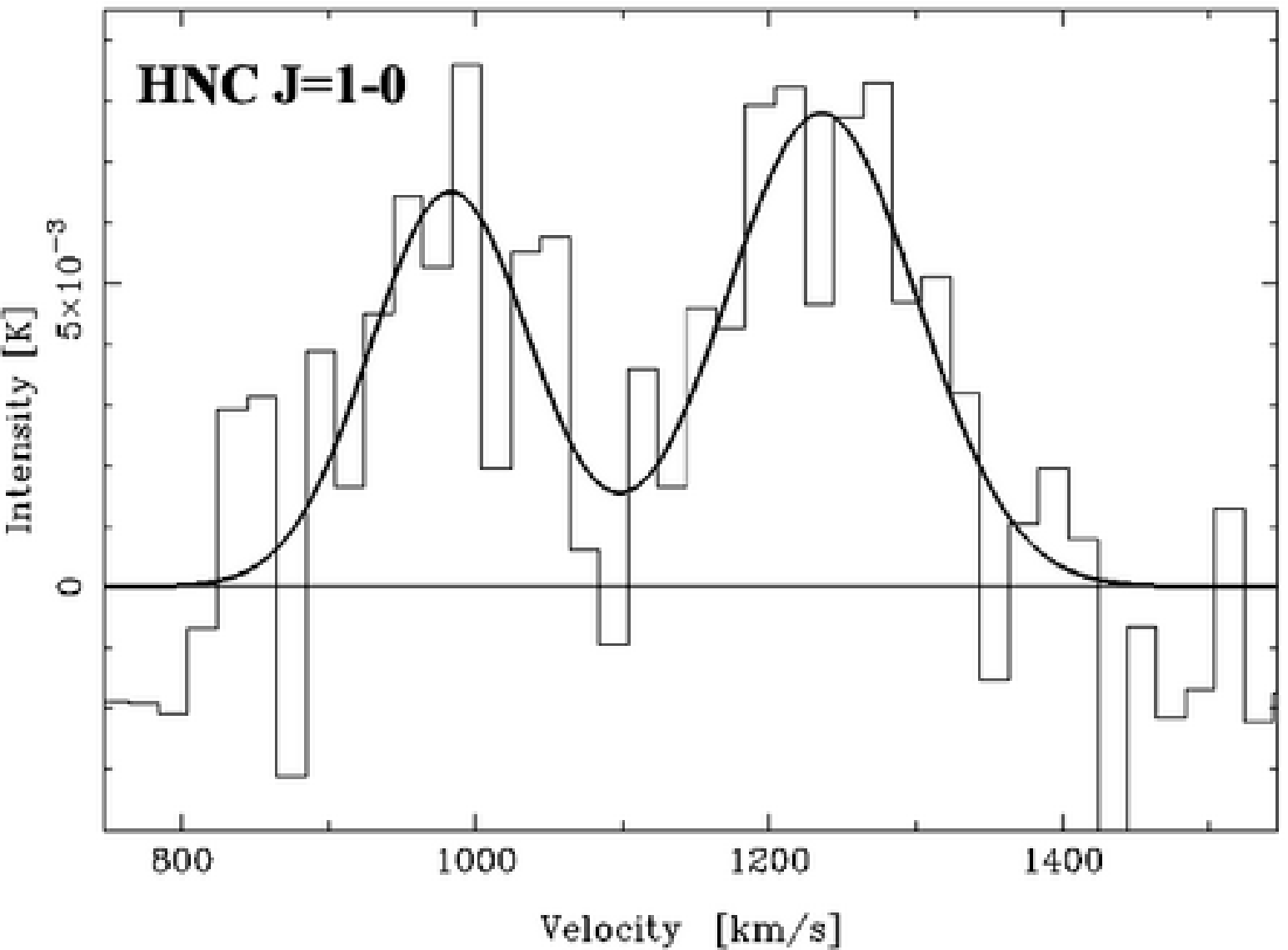}\hspace*{\fill}\\  
    
  \caption{{\footnotesize(\emph{Top-left panel}) Position-velocity (\emph{p-v}) 
  map of HCN 1--0 emission in \textbf{NGC~1068}, adapted from Tacconi et al. (\cite{tacconi94}). 
  The spectra below are the respective emission lines 
  of Figure~\ref{fig:1068-spectra}, re-rescaled to fit the velocity scale 
  of the HCN \emph{p-v} map. (\emph{Top-middle panel}) Position-velocity map of 
  CO 3--2 emission in \textbf{NGC~1365}, adapted from Sandqvist (\cite{sandqvist99}). 
  The spectra below are the re-scaled versions of the ones shown in 
  Figure~\ref{fig:1365-spectra}. (\emph{Top-right panel}) Position-velocity map of 
  CO 1--0 emission in \textbf{NGC~3079}, adapted from Koda et al. (\cite{koda02}). 
  The spectra below are the re-scaled versions of the ones shown in 
  Figure~\ref{fig:3079-spectra}.}
  }
  \label{fig:pvmaps}    
\end{figure*}

In NGC~1068 the two outer peaks of the CO 1--0 spectrum (Figure~\ref{fig:1068-spectra}) 
coincide with the maximum double peak emission seen in the CO position-velocity 
(\emph{p-v}) map obtained by Helfer \& Blitz (\cite{helfer95}), so they can 
be attributed to the emission emerging from the spiral arms. The center is 
attributed to the emission emerging from the CND, as can be inferred from
the CO 2--1 spectrum (Figure~\ref{fig:1068-spectra}). This 
center peak coincides with the maximum HCN emission at 1100 \kms seen in the 
HCN p-v map obtained by Tacconi et al., (\cite{tacconi94}). This \emph{p-v} 
map is shown in the \emph{left panel} of Figure~\ref{fig:pvmaps} for comparison, 
along with the corresponding scaled spectra observed in NGC~1068.

The main spingroup of the CN 1--0 line (Figure~\ref{fig:1068-spectra}) shows 
a shape similar to that of the CO 1--0 line. So contributions of emission coming 
from the CND, as well as from the spiral arms, can also be inferred. The main 
spingroup of the CN 2--1 spectrum also coincides with the maximum HCN emission 
around 1100 \kms.

The HCN spectra seem to contain two components, although we fit only one gaussian to the spectrum. The main 
component fits the region in the \emph{p-v} where the strongest HCN emission is coming from. 
The secondary component might be attributed to the secondary peak emission 
of HCN, observed around 1035 \kms. Another smaller peak is seen in the 
\emph{p-v} map around 1255 \kms, but this component is not detected in the 
HCN spectra.

In the HNC spectra, instead, two clear components are observed. The main component 
has a center velocity of 1073 \kms~which lies in between the two main peaks of the 
HCN emission observed in the corresponding \emph{p-v} map. With respect to the CO 
\emph{p-v} map (Helfer \& Blitz 1995), the main HNC component comes from a region 
where the CO emission is faint. Instead, the secondary component of HNC 
seems to emerge from a region around 1250 \kms, which corresponds to the secondary peak 
emission in the CO \emph{p-v} map, and roughly to the third peak 
of the HCN emission in the respective map. Although, the latter is uncertain due 
to the noise in the spectra.

\subsubsection{NGC~1365}

On top of the \emph{middle panel} of Figure~\ref{fig:pvmaps} we show the CO 3--2 
\emph{p-v} map of the central region of NGC~1365, from Sandqvist (\cite{sandqvist99}). 
The spectra below correspond to the re-scaled spectra shown in Figure~\ref{fig:1365-spectra}. 
The double peak structure of the spectra coincides fairly well with the double peak 
emission observed in the \emph{p-v} map. The left components of the spectra seem to 
emerge from a region around 1530 \kms.

The right components of most of the spectra could emerge from the region around 1710 
\kms, with the exception of the CN 1--0 line. The right component of CN 1--0 seems 
to emerge from around 1750 \kms which, in turn, coincides with the maximum emission 
level of the right peak in the CO 3--2 \emph{p-v} map.

The two peaks in the CO 3--2 \emph{p-v} map are of about the same intensity, which 
is reflected in the $^{12}$CO 1--0 spectrum, where both components have intensities 
$T_A^* \sim 0.3$ K. The $^{12}$CO 3--2 and $^{13}$CO 1--0 data obtained by Sandqvist 
(\cite{sandqvist99}) also have this feature. However, the CN, HCN and HNC spectra, 
exhibit a gradient between the intensities of their high and low velocity components. 
This can also be seen in the HCN 1--0 and HCO$^+$ 1--0 spectra 
obtained by Sandqvist (\cite{sandqvist99}). This intensity gradient could be due to either 
a larger abundance of the species, or to a higher excitation in the corresponding 
region of the lower velocities.

Since the observed double peak structure does not change with the beam size (one single feature), 
we think that the nucleus of NGC~1365 lacks a circumnuclear disk, i.e., it is consistent with 
a Seyfert 1 nucleus.

\subsubsection{NGC~3079}

The four-peak structure observed in NGC~3079 can be identified in 
the CO 1--0 \emph{p-v} map of Koda \etal~(2002). The \emph{top-right} panel 
of Figure~\ref{fig:pvmaps} shows an adaptation of the Fig.10 in Koda \etal~(2002)
with 0'' offset along the minor axis of NGC~3079. Since this galaxy is almost
edge-on, most of the CO emission is probably coming from the spiral arms, besides
the nuclear region. Instead, the high density tracers are expected to be mostly nuclear,
as in the case of NGC~1068. This can explain the double-peak structure of the line
shape of the CN, HCN and HNC molecules, in contrast to the CO lines. 

Note that in this case the CN spectra are slightly dominated by the lower-velocity peak, as
in the case of NGC~1068 and NGC~1365, whereas the HCN and HNC spectra of NGC~3079 are dominated 
by the higher-velocity peaks. The lower and higher velocity peaks are not perfectly aligned.
The maximum separation between the peaks is $\sim80$~\kms for the lower-velocity peaks and 
$\sim60$~\kms for the higher-velocity peaks. The lower-velocity peaks are centered around 
1000~\kms and the higher-velocity peaks around 1270~\kms, which means the high density tracers
tend to avoid the peaks of the CO 2--1 emission.

Since the line shapes of CN, HCN and HNC are similar, and they do not change with the beam size, 
their emissions likely emerge from the nuclear region in this galaxy, in contrast to the case of
NGC~1068, where the line shape of the HCN emission changes with the beam size.

\subsection{The {\rm HCN}/{\rm HNC} line ratios}

\begin{table*}[!t]
  \begin{minipage}{18cm}

    \centering   
      \caption[]{Intensity ratios between the high density tracers.}
         \label{tab:dense-ratios}
         \begin{tabular}{lcccccc}
            \hline
            \noalign{\smallskip}
            Galaxy & $\frac{\rm HCN}{\rm HNC}$~\tiny{1-0} & $\frac{\rm HCN}{\rm HNC}$~\tiny{3-2} & $\frac{\rm CN}{\rm HNC}$~\tiny{1-0} & $\frac{\rm CN~2-1}{\rm HNC~3-2}$ & $\frac{\rm CN}{\rm HCN}$~\tiny{1-0} & $\frac{\rm CN~2-1}{\rm HCN~3-2}$\\
            \noalign{\smallskip}
            \hline
            \noalign{\smallskip}
NGC~3079 & 2.15$\pm$0.67 & 1.60$\pm$0.75 & 0.64$\pm$0.20 & 1.25$\pm$0.58 & 0.30$\pm$0.09 & 0.78$\pm$0.25 \\

NGC~1068 & 2.01$\pm$0.65 & 6.48$\pm$1.95 & 1.98$\pm$0.58 & 2.58$\pm$0.78 & 0.98$\pm$0.29 & 0.40$\pm$0.29 \\
	
NGC~2623 & 1.4$^{\mathrm a}$ & $\lesssim$0.26 & $\lesssim$0.77$^{\mathrm a}$ & - & $\lesssim$0.5$^{\mathrm a}$ & - \\

NGC~1365 & 1.35$\pm$0.37 & - & 0.87$\pm$0.25 & - & 0.64$\pm$0.17 & 0.67$\pm$0.19 \\

NGC~7469$^{\mathrm b}$ & 1.50$\pm$0.57 & - & 0.85$\pm$0.28 & - & 0.57$\pm$0.21 & - \\
            \noalign{\smallskip}
            \hline
         \end{tabular}
\begin{list}{}{}
\item[${\mathrm{a}}$)] Ratios reported by Aalto et al. (\cite{aalto02}).
\item[${\mathrm{b}}$)] New ratios computed using the HCN 1-0 source size 
estimated from (Davies et al. \cite{davies04}).
\end{list}       
\end{minipage}
\end{table*}

\begin{figure}[!ht]
  \centering
  \includegraphics[width=7cm]{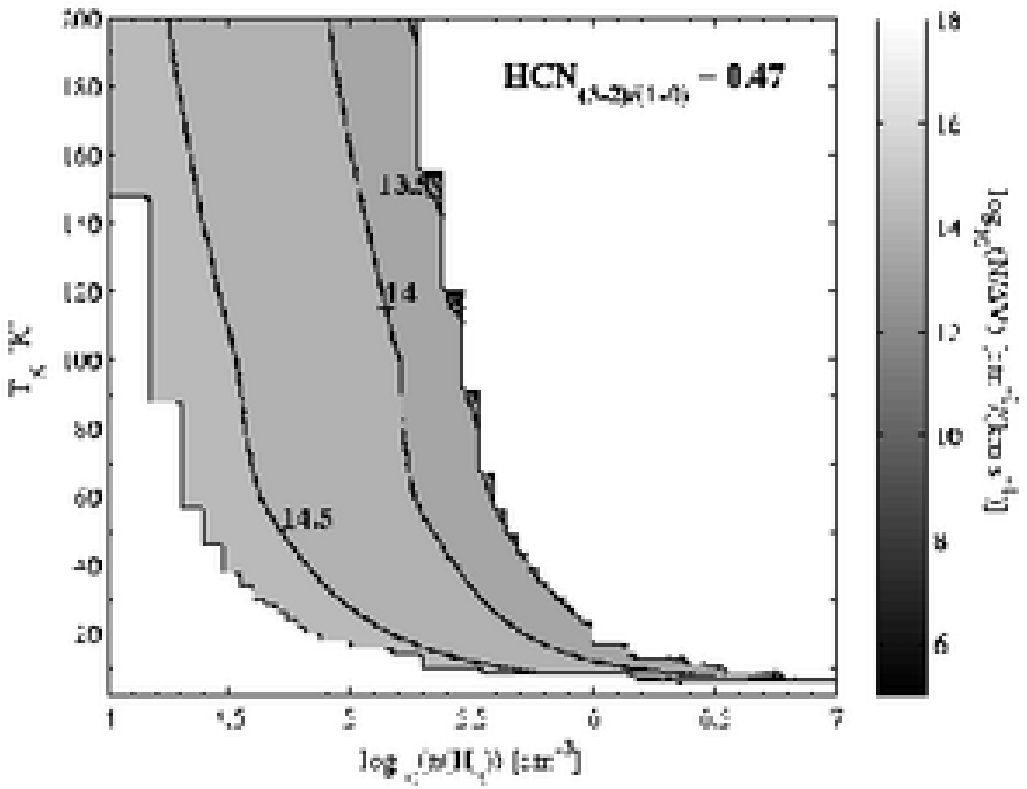}\\
  \includegraphics[width=7cm]{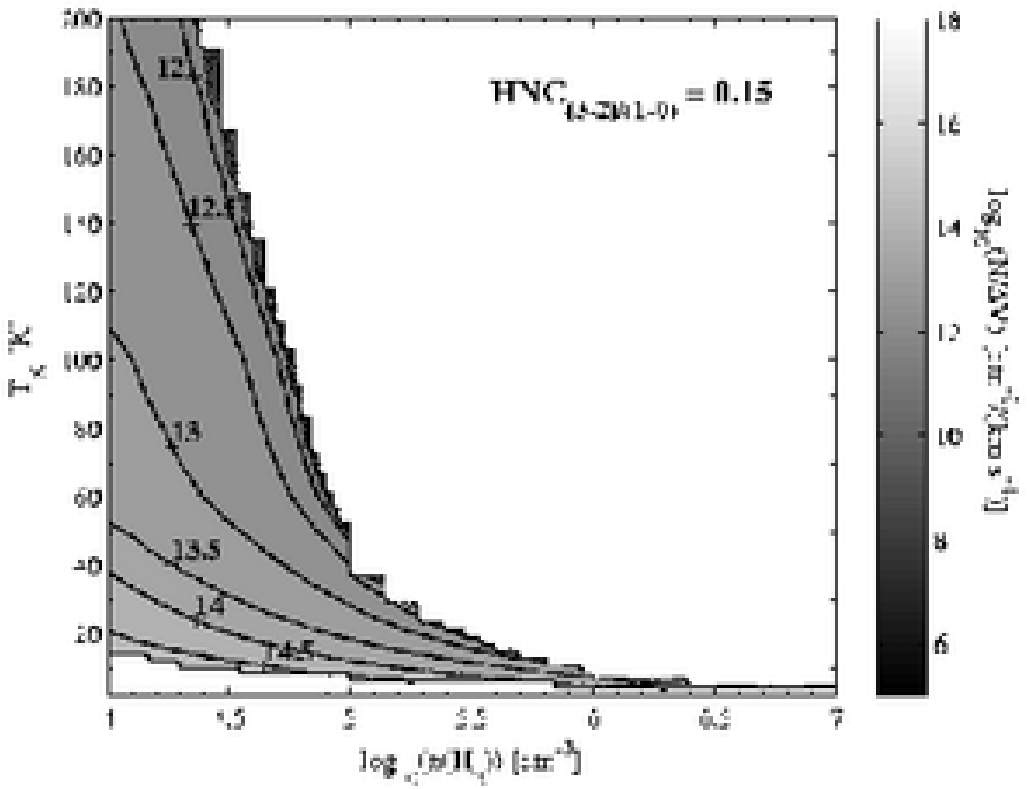}\\
  \includegraphics[width=7cm]{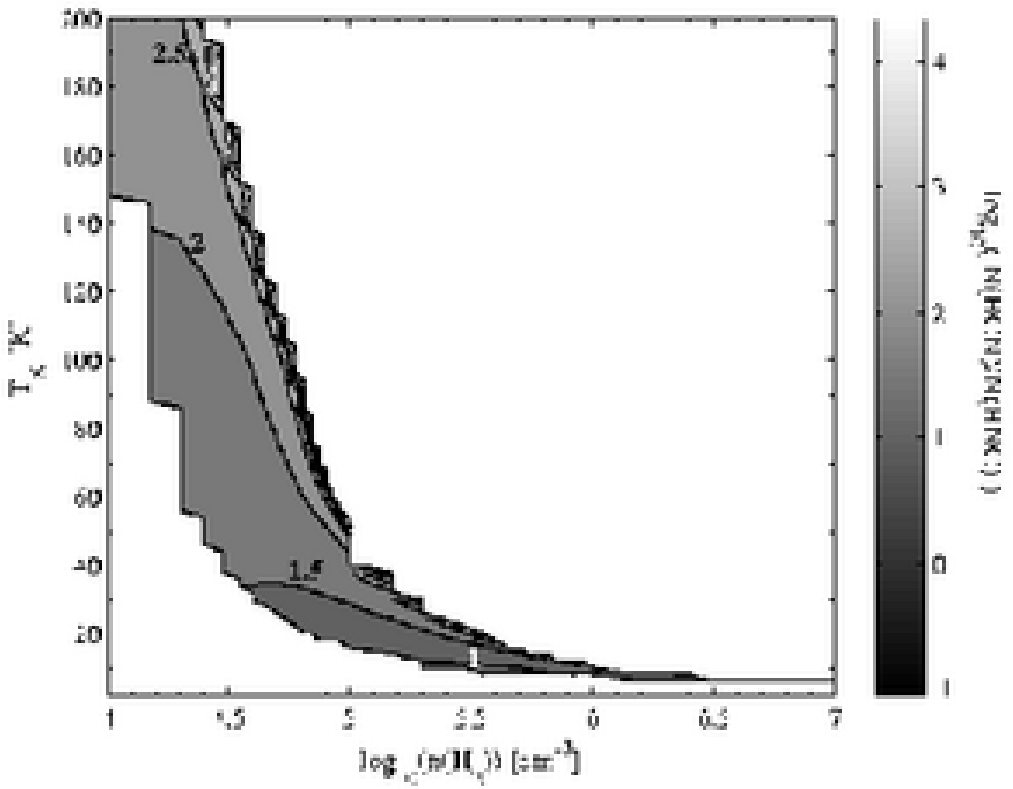}
  
  \caption{Excitation conditions modelled for the 3--2/1--0 line ratios
  of HCN (\textit{top}) and HNC (\textit{middle}) observed in \textbf{NGC 1068}. 
  The conditions required for the HCN and HNC molecules overlap in a narrow region.
  The relative column densities in the overlap zone of the excitation conditions of 
  these molecules is shown in the \textit{bottom} plot. 
  The optical depth in the whole region explored 
  for HCN ranges between 0.03 and 10 in the $J$=1--0 line, and between 0.32 and 30
  in the $J$=3--2 line. In the case of HNC the optical depth ranges between
  0.01 and 30 in the $J$=1--0 line, and between 0.003 and 30 in the the $J$=3--2 line. The optically
  thin limit of both molecules and lines is depicted by the right edge of the excitation conditions, whereas
  the optically thick limit correspond to the left edge of the figures above.}
  \label{fig:1068-xcmaps}
\end{figure}

{\large NGC~1068}: The HCN/HNC 3--2 \& 1--0 line intensity ratios increase 
towards the CND. The ratio varies from $\sim$2.0 for the lower transition 
lines, to $\sim$6.5 for the higher 
transition. The large $J$=1--0 beam picks up emission from both the 
CND and the starburst ring, whereas the $J$=3--2 beam picks up emission 
coming mainly from the CND. This could be interpreted 
either as that the abundance ratio differs between the starburst ring 
and the CND, or that the abundance is actually the same but the physical 
conditions are different in these two regions. On the other hand, there 
could also be optical depth effects since the difference in the ratios 
is consistent with a larger optical depth in the $J$=1--0 transition line
of HNC.  In section 4.3.1 we estimate the excitation conditions of
HCN and HNC, from which we can derive the corresponding optical depths.
As described in Figure~\ref{fig:1068-xcmaps}, the optical depth of the 
$J$=1--0 line of HNC is larger (starting from $\tau$=0.01) than that 
of the $J$=3--2 line (which starts from $\tau$=0.003) in almost 
all the possible excitation conditions. In the case of HCN, the situation 
is the opposite. The lowest optical depth of the HCN $J$=1--0 line is 0.03, 
whereas in the $J$=3--2 line is 0.32. In any case, X[HCN]/X[HNC] is 
at least 6.5 in the CND.
According to Meijerink \etal~(\cite{meijerink07}), this ratio can be 
found in gas of density $n_{\rm H}\sim10^5 \3cm$ and with PDR conditions,
at a distance from the source of $\sim10^{16}$ cm, if the Habing flux 
$G_0$ is about $10^4$, or at a slightly larger distance of 
$2\times10^{16}$ cm, if $G_0=10^5$. On the other hand, in an XDR environment, this
ratio would be found at a distance of $7\times10^{16}$ cm, if the radiation flux $F_{FUV}$ is
16 erg $\2cm$ s$^{-1}$, or at a much larger distance of about $2\times10^{18}$ cm, if 
$F_{FUV}=160$ erg $\2cm$ s$^{-1}$.\\
\\{\large NGC~3079}: The HCN/HNC ratio decreases for the higher transitions. 
The HNC 3--2 emission rivals that of HCN, making the HCN/HNC 3--2 line ratio only 1.6.
As described above, a similar analysis of the expected distribution of these ratios 
and molecules, in a PDR and XDR environments, can be done based on Meijerink \etal~(\cite{meijerink07}).\\
\\{\large NGC~2623}: In this galaxy we do not detect HCN 3--2. We estimate an 
upper limit of 0.26 for the HCN/HNC ratio of the $J$=3--2 line. We observe that
the ratio does not only decreases for the higher transitions, but also this is
the only galaxy in our sample where $I({\rm HCN}) < I({\rm HNC})$ in the $J$=3--2 line,
which makes it comparable to galaxies like Arp~220, Mrk~231 and NGC~4418, according 
to the recent work by Aalto \etal~(\cite{aalto07}). They propose that the overluminous 
HNC can be explained by a pumping effect due to mid-IR background radiation with
brightness temperatures $T_{\rm B}\gtrsim 50$ K and densities below critical, or due 
to the ISM chemistry being affected by X-rays. According to Schilke \etal~(\cite{schilke92}), 
shocks are also possible sources of explanation.\\

The HCN/HNC 1--0 ratio is about 2.0 in NGC~1068 
and NGC~3079, and $\sim$1.4 in the others. This indicates a brighter 
HNC emission in NGC~2623, NGC~1365 and NGC~7469. Luminous HNC in galaxies 
may have the following plausible explanations:
\begin{list}{}{}
\item[a)] {\bf \small Large masses of hidden cold gas and dust}.\\
If the HCN and HNC emission is emerging from gas of densities $n$(H$_2$) 
$> 10^5~\3cm$ then the HNC chemistry would be dominated by reactions like 
${\rm HNC}+{\rm O} \rightarrow {\rm CO} + {\rm NH}$ which would destroy 
HNC at higher temperatures. Thus, at high gas densities, a bright HNC 
line would imply a considerable amount of cold (T$_k< 24$ K) dense gas.
\item[b)] {\bf \small Chemistry dominated by ion-neutral reactions.}\\
If, however, the bulk of the HCN and HNC emission is emerging from gas of 
densities $\sim$10$^4~\3cm$ then the relative HNC abundance may be 
substantial, despite the high temperature. The reason for this is that, 
at lower densities, reactions with $\rm HCNH^+$ (HCN and HNC reacts with 
$\rm H_3^+$ to form $\rm HCNH^+$) become more important. The ion abundance 
is higher and once HCN and HNC become protonated, $\rm HCNH^+$ will 
recombine to produce either HCN or HNC with 50\% probability. This scenario 
is interesting since the electron and ion abundance is likely higher in PDRs 
Photon Dominated Regions (PDRs) (Tielens \& Hollenbach, 1985). 
Therefore, in a PDR chemistry, the connection 
between HNC and kinetic temperature may be weak since we expect the $\rm HCNH^+$ 
reactions to be important.
\item[c)] {\bf \small Chemistry dominated by hard X-rays.}\\
The X-ray irradiation of molecular gas leads to a so called X-ray dominated 
region (XDR) (e.g. Maloney \etal~1996) similar to PDRs associated with bright 
UV sources. The 
more energetic (1-100 keV) X-ray photons penetrate large columns 
($10^{22}-10^{24}~\2cm$) of gas and lead to a different ion-molecule chemistry. 
Models of XDRs by Meijerink \& Spaans (2005) and Meijerink \etal~(\cite{meijerink07})
indicate that the HNC/HCN column 
density ratio is elevated (and larger than unity) compared to PDRs and quiescent 
cloud regions for gas densities around $10^5~\3cm$.
\item[d)] {\bf \small HNC enhanced through mid-IR pumping.}\\
Both HCN and HNC may be pumped by an intense mid-IR radiation field boosting 
the emission also from low density regions where the lines would not be 
collisionally excited. For HNC the coupling to the field is even stronger 
than for HCN, thus increasing the probability for IR pumping in extreme galaxies, 
such as Mrk 231. Ultraluminous galaxies, such as Mrk 231 and Arp 220, have central 
mid-IR sources with optically thick radiation temperatures well in excess of those 
necessary to pump the HNC molecule (Soifer \etal~1999). Even if the HNC abundance 
is lower than HCN, the HNC emission may have a higher filling factor due to 
the IR pumping (i.e. IR pumped emission from gas clouds otherwise at too low density 
to excite the HNC molecule) (e.g., Aalto \etal~2002).
\end{list}

The study carried out so far allows us to distinguish between the above 
scenarios in some of the sources presented here. The number density 
$n(\rm H_2)$ required for alternative (b) is too low to efficiently 
excite the HNC (or HCN) 1--0 and 3--2 lines, and we should therefore 
expect subthermal HCN and HNC excitation in this case. An HNC 3--2/1--0 
line ratio of 0.3 or less is an indication that the gas densities 
are below $10^5~\3cm$, depending on temperature and column density, as 
it is discussed in section 4.3. A HNC 3--2/1--0 ratio lower than 0.3 is 
observed in NGC~3079, NGC~1068 and NGC~7469 (Table~\ref{tab:line-ratios}).

The ratio $\sim$0.4 found in NGC~2623 is in the limit between case (a) and 
(b). On the other hand, case (c) and (d) cannot easily be ruled out. More 
information is required in order to distinguish between the proposed scenarios, 
as suggested by Aalto \etal~\cite{aalto07}.

The HCN 3--2/1--0 line ratio is below 0.3 only in NGC~3079, whereas it is 
larger than 0.4 in NGC~1068 and NGC~1365. This result is interesting since it 
implies that in NGC~3079 both HCN and HNC emission emerge from the same gas, 
whereas in NGC~1068 the HNC emission has to emerge from a lower ($<10^5~\3cm$) density gas than 
HCN. It would be interesting to see if this result holds for NGC~1365. We expect 
to obtain the HNC 3--2 data for this galaxy in a future project.

The pumping scenario (c) should lead to a HNC 3--2/1--0 line ratio close to unity. 
This is not observed either in the HNC or in the HCN data we have. However,
it is not possible to rule out this scenario since low excitation may be the 
result of mid-IR pumping of low density gas. Detailed modelling is needed in 
this case.

\subsection{Excitation conditions of {\rm HCN} and {\rm HNC}}

We used the radiative transfer code RADEX\footnote{http://www.sron.rug.nl/$\sim$vdtak/radex/index.shtml} 
(Van der Tak et al.~\cite{vdtak07}) to explore a wide range of possible excitation conditions 
that can lead to the observed line ratios. This code is sensible to the column density of
a molecule per line width, and uses a constant temperature and density of the collision partner,
which in this case is H$_2$. Another limitation of RADEX is that it cannot handle large optical depths
($\tau<100$). Our analysis is not depth dependent and we assume a homogeneous sphere for the escape 
probability approach. Hence, our models aim to reproduce a sort of average cloud that represent
the physical conditions of the emitting gas, which is a well fitted starting model for single dish observations, where all the emissions detected are convolved with the telescope beams.

The grid consists of densities between $10^4$ 
and $10^7~\3cm$, temperatures between 4 and 200 K, and column densities per line width 
between $10^{10}$ and $10^{18}~\2cm~km^{-1}~s$.
Excitation maps were generated in order to obtain the total 
column density per line width ($N/\Delta\upsilon~\rm{cm^{-2}~km^{-1}~s}$) as function of 
the kinetic temperature ($T_k$~K) and the number density of molecular hydrogen 
($n(\rm{H_2})~\3cm$). The countour lines of the maps describe the dichotomy 
between temperature and density. This means that, for a given column density, 
the observed 3--2/1--0 line ratio can be obtained with high temperatures and 
low densities, or low temperatures and high densities.

The hyperfine structure of the HCN $J$=1--0 transition line
is not included in this study. The extrapolated HCN data of the LAMDA\footnote{http://www.strw.leidenuniv.nl/$\sim$moldata/} 
database were used instead (Sch\"oier \etal~\cite{schoier05}).
Note that the hyperfine components may 
be overlapping and may interact radiatively in AGN-like 
environments. This process requires further analysis and modelling that is
not included in RADEX. 
We are not able to generate an excitation map of 
CN due to the lack of collision data for this molecule. 
Although, an study to extrapolate the collision data for CN from
other know molecules (e.g. CS) is ongoing.

\begin{figure}[!ht]
  \centering
  \includegraphics[width=7cm]{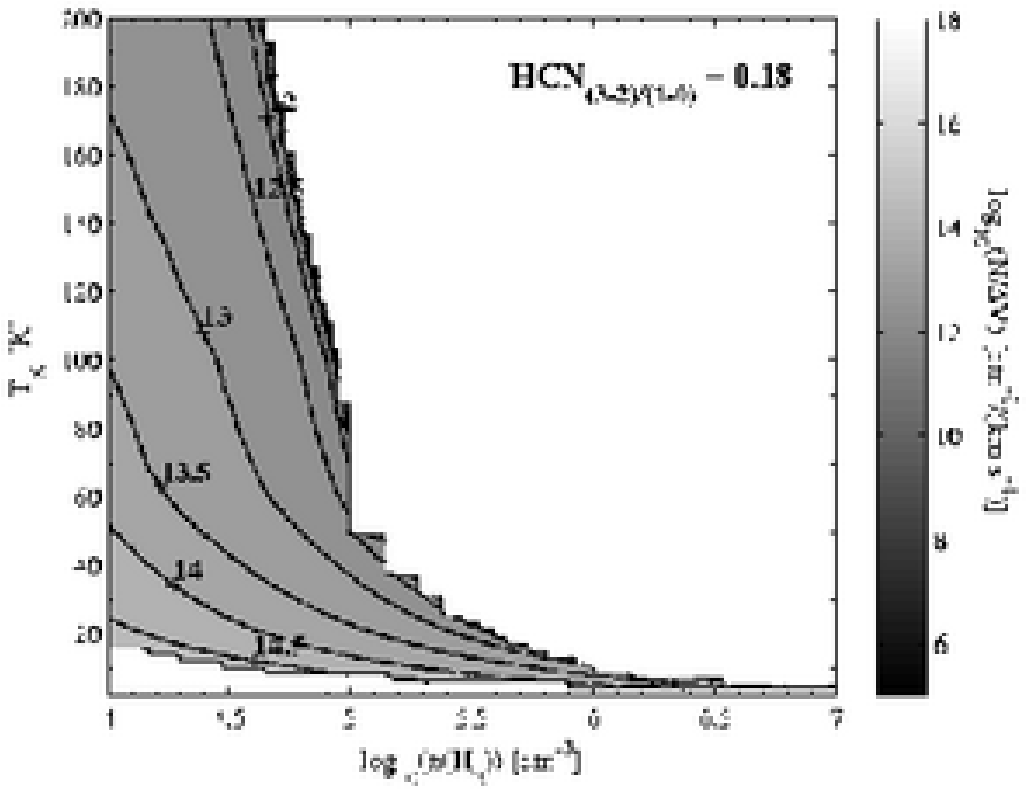}\\
  \includegraphics[width=7cm]{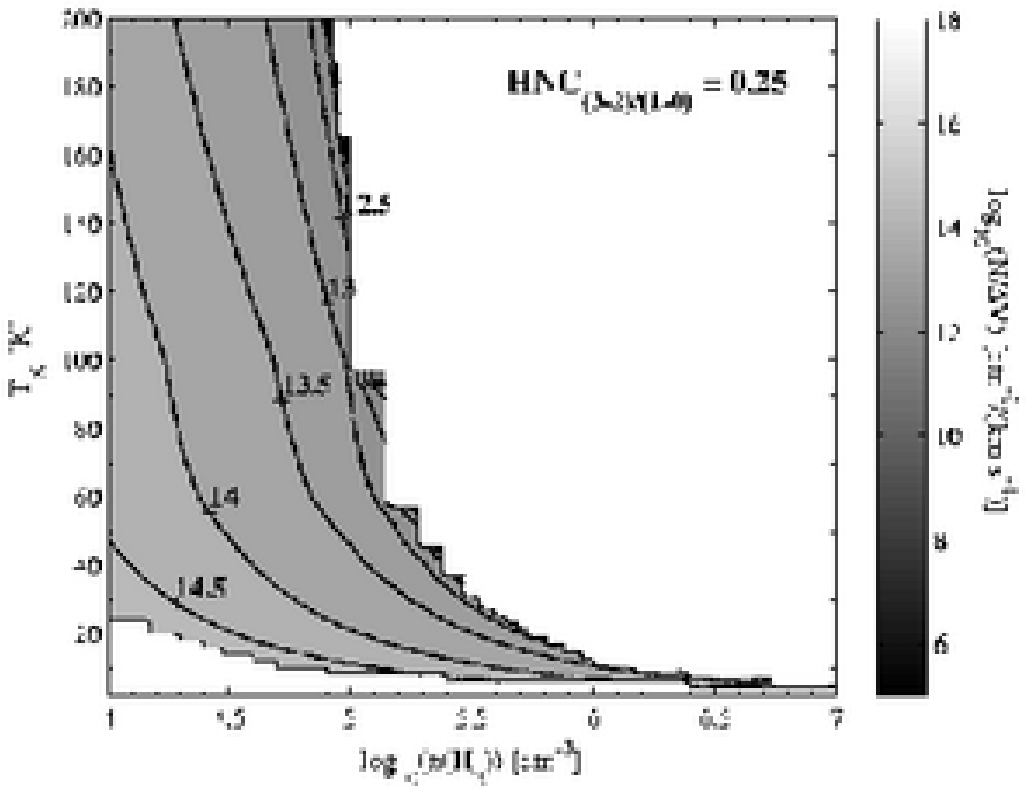}\\
  \includegraphics[width=7cm]{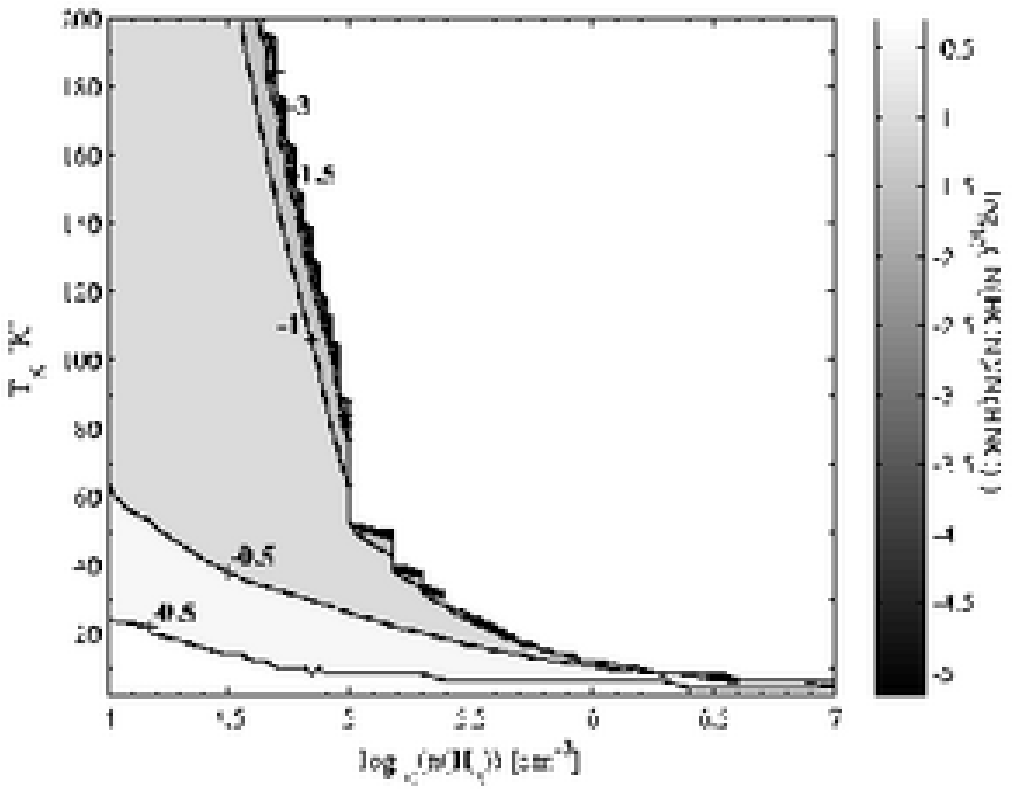}
    
  \caption{Excitation conditions modelled for the 3--2/1--0 line ratios
  of HCN (\textit{top}) and HNC (\textit{middle}) observed in \textbf{NGC 3079}. 
  The excitation conditions required for these 
  molecules overlap in most of the range explored. This suggest that 
  their emissions emerge from gas with the same physical conditions. 
  The \textit{bottom} plot shows the overlap zone. 
  There is a large zone of excitation conditions where the column density of
  HNC is between 3 and 10 times larger than that of HCN.
  The contour lines of -0.5 depict the zone of the physical conditions for which the 
  column density of HNC is about 3 times larger than that of HCN. 
  For both molecules and transition lines the optical depth ranges between 0.001 and 30.
  The optically thick limit is basically defined by the limit of convergence of RADEX 
  ($\tau_{\rm max}\sim100$).}
  \label{fig:3079-xcmaps}
\end{figure}

\subsubsection{NGC~1068}

Figure~\ref{fig:1068-xcmaps} shows the excitation condition maps for the HCN and 
HNC molecules, modeled from the average ratios observed in NGC~1068. These maps
show that HCN (\textit{top}) requires higher density gas and higher column density 
than HNC (\textit{middle}) in order to obtain the observed 3--2/1--0 line ratios. 

Considering a kinetic temperature $T_k=80$ K and a density $n(\rm H_2)\sim10^5~\3cm$, 
as estimated by Tacconi et al. (\cite{tacconi94}), we found that 
$N(\rm HCN)/\Delta\upsilon\sim1.5\times10^{14}~\rm{cm^{-2}~km^{-1}~s}$,
which agrees with the respective value found by Tacconi et al.

According to the recent PDR and XDR models by Meijerink \& Spaans (\cite{meijerink05})
and Meijerink \etal~(\cite{meijerink07}), under PDR conditions a temperature $T_k=80$ K 
can be reached at a total column density of $N_{\rm H}\approx3\times10^{21}~\2cm$, if the Habing 
flux is $G_0\sim10^3$ 
(a Habing flux $G_0 = 1$ corresponds to a far UV flux $F_{FUV} = 1.6\times10^{-3}$ erg $\2cm$ s$^{-1}$),
or at $N_{\rm H}\approx8\times10^{21}~\2cm$ if $G_0\sim10^5$. At those depths the density $n(\rm H_2)$ 
would be slightly higher than $10^5~\3cm$, though. On the other hand, in an XDR environment 
a temperature of 80 K and a density of $10^5~\3cm$ can be reached at a slightly larger column density of
$N_{\rm H}\approx10^{22}~\2cm$, if the impinging radiation field is $F_{FUV}\sim1.6$ erg $\2cm$ s$^{-1}$,
or at a much larger depth equivalent to $N_{\rm H}\approx2\times10^{24}~\2cm$
if the radiation flux is $F_{FUV}\sim160$ erg $\2cm$ s$^{-1}$.

However, at at a temperature of 80 K and density of $10^5~\3cm$~ 
there is no solution for HNC. This means that either 
the HNC emission arises from a gas with different physical conditions than HCN, 
or that both molecules trace a cooler ($T_k<80$ K) and lower density gas 
($n(\rm H_2)<10^5~\3cm$), if the emission of both molecules arise from the 
same gas.

If the HCN and HNC emissions actually trace the same gas, 
the range of possible excitation conditions can be constrained to the zone 
where the conditions for HCN and HNC overlap. The \textit{bottom} plot of 
Figure~\ref{fig:1068-xcmaps} shows the average overlap zone, where the contours 
correspond to the ratio between the total column densities of HCN and HNC. 
The overlap zone is defined by the optically thin limit (right edge of the excitation maps) 
of the HNC transition lines and the optically thick limit (left edge of the excitation maps)
of the HCN lines. The models show that $N(\rm HCN)$ is larger than $N(\rm HNC)$ in all the
overlap zone. According to Figure 10 in Meijerink \& Spaans (\cite{meijerink05}),
$N(\rm HNC)/N(\rm HCN)$ column density ratios lower than unity can be found mostly in PDRs, 
but also in XDR environments if the total column density $N(\rm H)$ is lower than $10^{24}~\2cm$.
If we still assume the same temperature (80 K) proposed by Tacconi et al. 
(\cite{tacconi94}), and we consider the contour line where $N(\rm HCN)$ is about
2.5 orders of magnitude larger than $N(\rm HNC)$, we find that the gas density should be 
$n(\rm H_2)\sim6\times10^4~\3cm$, which agrees with the observed HNC 3--2/1--0 line ratio discussed 
in section 4.2. However, this density would not be consistent with the result found by Tacconi et al. (1994)
nor with the observed HCN 3--2/1--0 line ratio, which implies that the HNC emissions arise 
from a more diffuse gas than HCN.

\subsubsection{NGC~3079}

In contrast to NGC~1068, the excitation conditions modelled for the HCN and HNC line 
ratios observed in NGC~3079, overlap in most of the range explored (Figure~\ref{fig:3079-xcmaps}). 
This suggest that the emission from both molecules likely arise from the same gas. The spectral line 
shapes of HCN and HNC, showed in Figure~\ref{fig:3079-spectra}, also hint that their distribution 
may be the same, although their line centers and widths are affected by noise. 
The \textit{bottom} plot of Figure~\ref{fig:3079-xcmaps} shows the overlap zone of the excitation 
conditions. In most of these conditions $N(\rm HNC)$ is between 0.5 and 1 order of magnitude 
larger than $N(\rm HCN)$. 

According with the PDR and XDR models by Meijerink \& Spaans (2005), 
a column density ratio $N(\rm HNC)/N(\rm HCN)$ larger than unity can be found only in XDR 
environments at a total column density $N(\rm H)>10^{24}~\2cm$, with radiation fields $F_x$ of about 
1.6 erg$~\2cm~s^{-1}$ (or equivalent Habing flux $G_0\sim10^3$) and total density $n(\rm H)\sim10^3~\3cm$. 
$N(\rm HNC)/N(\rm HCN)$ column density ratios larger than unity at $N(\rm H)>10^{24}~\2cm$ can also be 
found with stronger radiation fields ($F_x\sim160$ erg$~\2cm~s^{-1}$) and higher densities 
($n(\rm H)\sim3\times10^5~\3cm$). On the other hand, $N(\rm HNC)/N(\rm HCN)$ can be larger than unity 
at $N(\rm H)<10^{23}~\2cm$ if the radiation fields are on the order of 1.6 erg$~\2cm~s^{-1}$ and the 
total density is about $3\times10^5~\3cm$.

In our models we did not explore densities lower than $10^4~\3cm$ since the HCN and HNC molecules are
expected to trace higher densities. If we assume a kinetic temperature of 80 K, as in the case of NGC~1068, 
and a density $n(\rm H_2)\sim10^5~\3cm$, the column densities per line width would be about $10^{12}$ and $10^{13}~\2cm~{\rm km}^{-1}~{\rm s}$ for HCN and HNC, respectively. In the case of HNC both
transition lines are equally optically thick at 80 K in the whole range of densities, whereas HCN 
is more optically thin. At a density of $10^5~\3cm$ the optical depth of HNC is $\sim$1 in both 
lines, whereas $\tau$ is $\sim$0.3 and $\sim$0.1 in the $J$=1--0 and $J$=3--2 lines of HCN, respectively.
A density lower than $10^5~\3cm$ is more likely to be the right case in order to be consistent with 
the line ratios observed and discussed in section 4.2.

\subsubsection{NGC~2623 and NGC~1365}

Both NGC~2623 and NGC~1365 have similar excitation conditions for HNC and HCN,
respectively. As in the case of HCN in NGC~1068, both molecules could be embedded in 
gas with densities larger than $10^5~\3cm$ at 80 K, which is consistent with
line ratios larger than 0.3, as discussed in section 4.2.
The column densities per line width of these molecules would be about $6\times10^{13}~\rm{cm^{-2}~km^{-1}~s}$ 
for HNC and about $10^{14}~\rm{cm^{-2}~km^{-1}~s}$ for HCN, if the gas density is $10^5~\3cm$.

\begin{figure}[!t]
  \centering
  \includegraphics[width=7cm]{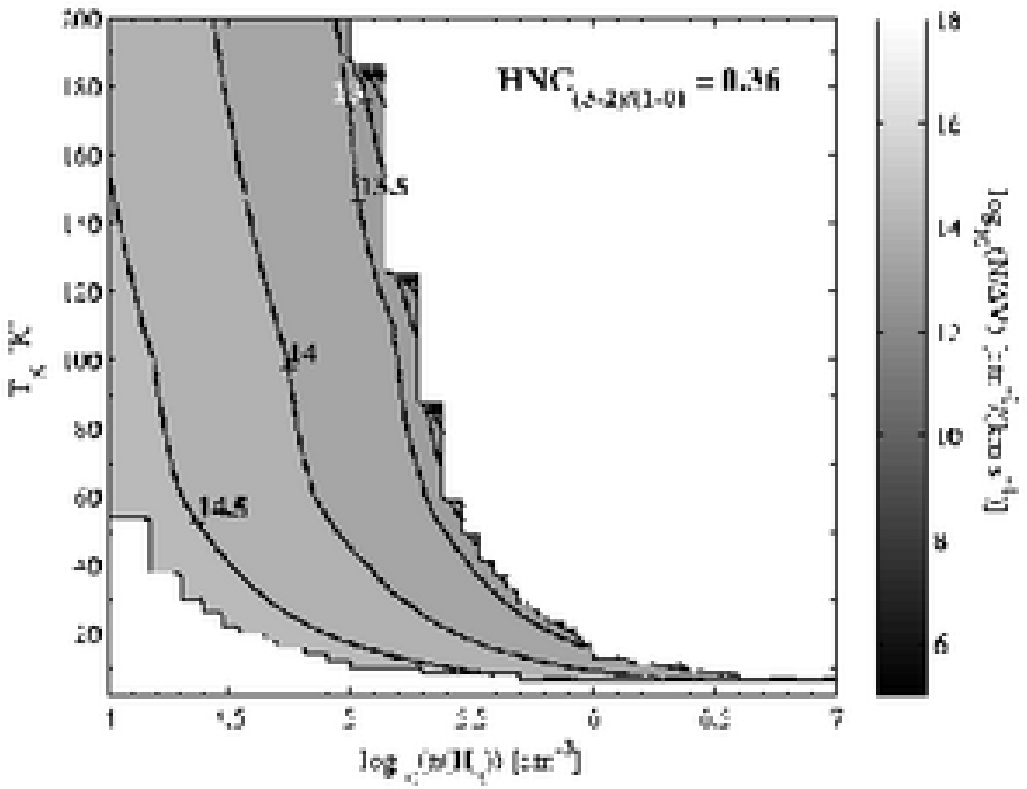}\\
  \includegraphics[width=7cm]{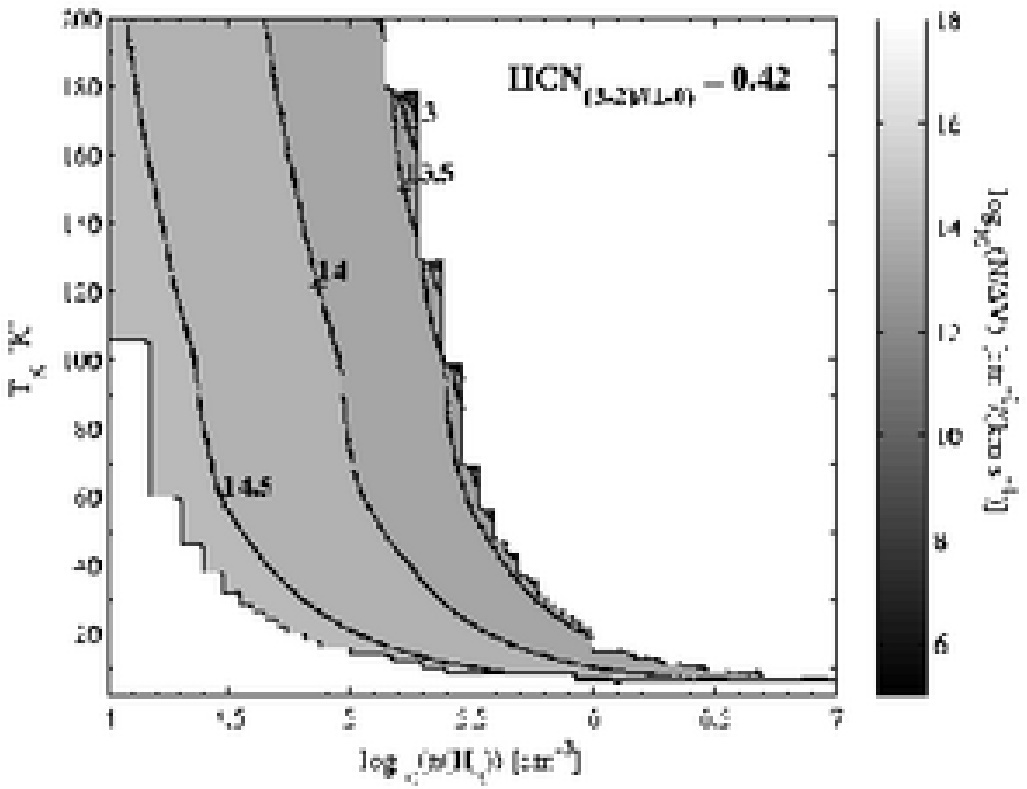}
      
  \caption{Excitation conditions modelled for the 3--2/1--0 line ratio of 
  HNC (\textit{top}) and HCN (\textit{bottom})  
  observed in \textbf{NGC~2623} and \textbf{NGC~1365}, respectively. 
  For NGC~2623 the optical depth of HNC 3--2 ranges between 0.003 and 10 in the J=1--0 line, 
  and between 0.32 and 30 in the J=3--2 line. In the case of NGC~1365
  the optical depth in the HCN 3--2 line ranges between 0.1 and 30, 
  and between 0.01 and 10 in the J=1--0 line.}
  \label{fig:2623-1365-xcmaps}
\end{figure}

\subsection{The {\rm CN/HNC} and {\rm CN/HCN} line ratios}

The CN/HNC ratio increases towards the CND in both galaxies, NGC~1068 and 
NGC~3079. The CN/HNC 1--0 ratio is lower than unity in all the galaxies, 
with the exception of NGC~1068. This can be interpreted as that HNC is more 
abundant than CN, assuming the same excitation conditions and that the emission 
emerge from the same gas and distribution in the galaxy. However, since we do not 
really know the source size of CN (and basically neither that of HNC) the beam 
dilution effects could be more (or less) severe than estimated here.

Although the beam dilution uncertainty of CN also applies to the CN/HCN ratio, 
this case is more interesting because none of the galaxies where we do 
have data show ratios larger than one, neither for the global ratio ($J$=1--0) 
nor for the nuclear part of the galaxies (higher transitions). All of these galaxies 
are considered active galaxies, so the presence of an AGN 
increases the chances of finding an X-ray dominated region (XDR) in their nuclear 
regions, as it seems to be the case of NGC~1068, according to Usero \etal~(2004). 

In an XDR the CN molecule is expected to be more abundant than 
HCN (e.g., Lepp \& Dalgarno \cite{lepp96}, Meijerink \& Spaans \cite{meijerink05} and 
Meijerink \etal~\cite{meijerink07}), and hence the CN/HCN intensity ratio could 
also be larger than unity, especially for the higher transition lines - CN 2--1 and 
HCN 3--2 - where the beam dilution is less important than for the 
$J$=1--0 line. However, this is not seen in any of the sources presented here, 
nor in the sample of galaxies shown by Aalto \etal~(2002), where results for 
AGN are presented along with starburst galaxies. 

According {to Meijerink \etal~(\cite{meijerink07})}, in a PDR environment the $N({\rm CN})/N(\rm HCN)$ 
column density ratio ranges between 2.0 and 0.5 for densities between $10^4$ and 
$10^6~\3cm$, respectively, whereas in an XDR this ratio varies from over a 1000 (at 
$n(\rm H_2)\sim10^4~\3cm$) to 40 (at $n(\rm H_2)\sim10^6~\3cm$). If the CN intensity
lines are proportional to the column density, and the estimate of the 
beam dilution is accurate enough, our results would favor a PDR scenario, rather than an XDR, 
with densities $<10^6~\3cm$ in the central regions of all these galaxies. However, if 
the CN molecule is indeed more abundant than HCN, then the weaker intensity lines 
could be also due to a stronger optical depth effect (in the escape probability sense) for CN.
A rigorous modeling and analysis of the CN molecule would be required in order to 
understand and predict the intensities of this molecule.

On the other hand, the PDR scenario would also be consistent with the results found for the HNC and HCN molecules 
described above. HCN/HNC line intensity ratios larger than unity are usually found
in PDR environments at total column densities $N(\rm H)$ lower than $10^{22}~\2cm$, while 
this ratio is larger for strong ($>10$ erg s$^{-1}~\2cm$) radiation fields and low ($\sim$10$^4~\3cm$)
densities in an XDR environment (Meijerink \etal~\cite{meijerink07}). 
The lower densities required to observe this ratios 
in an XDR environment tend to dismiss this alternative since our simulations in section 4.3 
favor densities $\sim$10$^5~\3cm$ for the HNC and HCN molecules, as in the case of NGC~1068.
In particular, there is evidence of recent starburst in the dense nuclear 
disk of NGC~1068 (Davies \etal~\cite{davies06}). These results are consistent 
with a model considering both, AGN and starburst components, required for modelling 
the UV to FIR atomic spectrum of NGC~1068 (Spinoglio \etal~\cite{spinoglio05}).

\section{Conclusions}

We have used the SEST and JCMT telescopes to carry out a survey of CN 2--1, 
HCN 3--2 and HNC 3--2 line emission in a sample of 4 Seyfert galaxies, plus NGC~3079 
which was observed with the IRAM 30m telescope. The conclusions we draw are as follows:
\begin{list}{}{}
\item[1)] We detected HNC 3--2 emission in 3 of the 5 galaxies, while 
we obtain an upper limit for one of them (NGC~7469). HCN 3--2 was also detected 
in 3 galaxies (NGC~3079, NGC~1068 and NGC~1365), while it was not detected in 
NGC~2623. CN 2--1, along with the spingroups (\textit{J} = 5/2 -- 3/2, \textit{F} 
= 7/2 -- 5/2) and (\textit{J} = 3/2 -- 1/2, \textit{F} = 5/2 -- 3/2) was also 
detected in NGC~3079, NGC~1068 and NGC~1365.
\item[2)] {The line shapes} observed in NGC~1365 and NGC~3079 
suggests that there is no circumnuclear disk in these galaxies. 
\item[3)] We find that in 3 of the galaxies the HNC 3--2/1--0 line ratios 
suggest that the HNC emissions emerge from gas of densities $n\lesssim10^5~\3cm$, 
where the chemistry is dominated by ion-neutral reactions. In NGC~2623 a model 
of large masses of hidden cold (10 K) gas and dust, as well as a chemistry dominated 
by ion-neutral reactions, are yet to be distinguished as the correct interpretation 
for the bright HNC observed in this galaxy.
\item[4)] The 3--2/1--0 line ratios and the modelled excitation conditions imply 
that the HNC emission emerges from a more diffuse ($n<10^5~\3cm$) gas region 
than the HCN emission ($n>10^5~\3cm$) in NGC~1068, whereas they emerge from the 
same lower density ($n\lesssim10^5~\3cm$) gas in NGC~3079. 
\item[5)] The HCN/HNC and CN/HCN line ratios tentatively favor a PDR scenario, 
rather than an XDR one, in the 3 Seyfert galaxies where we have CN, HNC and HCN data.
The $N({\rm HNC})/N({\rm HCN})$ column density ratios obtained for NGC~3079 can be found
only in XDR environments.
\end{list}

In order to complete the sample, we plan to observe HCN 3--2 and CN 2--1 in 
NGC~7469, CN 2--1 in NGC~2623 and HNC 3--2 in NGC~1365. We plan to perform 
high resolution observations to further study the distribution and source 
sizes of CN and HNC.

Modeling of the collision data for the CN molecule would be useful to estimate the
$N({\rm CN})/N({\rm HCN})$ column density ratio, which would complement the 
$N({\rm HNC})/N({\rm HCN})$ ratio in order to have a more sophisticated tool 
to estimate and distinguish the prevalent environment conditions of the high density
gas in the nuclear region of Seyfert galaxies.

The AGN contribution (through XDR effects) is typically of a small angular scale and
can be seriously affected by beam dilution at the transition lines studied in this work.
On the other hand, the starburst contribution is of a larger angular scale than the AGN,
and it effects can be contaminating our observations, and hence leading to the favored
PDR scenario found with our models. Hence, our suggested interpretations could change if
we zoom in on these sources. Therefore, high resolution maps of HNC and CN molecules 
are necessary to complement those of HCN, and to do a more accurate estimate of molecular 
abundances and line intensity ratios, which take source size into account. Observations 
of the higher transition lines (e.g. $J$=4--3) can also aid to disentangle the effects of 
the AGN and the starburst ring, due to the smaller beam size obtained at higher frequencies.

\acknowledgements{We are grateful to the SEST staff for their help during the observing 
run. We thank S. Curran and A. Polatidis for help with some of the SEST observations. 
We are grateful to S. H\"uttemeister, M. Spaans, and R. Meijerink for discussions.
We also thank F. van der Tak and J. Black for their help and discussions about RADEX.
Molecular databases that have been helpful include LAMBDA and NIST.}


\end{document}